\documentclass{ArticleModern}

\usepackage{SymbolsModern}
\usepackage{StylesModern}

\usepackage{multicol}
\usepackage{abc} 
\usepackage[linesnumbered,ruled,vlined]{algorithm2e}
\usepackage{algorithmic}
\usepackage{pgfplots}
\pgfplotsset{compat=newest}

\newcommand{\TreeT}{\mathfrak{t}}

\newcommand{\Notes}{\mathcal{N}}
\newcommand{\DegPat}{\mathbf{d}}
\newcommand{\RhyPat}{\mathbf{r}}
\newcommand{\Pat}{\mathbf{p}}
\newcommand{\MPat}{\mathbf{m}}
\newcommand{\CPat}{\mathbf{c}}
\newcommand{\DegreeInterpretation}{\rho}
\newcommand{\RhythmInterpretation}{\delta}

\newcommand{\Multiplicity}{\mathfrak{m}}
\newcommand{\Mirror}{\mathrm{mir}}

\newcommand{\HadamardProduct}{\boxtimes}
\newcommand{\Id}{\mathrm{I}}

\newcommand{\ConstructionT}{\mathbf{T}}
\newcommand{\ConstructionU}{\mathbf{U}}
\newcommand{\MonoidProduct}{\star}
\newcommand{\MonoidUnit}{\mathrm{e}}
\newcommand{\MonoidProductExtension}{\operatorname{\BBar{\MonoidProduct}}}
\newcommand{\MonoidProductExtensionLeft}{\operatorname{\overleftarrow{\MonoidProduct}}}
\newcommand{\MonoidProductExtensionRight}{\operatorname{\overrightarrow{\MonoidProduct}}}
\newcommand{\MonoidCyclic}{\mathbb{C}}
\newcommand{\MonoidMax}{\mathbb{M}}
\newcommand{\OperadDP}{\mathsf{DP}}
\newcommand{\OperadRP}{\mathsf{RP}}
\newcommand{\OperadP}{\mathsf{P}}
\newcommand{\OperadMP}{\OperadP}
\newcommand{\BudOperadMP}{\mathsf{B}\OperadMP}

\newcommand{\MorphismMul}{\mathrm{mul}}
\newcommand{\MorphismRed}{\mathrm{red}}
\newcommand{\MorphismIncr}{\mathrm{incr}}
\newcommand{\MorphismDil}{\mathrm{dil}}
\newcommand{\MorphismCopy}{\mathrm{cp}}
\newcommand{\DegPatToPat}[1]{\mathrm{dp}_{#1}}
\newcommand{\RhyPatToPat}[1]{\mathrm{rp}_{#1}}

\newcommand{\SetColors}{\mathfrak{C}}
\newcommand{\Color}{\mathtt{b}}
\newcommand{\ColoredOperad}{\mathcal{C}}
\newcommand{\Out}{\mathrm{out}}
\newcommand{\In}{\mathrm{in}}
\newcommand{\BudOperad}{\mathsf{B}}
\newcommand{\HomogeneousComposition}{\odot}
\newcommand{\SetRules}{\mathcal{R}}
\newcommand{\BudSystem}{\mathcal{B}}
\newcommand{\PartialDerivation}{\xrightarrow{\circ_i}}
\newcommand{\FullDerivation}{\xrightarrow{\circ}}
\newcommand{\HomogeneousDerivation}{\xrightarrow{\HomogeneousComposition}}
\newcommand{\Prune}{\mathrm{pr}}
\newcommand{\Random}{\textsc{random}}

\newcommand{\Concatenation}{\mathrm{conc}}
\newcommand{\Repetition}{\mathrm{rep}}
\newcommand{\Transposition}{\mathrm{tran}}
\newcommand{\Temporization}{\mathrm{temp}}
\newcommand{\Inverse}{\mathrm{inv}}
\newcommand{\RetrogradeInverse}{\mathrm{minv}}
\newcommand{\Harmonization}{\mathrm{har}}
\newcommand{\Arpeggiation}{\mathrm{arp}}

\newcommand{\BudSystemMix}{\BudSystem^\mathrm{mix}}
\newcommand{\BudSystemHorizontal}{\BudSystem^\mathrm{hor}}
\newcommand{\BudSystemVertical}{\BudSystem^\mathrm{ver}}
\newcommand{\BudSystemVariation}{\BudSystem^\mathrm{var}}

\DeclareMathOperator{\Rest}{
\raisebox{1pt}{%
\begin{tikzpicture}[Centering,scale=.16]
    \node[rectangle,draw=ColB!70,fill=ColB!2,thick,inner sep=1mm,minimum size=1.75mm]
        (0,0){};
\end{tikzpicture}}}

\DeclareMathOperator{\Beat}{
\raisebox{1pt}{%
\begin{tikzpicture}[Centering,scale=.16]
    \node[rectangle,draw=ColA!120,fill=ColA!60,thick,inner sep=1mm,minimum size=1.75mm]
        (0,0){};
\end{tikzpicture}}}

\newenvironment{MultiPattern}{%
    
    \setlength\arraycolsep{.2em}
    \begin{footnotesize}
        \begin{vmatrix}
        }{
        \end{vmatrix}
    \end{footnotesize}}

\newcommand{\Graph}[2]{%
    \noindent%
    \fcolorbox{ColA}{ColA!10}{%
    \begin{Page}{.45}%
        \vspace{0.5ex}
        \begin{center} \includegraphics[scale=0.45]{#1.png} \end{center}
        \begin{center} \footnotesize #2\end{center}
    \end{Page}}%
}

\tikzstyle{Circle}=[circle,draw=ColA!60,fill=ColA!6,inner sep=0,minimum size=14mm,thick,
font=\scriptsize]

\Title{%
    The music box operad: Random generation of musical phrases from patterns \vspace{-1ex}
}
\TitleShort{Music generation by operads}
\Author{Samuele Giraudo \vspace{-1ex}}
\AuthorShort{S. Giraudo}
\Subjects{00A65, 18M60, 68Q42}
\Date{\today}
\Address{%
    Université du Québec à Montréal, LACIM, \\
    Pavillon Président-Kennedy, 201 Avenue du Président-Kennedy, Montréal, H2X~3Y7, Canada.
    \vspace{-3ex}
}
\Email{giraudo.samuele@uqam.ca}
\Funding{None.}
\Abstract{%
    \footnotesize
    We introduce the notion of multi-patterns, a combinatorial abstraction of polyphonic
    musical phrases. The interest of this approach in encoding musical phrases lies in the
    fact that it becomes possible to compose multi-patterns in order to produce new ones.
    This composition is parameterized by a monoid structure on the scale degrees. This
    embeds the set of the musical phrases into an algebraic framework since the set of the
    multi-patterns is endowed with the structure of an operad. Operads are algebraic
    structures offering a formalization and an abstraction of the notion of operators and
    their compositions. Seeing musical phrases as operators allows us to perform
    computations on phrases and admits applications in generative music. Indeed, given a set
    of initial multi-patterns, we propose various algorithms to randomly generate a new and
    longer phrase emulating the style suggested by the inputted multi-patterns. The designed
    algorithms use types of grammars working with operads and colored operads, known as bud
    generating systems.
}

\begin{document}

\MakeFirstPage

\section{Introduction}
Generative music is a subfield of computational musicology in which the focus lies on the
automatic creation of musical material~\cite{DJ00,Col08,FV13}. This creation is based on
algorithms accepting inputs to influence the result obtained. Such algorithms could have a
randomized behavior in the sense that two executions of the algorithm with the same inputs
produce different results. One of the challenges to overcome in designing such an algorithm
consists in building a procedure offering a balance between freedom of what is created and
adequacy with the specified inputs. As a matter of fact, the expression {\em generative
music} is usually understood as a class of methods producing music by rules, with a possibly
deterministic behavior~\cite{LA85}.

Several very different approach exist, each with their own advantages and areas of
applications. For instance, some algorithms use Markov chains~\cite{Ame89,SH21}, others
genetic algorithms~\cite{GJC03,Mat10}, still others neural networks~\cite{BHP20}, or even
---and directly related to the present work--- formal
grammars~\cite{Roa79,Hol81,Ste84,Bel89,LJ96,McC96, GSGCS11,QH13,Eib18, HQ18}
(see~\cite{HMU06} for a general presentation of formal grammars). In the case of the Markov
chain approach, the input is a corpus of musical pieces. The algorithm essentially builds a
Markov chain for the probabilities of a note to be played according to some previous ones,
and then, uses it to build randomly a new musical piece which inherits in some sense from
the inputted ones. The genetic algorithm approach is based on a fitness function, a crucial
object to evaluate the quality of the generated phrases and make sure that, over the
iterations, the output converges to a satisfying one. Neural networks produce potentially
very interesting results provided that an adequate learning phase is carried out.
Nevertheless, the inherent disadvantage of this method is that it is a black box whose
internal mechanisms are difficult to understand. It is therefore in most of the cases hard
for the human composer using the computer as a source of inspiration to slightly modify the
generation parameters in order to direct the algorithm to generate phrases of a certain
kind. In the case of the formal grammar approach, the input data is a formal grammar
specifying the language of the possible results. The algorithm builds a musical piece by
performing a random generation of a word of the language specified by the grammar by
selecting, using various strategies, a rule of the grammar at random at each step.

The way in which such algorithms represent and manipulate musical data is crucial. Indeed,
the data structures used to represent musical phrases orient the nature of the operations we
can define on them. Operations that produce new phrases from old ones are an important
ingredient in specifying algorithms similar to those that manipulate formal grammars to
randomly generate music. A possible way for this purpose consists in giving at input some
musical phrases and the algorithm creates a new one by blending them through operations.
Therefore, the willingness to endow the infinite set of the musical phrases with operations
in order to obtain suitable algebraic structures is a promising approach. Such interactions
between music and algebra form a fruitful field of
investigation~\cite{War45,Hud04,Tym11,And18,Jed19}. For instance, in~\cite{Hud04} (see
also~\cite{HQ18}), musical phrases are encoded as treelike structures wherein phrases are
associated by concatenation or by superposition. In~\cite{Mor87}, an abstraction of phrases
is proposed wherein pitch classes are used instead of absolute notes. In~\cite{Eib18}, a
phrase is a sequence of tuples containing data including a pitch, a duration, and a
volume. Moreover, \cite{CCI01} reviews a wide range of other data structures.

In the present work, we propose to use tools coming from combinatorics and algebraic
combinatorics to represent musical phrases and operations on them, in order to introduce
generative music algorithms close to the family of those based on formal grammars. More
precisely, we present the music box model, a model to represent polyphonic phrases by
combinatorial objects called multi-patterns. Here, a ``multi-pattern'' (and its monophonic
version, a ``pattern'') should not be understood as, in the usual way, a fragment of a piece
admitting multiples occurrences of it. This terminology of multi-pattern and pattern should
be understood as follows: a multi-pattern is rather an abstraction of a non-fixed musical
phrase (that is, the same pattern can give rise to different musical phrases when it is seen
under different interpretations) which can be combined with others to form bigger phrases.
The music box model is designed to handle musical phrases written for the family of
lamellophones. Instruments of this family, like the kalimba, are capable of striking notes
whose duration is determined by how long they resonate, rather than being deliberately
chosen. This is precisely the musical framework of application of the tools designed in this
work.

The infinite set of the multi-patterns admits the structure of an operad. This specific
property is the starting point of this work. Operads originate from algebraic
topology~\cite{May72,BV73} and are used nowadays also in algebraic combinatorics and in
computer science~\cite{LV12,Men15,Gir18}. Roughly speaking, in these structures, the
elements are operations with several inputs and one output, and the composition law mimics
the usual composition of such operators. Since the set of multi-patterns forms an operad,
one can regard each multi-pattern as an operation. The fallout of this is that each
multi-pattern is, at the same time, a musical phrase (under some interpretation) and an
operation acting on musical phrases. In this way, our music box model and its associated
operad provide an algebraic and combinatorial framework to perform computations on musical
phrases. To the best of our knowledge, the present work is the first one to build a bridge
between operad theory and generative music.

The music box model and the constructed operads on multi-patterns admit direct applications
to design random generation algorithms since, as introduced by the author in~\cite{Gir19},
given an operad there exist algorithms to generate some of its elements. These algorithms
are based upon bud generating systems, which are general formal grammars based on colored
operads~\cite{Yau16}. In the present work, we introduce three different variations of these
algorithms to produce new musical phrases from old ones. Each algorithm works more or less
as follows. It takes as input a finite set of multi-patterns and an integer value to
influence the size of the output. It chooses iteratively a multi-pattern from the initial
collection in order to alter the current one by performing a composition using the operad
structure. As we shall explain, the initial multi-patterns are endowed with colors in order
to forbid certain compositions and, in this way, avoid specific musical intervals,
particular rhythmic motives, or impose a certain general structure. These colors play a
similar role to the one of nonterminal symbols of formal grammars. Even though this is
feasible, these generation algorithms are not intended to write complete musical pieces;
rather, they aim to generate a new, longer, and similar pattern from short primitive
multi-patterns, potentially bringing new ideas to the human composer. These algorithms can
be used to generate phrases satisfying rules similar to the ones of species counterpoint
(see~\cite{KS15} for the description of an automated composition method in this context), or
to create a piece by emulating and mixing the style of some input phrases.

This contribution presents some new interactions between music and mathematics, and a new
framework for automatic composition. Our method is not supposed to be better than the other
existing ones and has it is own advantages and disadvantages. As main features, the model
allows us to compute over musical phrases since, as mentioned before, each multi-pattern is
in fact an operation which admits as inputs other multi-patterns and produces a new
multi-pattern. Such operations are parameterizable by specifying how the scale degrees are
transformed by providing a monoid structure on the scale degrees (in other words, the operad
of the multi-patterns is parameterized by a monoid). A feature of the model is that we can
express some musical transformations (like transpositions, inversion, retrograde, {\em
etc.}) either as multi-patterns seen like operations or through operad morphisms. In all the
cases, these operations are defined by the same language as the one used by the music box
model. Besides, according to what is intended, it is possible to specify different monoids
in order to change the behavior of the operations and also of the generated patterns. The
other main features and limitations of the model are discussed in more details further.

This text is organized as follows. Section~\ref{sec:music_box_model} is devoted to present
the music box model. Degree patterns, rhythm patterns, patterns, and multi-patterns are
defined. We explain how a multi-pattern describes a musical phrase given an interpretation.
We discuss also the strengths and the weaknesses of this model. In
Section~\ref{sec:operads}, we begin by presenting a brief overview of operad theory and we
build step by step the music box operad. For this, we introduce first an operad on sequences
of scale degrees (depending as explained above on a monoid structure in order to encode how
to compute on degrees), an operad on rhythm patterns, and then an operad on monophonic
patterns to finish with the operad of multi-patterns. After providing some background on
colored operads and bud generating systems, three random generation algorithms for
multi-patterns are introduced in Section~\ref{sec:random_generation}.
Section~\ref{sec:applications} provides some concrete applications of the previous
algorithms. We first describe an implementation~\cite{Gir23} of the music box model (ready
to use by the reader) and the related algorithms. Then, we construct four particular bud
generating systems, each with a specific objective. The paper ends by presenting in
Section~\ref{sec:evaluation} an evaluation of the generation algorithms. This evaluation is
based on a questionnaire aimed at experts, intended to understand how pieces generated by
our methods are received.

This paper is an extended version of~\cite{Gir20} containing some new results (like the
complete study of the introduced operads) and their proofs. Moreover, we describe in this
present version a more general model than the one presented in the previous work: now, the
scale degrees of the multi-patterns are elements of a monoid.

\paragraph{General notations and conventions}
For any integers $i$ and $j$, $[i, j]$ denotes the set $\{i, i + 1, \dots, j\}$. For any
integer $i$, $[i]$ denotes the set $[1, i]$. A word is a finite sequence of elements. For
any set $A$, $A^*$ is the set of the words on $A$. Given a word $u$, $\Length(u)$ is the
length of $u$, and for any $i \in [\Length(u)]$, $u(i)$ is the $i$-th letter of $u$. For any
$i \leq j \in [\Length(u)]$, $u(i, j)$ is the word $u(i) u(i + 1) \dots u(j)$. If $a$ is a
letter and $n$ is a nonnegative integer, $a^n$ is the word consisting of $n$ occurrences of
$a$. In particular, $a^0$ is the empty word~$\epsilon$. If $u$ and $u'$ are two words, their
concatenation is the word denoted by $u \Conc u'$ or simply by $u u'$ when the context is
clear.

\section{The music box model} \label{sec:music_box_model}
The purpose of this section is to introduce multi-patterns, the main combinatorial objects
used in this work to propose abstractions of musical phrases. This encoding of musical
phrases forms the music box model.

\subsection{Patterns and multi-patterns}
We introduce here degree patterns, rhythm patterns, patterns, and finally multi-patterns
which are the components of music box model, described in the next section.

\subsubsection{Degree patterns} \label{subsubsec:degree_patterns}
A \Def{scale degree} (or, for short, a \Def{degree}) is any element of $\Z$.  Negative
degrees are denoted by putting a bar above their absolute value. For instance, the degree
$-3$ is denoted by $\bar{3}$. A \Def{degree pattern} is a word $\DegPat$ of degrees. The
\Def{arity} of $\DegPat$, also denoted by $|\DegPat|$, is the length of $\DegPat$ as a word.
For instance, $\DegPat := 0 \bar{1} 2 1 0$ is a degree pattern of arity~$5$.

\subsubsection{Rhythm patterns} \label{subsubsec:rhythm_patterns}
A \Def{rhythm pattern} $\RhyPat$ is a word on the alphabet $\Bra{\Rest, \Beat}$. The symbol
$\Rest$ is a \Def{rest} and the symbol $\Beat$ is a \Def{beat}. The \Def{length}
$\Length(\RhyPat)$ of $\RhyPat$ is the length of $\RhyPat$ as a word and the \Def{arity}
$|\RhyPat|$ of $\RhyPat$ is its number of occurrences of beats. The \Def{duration sequence}
of $\RhyPat$ is the unique word $\sigma$ of nonnegative integers and of length
$|\RhyPat| + 1$ such that
\begin{equation}
    \RhyPat
    =
    \Rest^{\sigma(1)}
    \; \Beat \;
    \Rest^{\sigma(2)}
    \; \dots \; \Beat \;
    \Rest^{\sigma\Par{|\RhyPat| + 1}}.
\end{equation}
For instance, $\RhyPat := \Rest \Rest \Beat \Rest \Beat \Beat \Rest$ is a rhythm pattern of
length $7$, arity~$3$, and of duration sequence~$2101$.

\subsubsection{Patterns} \label{subsubsec:patterns}
A \Def{pattern} is a pair $\Pat := (\DegPat, \RhyPat)$ such that $|\DegPat| = |\RhyPat|$.
The \Def{arity} $|\Pat|$ of $\Pat$ is the common arity of both $\DegPat$ and $\RhyPat$, and
the \Def{length} $\Length(\Pat)$ of $\Pat$ is the length $\Length(\RhyPat)$ of~$\RhyPat$.
For instance, $\Pat := \Par{1 \bar{1} 2, \Beat \Beat \Rest \Rest \Rest \Beat \Rest}$ is a
pattern of arity $3$ and of length~$7$.

In order to handle concise notations, we shall write any pattern $(\DegPat, \RhyPat)$ as a
word $\Pat$ on the alphabet $\Bra{\Rest} \cup \Z$ where the subword of $\Pat$ obtained by
removing all occurrences of $\Rest$ is the degree pattern $\DegPat$, and the word obtained
by replacing in $\Pat$ each integer by $\Beat$ is the rhythm pattern $\RhyPat$. For
instance,
\begin{equation} \label{equ:example_pattern}
    1 \Rest \Rest \bar{2} \Rest 1 2
\end{equation}
is the concise notation for the pattern
\begin{equation}
    \Par{1 \bar{2} 1 2, \Beat \Rest \Rest \Beat \Rest \Beat \Beat}.
\end{equation}
By following this convention, in the sequel we shall see and treat any pattern $\Pat$ as a
word on the alphabet $\Bra{\Rest} \cup \Z$. Therefore, for any $i \in [\Length(\Pat)]$,
$\Pat(i)$ is the $i$-th letter of $\Pat$ which is either $\Rest$ or a degree. Remark that
the length of $\Pat$ is the length of $\Pat$ as a word and that the arity of $\Pat$ is the
number of letters of $\Z$ it has.

\subsubsection{Multi-patterns} \label{subsubsec:multi_patterns}
A \Def{multi-pattern} $\MPat$ is a word of length $m \geq 1$ of patterns such that all
patterns $\MPat(i)$, $i \in [m]$, have the same arity and the same length. The \Def{arity}
$|\MPat|$ of $\MPat$ is the common arity of all the $\MPat(i)$, the \Def{length}
$\Length(\MPat)$ of $\MPat$ is the common length of all the $\MPat(i)$, and the
\Def{multiplicity} $\Multiplicity(\MPat)$ of $\MPat$ is~$m$. The \Def{stacking} of two
multi-patterns $\MPat_1$ and $\MPat_2$ having the same arity and the same length (but
possibly different multiplicities), is the multi-pattern $\MPat_1 \Conc \MPat_2$. This
multi-pattern has the same arity and length as the ones of $\MPat_1$ and $\MPat_2$, and its
multiplicity is $\Multiplicity\Par{\MPat_1} + \Multiplicity\Par{\MPat_2}$.

A multi-pattern $\MPat$ is represented by a matrix of dimension $\Multiplicity(\MPat) \times
\Length(\MPat)$, where the $i$-th row contains the pattern $\MPat(i)$ for any $i \in
[\Multiplicity(\MPat)]$. For any $i \in [\Multiplicity(\MPat)]$ and any $j \in
[\Length(\MPat)]$, $\MPat(i)(j)$ is hence the letter $\Pat(j)$ where $\Pat$ is the pattern
$\MPat(i)$. For instance,
\begin{equation}
    \MPat :=
    \begin{MultiPattern}
        0 & \Rest & 1 & \Rest & 1 \\
        \Rest & \bar{2} & \bar{3} & \Rest & 0
    \end{MultiPattern}
\end{equation}
is a multi-pattern of multiplicity $2$, having $3$ as arity, and $5$ as length. It satisfies
$\MPat(1)(1) = 0$, $\MPat(1)(4) = \Rest$, and $\MPat(2)(3) = \bar{3}$.

In the sequel, for obvious reasons of practicality, we shall consider that a pattern is a
multi-pattern of multiplicity $1$ and conversely.

\subsection{Interpretations}
Let us now explain how to convert multi-patterns into musical phrases. We introduce also the
music box model and discuss some of its drawbacks and advantages.

\subsubsection{From multi-patterns to musical phrases}
Let $\Notes$ be any set of musical notes. A \Def{degree interpretation} is a map
$\DegreeInterpretation : \Z \to \Notes$. A \Def{rhythm interpretation} is a duration
$\RhythmInterpretation$. An \Def{interpretation} is a pair $(\DegreeInterpretation,
\RhythmInterpretation)$ where $\DegreeInterpretation$ is a degree interpretation and
$\RhythmInterpretation$ is a rhythm interpretation. The \Def{$(\DegreeInterpretation,
\RhythmInterpretation)$-interpretation} of a pattern $\Pat$ is the musical phrase composed
of the sequence of rests and notes obtained by translating each rest of $\Pat$ as a rest and
each degree $d$ of $\Pat$ as the note $\DegreeInterpretation(d)$, both lasting the duration
prescribed by $\RhythmInterpretation$. The \Def{$(\DegreeInterpretation,
\RhythmInterpretation)$-interpretation} of a multi-pattern $\MPat$ is the musical phrase
obtained by superimposing the $(\DegreeInterpretation,
\RhythmInterpretation)$-interpretations of each pattern $\MPat(i)$, $i \in
\Han{\Multiplicity(\MPat)}$.

\subsubsection{Standard interpretation}
The \Def{standard interpretation} is the interpretation $(\DegreeInterpretation,
\RhythmInterpretation)$ where $\DegreeInterpretation$ sends the degree $2$ to the ``middle
C'' and the other degrees accordingly with the diatonic scale (for instance, the degree $0$
is interpreted as the note A located three semitones below the middle C, the degree $9$ is
interpreted as the note C located one octave above the middle C, and the degree $\bar{1}$
is interpreted as the note G located five semitones below the middle C) and
$\RhythmInterpretation$ is the duration of $\frac{1}{2}$ seconds.

For instance, the multi-pattern
\begin{equation}
    \MPat :=
    \begin{MultiPattern}
        0 & \Rest & 1 & 2 & 1 & \Rest & 1 & \Rest & 2 & 3 & 2 & \Rest \\
        \Rest & 2 & 3 & 4 & 3 & \Rest & 3 & 4 & 5 & 4 & \Rest & \Rest \\
        \bar{7} & \Rest & \bar{6} & \bar{5} & \bar{6} & \Rest & \bar{6} & \Rest & \bar{5}
            & \bar{4} & \bar{5} & \Rest
    \end{MultiPattern}
\end{equation}
is interpreted through the standard interpretation as the musical phrase
\begin{abc}[name=ExampleInterpretation,width=.85\abcwidth]
X:
T:
C:
M: 4/4
K: Am
Q: 1/4=120
V: 1 clef=treble
L: 1/4
A, z B, C B, z B, z C D C z
V: 2 clef=treble
L: 1/4
z C D E D z D E F E z z
V: 3 clef=bass
L: 1/4
A,, z B,, C, B,, z B,, z C, D, C, z
\end{abc}

Unless otherwise stated, all the next multi-patterns of this paper are interpreted through
the standard interpretation.

\subsubsection{Music box model}
Due to the fact that multi-patterns evoke paper tapes of a programmable music box, we call
the model just described the \Def{music box model} to represent musical phrases by
multi-patterns within the context of interpretations. Interpretations of this model are
musical phrases that may be particularly suitable to be played by the family of
lamellophones (including the mbira and the kalimba) or keyboard percussive instruments
(including the marimba and the xylophone).

\subsubsection{Limitations of the Music box model}
\label{subsubsec:limitations_music_box_model}
Despite its simplicity, this model suffers from at least the following three main
apparent limitations:
\begin{enumerate}
    \item all patterns of a multi-pattern must have the same length;
    \item all patterns of a multi-pattern must have the same arity;
    \item the interpretation of a multi-pattern uses only rests and notes all having the
    same duration.
\end{enumerate}

The first limitation, which requires that all patterns within a multi-pattern have the same
length, ensures that all patterns of a multi-pattern have the same total duration. This
limitation can be lifted by adding final rests to the shorter patterns within a
multi-pattern so that they have virtually the same length as the longest.

The second limitation, which requires that all patterns within a multi-pattern have the same
arity, is a particularity of our model and comes from algebraic reasons which will be
clarified later in the text. Here this limitation can be lifted by replacing, in a pattern
with a smaller arity, some of its rests by degrees of any other pattern of the
multi-pattern. For instance, consider the phrase
\begin{abc}[name=ExampleArityCompletion1,width=.40\abcwidth]
X:
T:
C:
M: 4/4
K: Am
Q: 1/4=120
V: 1 clef=treble
L: 1/4
A, C E z A, c E
V: 2 clef=bass
L: 1/4
z A, z A,, z z z
\end{abc}
As it stands, no multi-pattern interprets into this phrase because the first voice is
composed of six quarter notes while the second of only two. Nevertheless, as we have just
explained, the multi-pattern
\begin{equation}
    \begin{MultiPattern}
        0 & 2 & 4 & \Rest & 0 & 9 & 4 \\
        0 & 0 & 4 & \bar{7} & 0 & \Rest & 4
    \end{MultiPattern}
\end{equation}
interprets as
\begin{abc}[name=ExampleArityCompletion2,width=.40\abcwidth]
X:
T:
C:
M: 4/4
K: Am
Q: 1/4=120
V: 1 clef=treble
L: 1/4
A, C E z A, c E
V: 2 clef=bass
L: 1/4
A, A, E A,, A, z E
\end{abc}
and this phrase sounds like the previous one.

The last presented limitation, which restricts the interpretation of multi-patterns to rests
and notes of the same duration, can be lifted through different techniques. One approach is
to make each note of the interpretation of a multi-pattern continue to sound until the next
note of the same pattern begins. This means for instance that if a degree is followed by
three rests, this degree and the three rests are interpreted as a note of duration $1 + 3$
times the duration prescribed by $\RhythmInterpretation$, where $\RhythmInterpretation$ is a
rhythm interpretation. A more sophisticated solution involves using a map $\tau : \N
\setminus \{0\} \to \N^2$ that assigns any $\alpha \in \N \setminus \{0\}$ to a pair
$\Par{\alpha_1, \alpha_2}$ where $\alpha = \alpha_1 + \alpha_2$. This approach interprets
any degree followed by $\alpha - 1$ rests as a note lasting $\alpha_1$ times the duration
prescribed by $\delta$ and then a rest lasting $\alpha_2$ times the duration prescribed by
$\delta$, where $\tau(\alpha) = \Par{\alpha_1, \alpha_2}$. For instance, by considering such
a map $\tau$ satisfying $\tau(1) = (1, 0)$, $\tau(2) = (1, 1)$, $\tau(3) = (2, 1)$, and
$\tau(4) = (4, 0)$, the multi-pattern
\begin{equation}
    \begin{MultiPattern}
        2 & \Rest & 1 & \Rest & \Rest & 0 & 0 & \Rest \\
        4 & \Rest & \Rest & \Rest & 2 & 2 & \Rest & 7
    \end{MultiPattern}
\end{equation}
is interpreted through the standard interpretation as the musical phrase
\begin{abc}[name=ExampleSophisticatedDurations,width=.40\abcwidth]
X:
T:
C:
M: 4/4
K: Am
Q: 1/4=120
V: 1 clef=treble
L: 1/4
C z B2 z A, A, z
V: 2 clef=treble
L: 1/4
E4 C C z C'
\end{abc}

\subsubsection{Strengths of the music box model} \label{subsubsec:strengths_music_box_model}
Let us now discuss some of the main strengths of the music box model. Multi-patterns have
the following features:
\begin{enumerate}
    \item the translation of multi-patterns to musical phrases through interpretations is
    almost transparent and is flexible;
    \item more than just representing musical phrases, multi-patterns are in fact operations
    on musical phrases.
\end{enumerate}

First point is obvious: with a little training, and with an interpretation in mind, it is
easy to imagine how a multi-pattern sounds. Moreover, this representation is flexible in the
sense that a multi-pattern does not specify a fixed sequence of rests and notes but rather a
scheme (whence the appellation of ``pattern'' and ``multi-pattern''). For instance, the
pattern
\begin{math}
    \begin{MultiPattern}
        0 & 2 & 4 & 2 & 4
    \end{MultiPattern}
\end{math}
interprets as a minor arpeggio in the standard interpretation but interprets also as a major
arpeggio in any major scale, among other possibilities.

The second point will be clarified in the sequel. It basically says that multi-patterns are
not limited to simply representing musical phrases since they are in fact operations on
musical phrases. This feature is perhaps for us the most compelling aspect of the model as
it allows for the construction of larger multi-patterns by using a natural notion of
composition. As we shall see, this composition is also highly flexible, as it can be
parameterized according to the specific needs and desired musical outcomes.

\section{Operad structures} \label{sec:operads}
The purpose of this section is to introduce an operad structure on multi-patterns, called
music box operad. The main interest of endowing the set of multi-patterns with the structure
of an operad is that this leads to an algebraic framework to perform parameterizable
computations on musical phrases.

\subsection{A primer on operads}
We begin by setting here the elementary notions of operad theory used in the sequel. Most of
them come from~\cite{Gir18}.

\subsubsection{Graded sets}
A \Def{graded set} is a set $\Operad$ decomposing as a disjoint union
\begin{equation}
    \Operad := \bigsqcup_{n \in \N} \Operad(n),
\end{equation}
where the $\Operad(n)$, $n \in \N$, are sets. For any $x \in \Operad$, there is by
definition a unique $n \in \N$ such that $x \in \Operad(n)$. This integer $n$ is the
\Def{arity} of $x$ and is denoted by~$|x|$. Let $\Operad'$ be a second graded set. A map
$\phi : \Operad \to \Operad'$ is a \Def{graded set morphism} if for any $x \in \Operad$,
$\phi(x) \in \Operad'(|x|)$. The identity graded set morphism is denoted by $\Id$. Besides,
$\Operad'$ is a \Def{graded subset} of $\Operad$ if $\Operad'(n) \subseteq \Operad(n)$ for
any $n \in \N$.

The \Def{Hadamard product} of two graded sets $\Operad$ and $\Operad'$ is the graded set
$\Operad \HadamardProduct \Operad'$ defined, for any $n \in \N$, by $\Par{\Operad
\HadamardProduct \Operad'}(n) := \Operad(n) \times \Operad'(n)$. Observe that if $\Operad''$
is another graded set, the graded sets $\Par{\Operad \HadamardProduct \Operad'}
\HadamardProduct \Operad''$ and $\Operad \HadamardProduct \Par{\Operad' \HadamardProduct
\Operad''}$ are isomorphic, so that $\HadamardProduct$ is an associative operation. By a
slight abuse of notation, for any other graded sets $\Operad_1$ and $\Operad_1'$, and any
graded set morphisms $\phi : \Operad \to \Operad_1$ and $\phi' : \Operad' \to \Operad_1'$,
we denote by $\phi \HadamardProduct \phi'$ the map from $\Operad \HadamardProduct \Operad'$
to $\Operad_1 \HadamardProduct \Operad_1'$ defined, for any $\Par{x, x'} \in \Operad
\HadamardProduct \Operad'$, by
\begin{math}
    \Par{\phi \HadamardProduct \phi'}\Par{\Par{x, x'}} := \Par{\phi(x), \phi'\Par{x'}}.
\end{math}
This map is a graded set morphism.

\subsubsection{Nonsymmetric operads}
A \Def{nonsymmetric operad}, or an \Def{operad} for short, is a triple $\Par{\Operad,
\circ_i, \Unit}$ such that $\Operad$ is a graded set, $\circ_i$ is a map
\begin{equation}
    \circ_i : \Operad(n) \times \Operad(m) \to \Operad(n + m - 1),
    \qquad 1 \leq i \leq n,
\end{equation}
called \Def{partial composition} map, and $\Unit$ is a distinguished element of
$\Operad(1)$, called \Def{unit}. This data has to satisfy, for any $x, y, z \in \Operad$,
the three relations
\begin{subequations}
\begin{equation} \label{equ:operad_axiom_1}
    \Par{x \circ_i y} \circ_{i + j - 1} z = x \circ_i \Par{y \circ_j z},
    \qquad 1 \leq i \leq |x|, \enspace 1 \leq j \leq |y|,
\end{equation}
\begin{equation} \label{equ:operad_axiom_2}
    \Par{x \circ_i y} \circ_{j + |y| - 1} z = \Par{x \circ_j z} \circ_i y,
    \qquad 1 \leq i < j \leq |x|,
\end{equation}
\begin{equation} \label{equ:operad_axiom_3}
    \Unit \circ_1 x = x = x \circ_i \Unit,
    \qquad 1 \leq i \leq |x|.
\end{equation}
\end{subequations}

\subsubsection{Abstract operators and intuition} \label{subsubsec:abstract_operators}
From an intuitive point of view, an operad is an algebraic structure wherein each element
$x$ can be seen as an operator having $|x|$ inputs and one output. Such an operator is
depicted as
\begin{equation}
    \begin{tikzpicture}
        [Centering,xscale=.25,yscale=.3,font=\scriptsize]
        \node[NodeST](x)at(0,0){$x$};
        \node(r)at(0,2){};
        \node(x1)at(-3,-2){};
        \node(xn)at(3,-2){};
        \node[below of=x1,node distance=1mm](ex1){$1$};
        \node[below of=xn,node distance=1mm](exn){$|x|$};
        \draw[Edge](r)--(x);
        \draw[Edge](x)--(x1);
        \draw[Edge](x)--(xn);
        \node[below of=x,node distance=7mm]{$\dots$};
    \end{tikzpicture}
\end{equation}
where the inputs are at the bottom and numbered from $1$ to $|x|$, and the output is at the
top. Given two operations $x$ and $y$ of $\Operad$, the partial composition $x \circ_i y$ is
a new operator obtained by composing $y$ onto the $i$-th input of $x$. Pictorially, this
partial composition expresses as
\begin{equation} \label{equ:partial_compostion_on_operators}
    \begin{tikzpicture}
        [Centering,xscale=.25,yscale=.3,font=\scriptsize]
        \node[NodeST](x)at(0,0){$x$};
        \node(r)at(0,2){};
        \node(x1)at(-3,-2){};
        \node(xn)at(3,-2){};
        \node(xi)at(0,-2){};
        \node[below of=x1,node distance=1mm](ex1){$1$};
        \node[below of=xn,node distance=1mm](exn){$|x|$};
        \node[below of=xi,node distance=1mm](exi){$i$};
        \draw[Edge](r)--(x);
        \draw[Edge](x)--(x1);
        \draw[Edge](x)--(xn);
        \draw[Edge](x)--(xi);
        \node[right of=ex1,node distance=4mm]{$\dots$};
        \node[left of=exn,node distance=4mm]{$\dots$};
    \end{tikzpicture}
    \enspace \circ_i \enspace
    \begin{tikzpicture}
        [Centering,xscale=.25,yscale=.3,font=\scriptsize]
        \node[NodeST](x)at(0,0){$y$};
        \node(r)at(0,2){};
        \node(x1)at(-3,-2){};
        \node(xn)at(3,-2){};
        \node[below of=x1,node distance=1mm](ex1){$1$};
        \node[below of=xn,node distance=1mm](exn){$|y|$};
        \draw[Edge](r)--(x);
        \draw[Edge](x)--(x1);
        \draw[Edge](x)--(xn);
        \node[below of=x,node distance=7mm]{$\dots$};
    \end{tikzpicture}
    \enspace = \enspace
    \begin{tikzpicture}
        [Centering,xscale=.5,yscale=.4,font=\scriptsize]
        \node[NodeST](x)at(0,0){$x$};
        \node(r)at(0,1.5){};
        \node(x1)at(-3,-2){};
        \node(xn)at(3,-2){};
        \node[below of=x1,node distance=1mm](ex1){$1$};
        \node[below of=xn,node distance=1mm](exn){$|x| + |y| - 1$};
        \node[right of=ex1,node distance=8mm]{$\dots$};
        \node[left of=exn,node distance=9mm]{$\dots$};
        \draw[Edge](r)--(x);
        \draw[Edge](x)--(x1);
        \draw[Edge](x)--(xn);
        \node[NodeST](y)at(0,-2.5){$y$};
        \node(y1)at(-1.6,-4.5){};
        \node(yn)at(1.6,-4.5){};
        \node[below of=y1,node distance=1mm](ey1){$i$};
        \node[below of=yn,node distance=1mm](eyn){$i + |y| - 1$};
        \draw[Edge](y)--(y1);
        \draw[Edge](y)--(yn);
        \node[below of=y,node distance=9mm]{$\dots$};
        \draw[Edge](x)--(y);
    \end{tikzpicture}.
\end{equation}
Relations~\eqref{equ:operad_axiom_1}, \eqref{equ:operad_axiom_2},
and~\eqref{equ:operad_axiom_3} become clear when they are interpreted into this context of
abstract operators and rooted trees (also called syntax trees).

In a complementary manner, an interesting way to see the elements of an operad $\Operad$
consists in regarding any $x \in \Operad(n)$, $n \in \N$, as a combinatorial object with
$n$ substitute sectors labeled from $1$ to $n$. The partial composition $x \circ_i y$
consists in replacing the substitute sector of $x$ having $i$ as label by $y$ and by
shifting the labels accordingly. For instance, here is a schematic composition of an element
of arity $3$ at the $2$-nd position inside an element of arity~$5$:
\begin{equation*}
    \begin{tikzpicture}[scale=.25,Centering]
        \node[Circle,minimum size=19mm](x)at(0,0){};
        \node[Circle,minimum size=4mm]at(-1.75,-1){$1$};
        \node[Circle,MarkB,minimum size=4mm]at(1.5,-1){$2$};
        \node[Circle,minimum size=4mm]at(.5,2.5){$3$};
        \node[Circle,minimum size=4mm]at(-1.75,1.5){$4$};
        \node[Circle,minimum size=4mm]at(-.5,-2.75){$5$};
    \end{tikzpicture}
    \enspace \circ_2 \enspace
    \begin{tikzpicture}[scale=.23,Centering]
        \node[Circle,MarkB,minimum size=13mm](x)at(0,0){};
        \node[Circle,MarkB,minimum size=4mm](xa)at(-1,-1){$1$};
        \node[Circle,MarkB,minimum size=4mm](xb)at(1,0){$2$};
        \node[Circle,MarkB,minimum size=4mm](xb)at(-1,1){$3$};
    \end{tikzpicture}
    \enspace = \enspace
    \begin{tikzpicture}[scale=.35,Centering]
        \node[Circle,minimum size=27mm](x)at(0,0){};
        \node[Circle,minimum size=4mm]at(-1.75,-1){$1$};
        \node at(1.5,-1){
            \begin{tikzpicture}[scale=.23,Centering]
                \node[Circle,MarkB,minimum size=13mm](x)at(0,0){};
                \node[Circle,MarkB,minimum size=4mm](xa)at(-1,-1){$1$};
                \node[Circle,MarkB,minimum size=4mm](xb)at(1,0){$2$};
                \node[Circle,MarkB,minimum size=4mm](xb)at(-1,1){$3$};
            \end{tikzpicture}};
        \node[Circle,minimum size=4mm]at(.5,2.5){$3$};
        \node[Circle,minimum size=4mm]at(-1.75,1.5){$4$};
        \node[Circle,minimum size=4mm]at(-.5,-2.75){$5$};
    \end{tikzpicture}
    \enspace = \enspace
    \begin{tikzpicture}[scale=.35,Centering]
        \node[Circle,minimum size=25mm](x)at(0,0){};
        \node[Circle,minimum size=4mm]at(-1.75,-1){$1$};
        \node[Circle,minimum size=4mm](xa)at(1,-1.5){$2$};
        \node[Circle,minimum size=4mm](xb)at(2.25,-.75){$3$};
        \node[Circle,minimum size=4mm](xb)at(1,0){$4$};
        \node[Circle,minimum size=4mm]at(.5,2.5){$5$};
        \node[Circle,minimum size=4mm]at(-1.75,1.5){$6$};
        \node[Circle,minimum size=4mm]at(-.5,-2.75){$7$};
    \end{tikzpicture}.
\end{equation*}
In the context of this work, the combinatorial objects on which an operad structure will be
defined are the multi-patterns. The substitute sectors of a multi-pattern $\MPat$ are its
degrees, from the first labeled by $1$ to the last labeled by $n$, where $n$ is the arity of
$\MPat$. Of course, this operad will be presented in detail in the following.

The operad axioms can be understood on multi-patterns in the following way (even if at this
stage, the operad structure on these objects has not yet been defined).
Relation~\eqref{equ:operad_axiom_1} says that, given three multi-patterns $\MPat_1$,
$\MPat_2$, and $\MPat_3$, composing $\MPat_2$ into $\MPat_1$ and then, $\MPat_3$ into the
area occupied by $\MPat_2$ of the obtained multi-pattern is the same as composing $\MPat_3$
into $\MPat_2$ and then, the obtained multi-pattern into $\MPat_1$.
Relation~\eqref{equ:operad_axiom_2} says that, given three multi-patterns $\MPat_1$,
$\MPat_2$, and $\MPat_3$, composing $\MPat_2$ into $\MPat_1$ and then $\MPat_3$ to the right
of the area occupied by $\MPat_2$ of the obtained multi-pattern is the same as composing
$\MPat_3$ into $\MPat_1$ and then, $\MPat_3$ to the left of the area occupied by $\MPat_2$
of the obtained multi-pattern. Finally, Relation~\eqref{equ:operad_axiom_3} says simply that
there exists a multi-pattern playing the role of a unit element. These relations are the
right ones to be in position to see multi-patterns as operations and to compose them as
such.

\subsubsection{Elementary definitions} \label{subsubsec:elementary_definitions_operads}
Let $\Par{\Operad, \circ_i, \Unit}$ be an operad. The \Def{full composition} map of
$\Operad$ is the map
\begin{equation}
    \circ :
    \Operad(n) \times \Operad\Par{m_1} \times \dots \times \Operad\Par{m_n}
    \to \Operad\Par{m_1 + \dots + m_n}
\end{equation}
defined for any $x \in \Operad(n)$ and $y_1, \dots, y_n \in \Operad$ by
\begin{equation} \label{equ:full_composition_maps}
    x \circ \Han{y_1, \dots, y_n}
    := \Par{\dots \Par{\Par{x \circ_n y_n} \circ_{n - 1}
        y_{n - 1}} \dots} \circ_1 y_1.
\end{equation}
Intuitively, $x \circ \Han{y_1, \dots, y_n}$ is obtained by grafting simultaneously the
outputs of all the $y_i$ onto the $i$-th inputs of~$x$. Additionally, the \Def{homogeneous
composition} map of $\Operad$ is the map
\begin{equation}
    \HomogeneousComposition : \Operad(n) \times \Operad(m) \to \Operad(nm),
\end{equation}
defined, for any $x \in \Operad(n)$ and $y \in \Operad(m)$ by
\begin{equation}
    x \HomogeneousComposition y 
    := x \circ \underbrace{\Han{y, \dots, y}}_{\footnotesize n \mbox{ elements}}.
\end{equation}

Let $\Par{\Operad', \circ'_i, \Unit'}$ be a second operad. A graded set morphism $\phi :
\Operad \to \Operad'$ is an \Def{operad morphism} if $\phi(\Unit) = \Unit'$ and for any $x,
y \in \Operad$ and $i \in [|x|]$,
\begin{equation} \label{equ:operad_morphisms}
    \phi\Par{x \circ_i y} = \phi(x) \circ'_i \phi(y).
\end{equation}
If instead~\eqref{equ:operad_morphisms} holds by replacing the second occurrence of $i$ by
$|x| - i + 1$, then $\phi$ is an \Def{operad antimorphism}. We say that $\Operad'$ is a
\Def{suboperad} of $\Operad$ if $\Operad'$ is a graded subset of $\Operad$, $\Unit =
\Unit'$, and for any $x, y \in \Operad'$ and $i \in [|x|]$, $x \circ_i y = x \circ'_i y$.
For any subset $\GeneratingSet$ of $\Operad$, the \Def{operad generated} by $\GeneratingSet$
is the smallest suboperad $\Operad^{\GeneratingSet}$ of $\Operad$ containing
$\GeneratingSet$. When $\Operad^{\GeneratingSet} = \Operad$ and $\GeneratingSet$ is minimal
with respect to the inclusion among the subsets of $\GeneratingSet$ satisfying this
property, $\GeneratingSet$ is a \Def{minimal generating set} of $\Operad$ and its elements
are \Def{generators} of~$\Operad$.

The \Def{Hadamard product} of $\Operad$ and $\Operad'$ is the graded set $\Operad
\HadamardProduct \Operad'$ endowed with the partial composition map $\circ''_i$ defined, for
any $\Par{x, x'}, \Par{y, y'} \in \Operad \HadamardProduct \Operad'$ and $i \in
\Han{\Brr{\Par{x, x'}}}$, by
\begin{equation}
    \Par{x, x'} \circ''_i \Par{y, y'} := \Par{x \circ_i y, x' \circ'_i y'},
\end{equation}
and having $\Par{\Unit, \Unit'}$ as unit. This graded set $\Operad \HadamardProduct
\Operad'$ is an operad. Moreover, for any other operads $\Operad_1$ and $\Operad_1'$, and
any operads morphisms (resp.\ antimorphisms) $\phi : \Operad \to \Operad_1$ and $\phi' :
\Operad' \to \Operad_1'$, the graded set morphism $\phi \HadamardProduct \phi'$ is also an
operad morphism (resp.\ antimorphism) from $\Operad \HadamardProduct \Operad'$ to $\Operad_1
\HadamardProduct \Operad_1'$.

\subsection{Music box operads}
We build operads on multi-patterns step by step by introducing operads on degree patterns
and an operad on rhythm patterns. The operads of patterns are constructed as the Hadamard
product of the two previous ones. Finally, the operads of multi-patterns are suboperads of
iterated Hadamard products of operads of patterns with themselves. These operads (except the
operad of rhythm patterns) depend on a monoid structure on $\Z$ in order to encode how to
compute on degrees.

\subsubsection{Operads of degree patterns} \label{subsubsec:operad_degree_patterns}
The construction of an operad on degree patterns is based upon the \Def{construction
$\ConstructionT$}, a construction from monoids to operads introduced in~\cite{Gir15} which
we recall now. Given a monoid $\Par{\Monoid, \MonoidProduct, \MonoidUnit}$, where
$\MonoidProduct$ is an associative operation admitting $\MonoidUnit$ as unit, let
$\ConstructionT(\Monoid)$ be the graded set of the words on $\Monoid$, where the arity of a
word is its length. Let
\begin{math}
    \MonoidProductExtension :
    \Monoid \times \ConstructionT(\Monoid) \to \ConstructionT(\Monoid)
\end{math}
be the map defined for any $a \in \Monoid$ and $u \in \ConstructionT(\Monoid)$ by
\begin{equation}
    a \MonoidProductExtension u
    := a \MonoidProduct u(1) \Conc \dots \Conc a \MonoidProduct u(\Length(u)).
\end{equation}
The graded set $\ConstructionT(\Monoid)$ is endowed with a partial composition map $\circ_i$,
defined for any $u, u' \in \ConstructionT(\Monoid)$ and $i \in [|u|]$, by
\begin{equation}
    u \circ_i u' :=
    u(1, i - 1)
    \Conc \; u(i) \MonoidProductExtension u' \; \Conc
    u(i + 1, \Length(u)).
\end{equation}
It is shown in~\cite{Gir15} that $\Par{\ConstructionT(\Monoid), \circ_i, \Unit}$, where
$\Unit$ is the element $\MonoidUnit$ of $\ConstructionT(\Monoid)(1)$, is an operad. To be
perfectly precise, in~\cite{Gir15}, the presented construction is so that in
$\ConstructionT(\Monoid)$ there is no element of arity $0$. We consider here the same
construction with the difference that $\ConstructionT(\Monoid)(0)$ is the singleton
$\{\epsilon\}$. This is justified in the present work because we wish to manipulate degree
patterns of arity~$0$.

Besides, by extending some results of~\cite{Gir15}, it is possible to show that
$\ConstructionT(\Monoid)$ admits as a minimal generating set the set
\begin{equation} \label{equ:generating_set_construction_T}
    \Bra{\epsilon, \MonoidUnit \MonoidUnit} \cup \GeneratingSet_\Monoid
\end{equation}
where $\GeneratingSet_\Monoid$ is a minimal generating set of $\Monoid$ as a monoid.
Besides, if $\Monoid'$ is another monoid and $\phi : \Monoid \to \Monoid'$ is a monoid
morphism, let
\begin{math}
    \ConstructionT(\phi) : \ConstructionT(\Monoid) \to \ConstructionT\Par{\Monoid'}
\end{math}
be the map defined, for any $u \in \ConstructionT(\Monoid)$, by
\begin{equation}
    \ConstructionT(\phi) (u) := \phi\Par{u(1)} \Conc \dots \Conc \phi\Par{u(\Length(u))}.
\end{equation}
It is also shown in~\cite{Gir15} that $\ConstructionT(\phi)$ is an operad morphism
preserving injections and surjections.

Let $\Mirror : \ConstructionT(\Monoid) \to \ConstructionT(\Monoid)$ be the map defined, for
any $u \in \ConstructionT(\Monoid)$, by
\begin{equation}
    \Mirror(u) := u(\Length(u)) \Conc \dots \Conc u(1).
\end{equation}
The word $\Mirror(u)$ is the \Def{mirror} of $u$.

\begin{Statement}{Proposition}{prop:construction_T_mirror_anti_automorphism}
    For any monoid $\Monoid$, the map $\Mirror$ is an operad anti-auto\-mor\-phism
    of~$\ConstructionT(\Monoid)$.
\end{Statement}
\begin{Proof}
    This is a straightforward verification based upon the fact that for any $u \in
    \ConstructionT(\Monoid)$, the $i$-th letter of $\Mirror(u)$ is $u(\Length(u) - i + 1)$.
\end{Proof}

A \Def{degree monoid} is any monoid $\Par{D, \MonoidProduct, \MonoidUnit}$ such that $D
\subseteq \Z$. By construction, for any $n \in \N$, each $\DegPat \in \ConstructionT(D)(n)$
is a sequence of integers of length $n$, and thus $\DegPat$ is a degree pattern of
arity~$n$. For this reason, $\ConstructionT(D)$ is an operad having as underlying graded set
the graded set of the degree patterns having elements of $D$ as degrees. We denote by
\begin{equation}
    \OperadDP^D := \ConstructionT(D)
\end{equation}
this operad, called \Def{$D$-degree pattern operad}.

Let us consider three important examples depending on different natural degree monoids.
\begin{enumerate}
    \item By denoting by $\Z$ the additive monoid $(\Z, +, 0)$, $\OperadDP^\Z$ contains all
    degree patterns. In $\OperadDP^\Z$, we have
    \begin{equation}
        \bar{1}01\bar{3}2 \circ_4 211
        = \bar{1} 0 1 \Par{-3 + 2} \Par{-3 + 1} \Par{-3 + 1} 2
        = \bar{1}01 \bar{1}\bar{2}\bar{2} 2.
    \end{equation}
    Moreover, since $\Z$ admits $\{-1, 1\}$ as a minimal generating set, the operad
    $\OperadDP^\Z$ admits $\Bra{\epsilon, \bar{1}, 1, 00}$ as a minimal generating set.
    \item By denoting, for any $k \geq 1$, by $\MonoidCyclic_k$ the cyclic monoid $\Par{\Z /
    k \Z, +, 0}$, $\OperadDP^{\MonoidCyclic_k}$ contains the degree patterns having degrees
    between $0$ and $k - 1$. In $\OperadDP^{\MonoidCyclic_3}$ we have
    \begin{equation}
        20 1 01 \circ_3 2120 = 20 0201 01.
    \end{equation}
    Moreover, since $\MonoidCyclic_k$ admits $\{1\}$ as a minimal generating set, the operad
    $\OperadDP^{\MonoidCyclic_k}$ admits $\Bra{\epsilon, 1, 00}$ as a minimal generating
    set.
    \item By denoting, for any subset $Z$ of $\Z$ having a lower bound $z$, by
    $\MonoidMax_Z$ the monoid $\Par{Z, \max, z}$, $\OperadDP^{\MonoidMax_Z}$ contains all
    degree patterns having degrees in $Z$. In $\OperadDP^{\MonoidMax_{[0, 2]}}$ we have
    \begin{equation}
        200 1 0 \circ_4 2120 = 200 2121 0.
    \end{equation}
    Moreover, since $\MonoidMax_Z$ admits $Z \setminus \{z\}$ as a minimal generating set,
    the operad $\OperadDP^{\MonoidMax_Z}$ admits
    \begin{math}
        \{\epsilon\} \cup \Par{Z \setminus \{z\}} \cup \{00\}
    \end{math}
    as a minimal generating set.
\end{enumerate}

By Proposition~\ref{prop:construction_T_mirror_anti_automorphism}, for any degree monoid
$D$, the map $\Mirror : \OperadDP^D \to \OperadDP^D$ is an operad anti-automorphism.
Moreover, we can consider on the operads $\OperadDP^\Z$, $\OperadDP^{\MonoidCyclic_k}$, and
$\OperadDP^{\MonoidMax_Z}$ the following other morphisms.
\begin{enumerate}
    \item For any $\alpha \in \Z$, let $\MorphismMul_\alpha : \OperadDP^\Z \to \OperadDP^\Z$
    be the map defined by $\MorphismMul_\alpha := \ConstructionT(\phi)$ where $\phi$ is the
    monoid morphism satisfying $\phi(d) = \alpha d$ for any $d \in \Z$. For instance,
    \begin{equation}
        \MorphismMul_{-2}\Par{1 \bar{2} 0 0 3} = \bar{2} 4 0 0 \bar{6}.
    \end{equation}
    Since $\phi$ is a monoid morphism, $\MorphismMul_\alpha$ is an operad endomorphism.
    Moreover, when $\alpha \ne 0$, $\MorphismMul_\alpha$ is injective, and when $\alpha \in
    \{-1, 1\}$, $\MorphismMul_\alpha$ is bijective.
    \item For any $k \geq 1$, let also $\MorphismRed_k : \OperadDP^\Z \to
    \OperadDP^{\MonoidCyclic_k}$ be the map defined by $\MorphismRed_k :=
    \ConstructionT(\phi)$ where $\phi$ is the monoid morphism satisfying $\phi(d) = d \bmod
    k$ for any $d \in \Z$. For instance,
    \begin{equation}
        \MorphismRed_3\Par{1 \bar{2} 0 0 3} = 1 1 0 0 0.
    \end{equation}
    Since $\phi$ is a surjective monoid morphism, $\MorphismRed_\alpha$ is a surjective
    operad morphism.
    \item For any subsets $Z$ and $Z'$ of $\Z$ having respective lower bounds $z$ and $z'$,
    a map $\theta : Z \to Z'$ is a \Def{rooted weakly increasing map} if $\theta(z) = z'$
    and, for any $d, d' \in Z$, $d \leq d'$ implies $\theta(d) \leq \theta\Par{d'}$. If
    $\theta$ is such a map, let $\MorphismIncr_\theta : \OperadDP^{\MonoidMax_Z} \to
    \OperadDP^{\MonoidMax_Z'}$ be the map defined by $\MorphismIncr_\theta :=
    \ConstructionT(\theta)$. For instance, for $Z := [0, 3]$ and $Z' := [2, 5]$, and
    $\theta$ satisfying $\theta(d) = d + 2$ for any $d \in Z$,
    \begin{equation}
        \MorphismIncr_\theta\Par{1 2 0 0 3} = 3 4 2 2 5.
    \end{equation}
    Since $\theta$ is a monoid morphism, $\MorphismIncr_\theta$ is an operad morphism. This
    morphism is not necessarily injective nor surjective.
\end{enumerate}

\subsubsection{Operad of rhythm patterns} \label{subsubsec:operad_rhythm_patterns}
Let us introduce a new construction from monoids to operads, similar to the construction
$\ConstructionT$ recalled in Section~\ref{subsubsec:operad_degree_patterns}. Given a monoid
$\Par{\Monoid, \MonoidProduct, \MonoidUnit}$, let $\ConstructionU(\Monoid)$ be the graded
set of the nonempty words on $\Monoid$, where the arity of a word is its length minus $1$.
Let
\begin{math}
    \MonoidProductExtensionLeft :
    \Monoid \times \ConstructionU(\Monoid) \to \ConstructionU(\Monoid)
\end{math}
and
\begin{math}
    \MonoidProductExtensionRight :
    \ConstructionU(\Monoid) \times \Monoid \to \ConstructionU(\Monoid)
\end{math}
be the maps defined for any $a \in \Monoid$ and $u \in \ConstructionU(\Monoid)$ by
\begin{equation}
    a \MonoidProductExtensionLeft u
    := a \MonoidProduct u(1) \; \Conc u(2, \Length(u))
\end{equation}
and
\begin{equation}
    u \MonoidProductExtensionRight a
    := u(1, \Length(u) - 1) \Conc \; u(\Length(u)) \MonoidProduct a.
\end{equation}
The graded set $\ConstructionU(\Monoid)$ is endowed with a partial composition map
$\circ_i$, defined for any $u, u' \in \ConstructionU(\Monoid)$ and $i \in [|u|]$, by
\begin{equation} \label{equ:partial_composition_construction_u}
    u \circ_i u' :=
    u(1, i - 1) \Conc
    \; u(i) \MonoidProductExtensionLeft u' \MonoidProductExtensionRight u(i + 1) \;
    \Conc u(i + 2, \Length(u)).
\end{equation}
Note that the associativity of $\MonoidProduct$ ensures
that~\eqref{equ:partial_composition_construction_u} is well-defined in particular when
$\Length\Par{u'} = 1$. Let us also denote by $\Unit$ the element $\MonoidUnit \MonoidUnit$
of $\ConstructionU(\Monoid)(1)$. Moreover, if $\Monoid'$ is another monoid and $\phi :
\Monoid \to \Monoid'$ is a monoid morphism, let $\ConstructionU(\phi) :
\ConstructionU(\Monoid) \to \ConstructionU\Par{\Monoid'}$ be the map defined, for any $u \in
\ConstructionU(\Monoid)$, by
\begin{equation}
    \ConstructionU(\phi)(u) := \phi(u(1)) \Conc \dots \Conc \phi(u(\Length(u))).
\end{equation}

\begin{Statement}{Proposition}{prop:construction_U}
    For any monoid $\Monoid$, $\ConstructionU(\Monoid)$ is an operad. Moreover, for any
    monoids $\Monoid$ and $\Monoid'$, and any monoid morphism $\phi : \Monoid \to \Monoid'$,
    $\ConstructionU(\phi)$ is an operad morphism.
\end{Statement}
\begin{Proof}
    By using the fact that the product of $\Monoid$ is associative and that $\Monoid$ has a
    unit, it follows by a straightforward but technical verification that
    Relations~\eqref{equ:operad_axiom_1}, \eqref{equ:operad_axiom_2},
    and~\eqref{equ:operad_axiom_3} are satisfied. Finally, the fact that $\phi$ is a monoid
    morphism says that $\phi$ commutes with the product of $\Monoid$. As a straightforward
    computation shows, this implies that $\ConstructionU(\phi)$ is an operad morphism.
\end{Proof}

Let $\Mirror : \ConstructionU(\Monoid) \to \ConstructionU(\Monoid)$ be the map defined, for
any $u \in \ConstructionU(\Monoid)$, by
\begin{equation}
    \Mirror(u) := u(\Length(u)) \Conc \dots \Conc u(1).
\end{equation}
The word $\Mirror(u)$ is the \Def{mirror} of $u$.

\begin{Statement}{Proposition}{prop:construction_U_mirror_anti_automorphism}
    For any commutative monoid $\Monoid$, the map $\Mirror$ is an operad
    anti-auto\-mor\-phism of~$\ConstructionU(\Monoid)$.
\end{Statement}
\begin{Proof}
    This is a straightforward verification based upon the fact that for any $u \in
    \ConstructionU(\Monoid)$, the $i$-th letter of $\Mirror(u)$ is $u(\Length(u) - i + 1)$.
    The commutativity of $\Monoid$ is important here.
\end{Proof}

By Proposition~\ref{prop:construction_U},
\begin{equation}
    \OperadRP := \ConstructionU(\N)
\end{equation}
where $\N$ is the additive monoid $(\N, +, 0)$ is an operad. By construction, for any $n \in
\N$, each $\sigma \in \OperadRP(n)$ is a sequence of nonnegative integers of length $n + 1$,
and thus, $\sigma$ is a duration sequence of a rhythm pattern of arity $n$ (see
Section~\ref{subsubsec:rhythm_patterns}). For this reason, $\OperadRP$ can be seen as an
operad having as underlying graded set the graded set of the rhythm patterns. This operad is
the \Def{rhythm pattern operad}. We have for instance,
\begin{subequations}
\begin{equation} \label{equ:example_composition_operad_rhythm_patterns_1}
    00121 \circ_3 110 = 002121,
\end{equation}
\begin{equation} \label{equ:example_composition_operad_rhythm_patterns_2}
    110 \circ_1 12 = 230,
\end{equation}
\begin{equation} \label{equ:example_composition_operad_rhythm_patterns_3}
    211 \circ_1 2 = 51.
\end{equation}
\end{subequations}
Directly on rhythm patterns, the partial composition rephrases as follows. For any rhythm
patterns $\RhyPat$ and $\RhyPat'$, and any integer $i \in [|\RhyPat|]$, $\RhyPat \circ_i
\RhyPat'$ is obtained by replacing the $i$-th occurrence of $\Beat$ in $\RhyPat$ by
$\RhyPat'$. For instance, on rhythm patterns,
\eqref{equ:example_composition_operad_rhythm_patterns_1},
\eqref{equ:example_composition_operad_rhythm_patterns_2},
and~\eqref{equ:example_composition_operad_rhythm_patterns_3} translate respectively as
\begin{subequations}
\begin{equation}
    \Beat \Beat \Rest \Beat \Rest \Rest \Beat \Rest \circ_3 \Rest \Beat \Rest \Beat
    =
    \Beat \Beat \Rest \Rest \Beat \Rest \Beat \Rest \Rest \Beat \Rest,
\end{equation}
\begin{equation}
    \Rest \Beat \Rest \Beat \circ_1 \Rest \Beat \Rest \Rest
    =
    \Rest \Rest \Beat \Rest \Rest \Rest \Beat,
\end{equation}
\begin{equation}
    \Rest \Rest \Beat \Rest \Beat \Rest \circ_1 \Rest \Rest
    =
    \Rest \Rest \Rest \Rest \Rest \Beat \Rest.
\end{equation}
\end{subequations}

\begin{Statement}{Proposition}{prop:generating_set_operad_rhythm_pattern}
    The operad $\OperadRP$ admits $\Bra{\epsilon, \Rest, \Beat \Beat}$ as a minimal
    generating set.
\end{Statement}
\begin{Proof}
    Let us denote by $\GeneratingSet$ the candidate minimal generating set for $\OperadRP$
    described in the statement of the proposition. Any rhythm pattern $\RhyPat \in
    \OperadRP$ such that $\Length(\RhyPat) \geq 2$ decomposes as
    \begin{equation}
        \RhyPat =
        \underbrace{\Par{\Beat \Beat \circ_1 \dots \circ_1 \Beat \Beat}}
            _{\footnotesize \Length(\RhyPat) - 1 \mbox{ terms}}
        \circ \Han{\RhyPat(1), \dots, \RhyPat(\Length(\RhyPat))}.
    \end{equation}
    Since for any $i \in [\Length(\RhyPat)]$, $\RhyPat(i) \in \Bra{\Beat, \Rest}$ and
    $\Beat$ is the unit of $\OperadRP$ and $\Rest$ belongs to $\GeneratingSet$, $\RhyPat$ is
    an element of $\OperadRP^\GeneratingSet$. Additionally, all rhythm patterns of lengths
    $0$ or $1$ belong to $\OperadRP^\GeneratingSet$ because $\epsilon \in \GeneratingSet$,
    $\Rest \in \GeneratingSet$, and $\Beat$ is the unit of $\OperadRP$. This shows that
    $\GeneratingSet$ is a generating set of $\OperadRP$. Finally, this generating set is
    minimal since no element of $\GeneratingSet$ can be expressed by partial compositions of
    some different other ones.
\end{Proof}

For the next lemma, recall that when $\RhyPat$ is a rhythm pattern, $\Length(\RhyPat)$ is
the length of $\RhyPat$ and this quantity is the number of occurrences of $\Rest$ plus the
number of occurrences $\Beat$ in $\RhyPat$.

\begin{Statement}{Lemma}{lem:length_operad_rhythm_patterns}
    For any two rhythm patterns $\RhyPat$ and $\RhyPat'$, and $i \in [|\RhyPat|]$, in the
    operad $\OperadRP$, $\Length\Par{\RhyPat \circ_i \RhyPat'} = \Length(\RhyPat) +
    \Length\Par{\RhyPat'} - 1$.
\end{Statement}
\begin{Proof}
    This is a direct consequence of the interpretation of the partial composition of
    $\OperadRP$ in terms of rhythm patterns: the rhythm pattern $\RhyPat \circ_i \RhyPat'$
    is obtained by replacing a $\Beat$ of $\RhyPat$ by~$\RhyPat'$.
\end{Proof}

By Proposition~\ref{prop:construction_U_mirror_anti_automorphism}, since $\N$ is a
commutative monoid, the map $\Mirror : \OperadRP \to \OperadRP$ is an operad
anti-automorphism. On rhythm patterns, given $\RhyPat \in \OperadRP$, $\Mirror(\RhyPat)$ is
the rhythm pattern obtained by reading $\RhyPat$ from right to left. For instance,
\begin{equation}
    \Mirror(\Rest \Beat \Beat \Rest \Beat \Beat \Beat) =
    \Beat \Beat \Beat \Rest \Beat \Beat \Rest.
\end{equation}

For any $\beta \in \N$, let $\MorphismDil_\beta : \OperadRP \to \OperadRP$ be the map
defined by $\MorphismDil_\beta := \ConstructionU(\phi)$ where $\phi$ is the monoid morphism
satisfying $\phi(s) = \beta s$ for any $s \in \N$. Since $\phi$ is a monoid morphism of
$\N$, by Proposition~\ref{prop:construction_U_mirror_anti_automorphism},
$\MorphismDil_\beta$ is an operad endomorphism. When $\beta \ne 0$, $\MorphismDil_\beta$ is
injective and $\MorphismDil_\beta$ is surjective if and only if $\beta = 1$. Interpreted on
rhythm patterns, $\MorphismDil_\beta(\RhyPat)$ is obtained by replacing each occurrence of
$\Rest$ in $\RhyPat$ by $\Rest^\beta$. For instance,
\begin{subequations}
\begin{equation}
    \MorphismDil_2\Par{\Rest \Beat \Beat \Rest \Rest} =
    \Rest \Rest \Beat \Beat \Rest \Rest \Rest \Rest,
\end{equation}
\begin{equation}
    \MorphismDil_0\Par{\Rest \Beat \Beat \Rest \Rest} = \Beat \Beat.
\end{equation}
\end{subequations}

\subsubsection{Operads of patterns} \label{subsubsec:operad_patterns}
For any degree monoid $(D, \MonoidProduct, \MonoidUnit)$, let $\OperadP^D$ be the operad
defined as
\begin{equation}
    \OperadP^D := \OperadDP^D \HadamardProduct \OperadRP.
\end{equation}
By construction, for any $n \in \N$, each $\Pat \in \OperadP^D(n)$ is a pair $(\DegPat,
\RhyPat)$ such that $\DegPat$ is a degree pattern of arity $n$ having elements of $D$ as
degrees and $\RhyPat$ is a rhythm pattern of arity $n$. For this reason, $\OperadP^D$ is an
operad having as underlying graded set the graded set of the patterns having elements of $D$
as degrees. We call $\OperadP^D$ the \Def{$D$-pattern operad}.
In $\OperadP^\Z$, we have for instance
\begin{subequations}
\begin{equation} \label{equ:example_composition_operad_patterns_1}
    \Par{\bar{2} 3 1, \Rest \Beat \Beat \Rest \Beat}
    \circ_2 \Par{0 \bar{1}, \Beat \Rest \Beat}
    = \Par{\bar{2} 3 2 1, \Rest \Beat \Beat \Rest \Beat \Rest \Beat},
\end{equation}
\begin{equation} \label{equ:example_composition_operad_patterns_2}
    \Par{1 1 2, \Beat \Rest \Beat \Rest \Beat} \circ_1 \Par{\bar{1}, \Rest \Rest \Beat}
    = \Par{0 1 2, \Rest \Rest \Beat \Rest \Beat \Rest \Beat}.
\end{equation}
\end{subequations}
Directly by using the concise notation for patterns described in
Section~\ref{subsubsec:patterns}, the partial composition rephrases as follows. For any
patterns $\Pat$ and $\Pat'$, and any integer $i \in [|\Pat|]$, $\Pat \circ_i \Pat'$ is
obtained by replacing the $i$-th degree $d$ of $\Pat$ by $\Pat''$ where $\Pat''$ is the
pattern obtained by replacing each degree $d'$ of $\Pat'$ by $d \MonoidProduct d'$. For instance,
by using the concise notation for patterns,
\eqref{equ:example_composition_operad_patterns_1}
and~\eqref{equ:example_composition_operad_patterns_2} translate as
\begin{subequations}
\begin{equation}
    \Rest \bar{2} 3 \Rest 1 \circ_2 0 \Rest \Bar{1} = \Rest \bar{2} 3 \Rest 2 \Rest 1,
\end{equation}
\begin{equation}
    1 \Rest 1 \Rest 2 \circ_1 \Rest \Rest \bar{1} = \Rest \Rest 0 \Rest 1 \Rest 2.
\end{equation}
\end{subequations}

\begin{Statement}{Proposition}{prop:generating_set_operad_patterns}
    Let $\Par{D, \MonoidProduct, \MonoidUnit}$ be a degree monoid admitting
    $\GeneratingSet_D$ as a minimal generating set. The operad $\OperadP^D$ admits
    \begin{math}
        \Bra{\epsilon, \Rest, \MonoidUnit \MonoidUnit} \cup \GeneratingSet_D 
    \end{math}
    as a minimal generating set.
\end{Statement}
\begin{Proof}
    Let us denote by $\GeneratingSet$ the candidate minimal generating set for $\OperadP^D$
    described in the statement of the proposition. Any pattern $\Pat \in \OperadP^D$
    such that $\Length(\Pat) \geq 2$ decomposes as
    \begin{equation}
        \Pat =
        \underbrace{
            \Par{\MonoidUnit \MonoidUnit \circ_1 \dots \circ_1 \MonoidUnit \MonoidUnit}}
            _{\footnotesize \Length(\Pat) - 1 \mbox{ terms}}
        \circ \Han{\Pat(1), \dots, \Pat(\Length(\Pat))}.
    \end{equation}
    By definition of patterns, for any $i \in [n]$, $\Pat(i)$ is either $\Rest$ or an
    element of $D$. Therefore, since $\GeneratingSet$ contains $\GeneratingSet_D$, each
    $\Pat(i)$ which is an element of $D$ can be expressed by composing some elements of
    $\GeneratingSet_D$. Since moreover $\Rest$ belongs to $\GeneratingSet$, this shows that
    $\Pat$ is an element of ${\OperadP^D}^\GeneratingSet$. Additionally, all patterns of
    length $0$ or $1$ belong to ${\OperadP^D}^\GeneratingSet$ because $\epsilon \in
    \GeneratingSet$, $\Rest \in \GeneratingSet$, and since $\GeneratingSet_D$ is a
    generating set of $D$, each pattern $d \in D$ of length $1$ belongs to
    ${\OperadP^D}^\GeneratingSet$. Therefore, $\GeneratingSet$ is a generating set of
    $\OperadP^D$. Finally, this generating set is minimal since no element of
    $\GeneratingSet$ can be expressed by partial compositions of some different other ones.
\end{Proof}

By Proposition~\ref{prop:generating_set_operad_patterns},
\begin{enumerate}
    \item the operad $\OperadP^\Z$ admits $\Bra{\epsilon, \Rest, \bar{1}, 1, 00}$ as a
    minimal generating set;

    \item for any $k \geq 1$, the operad $\OperadP^{\MonoidCyclic_k}$, admits
    $\Bra{\epsilon, \Rest, 1, 00}$ as a minimal generating set;

    \item for any subset $Z$ of $\Z$ having a lower bound $z$, $\OperadP^{\MonoidMax_Z}$
    admits $\Bra{\epsilon, \Rest, zz} \cup \Par{Z \setminus \{z\}}$ as a minimal generating
    set.
\end{enumerate}

Due to the construction of $\OperadP^D$ as the Hadamard product of $\OperadDP^D$ and
$\OperadRP$, by Propositions~\ref{prop:construction_T_mirror_anti_automorphism}
and~\ref{prop:construction_U_mirror_anti_automorphism}, the map $\Mirror \HadamardProduct
\Mirror$ is an operad anti-automorphism of $\OperadP^D$. Moreover, again due to the
construction of $\OperadP^D$ and the existence of the operad morphisms involving
$\OperadDP^D$ and $\OperadRP$ presented in Sections~\ref{subsubsec:operad_degree_patterns}
and~\ref{subsubsec:operad_rhythm_patterns}, one can consider on the operads $\OperadP^\Z$,
$\OperadP^{\MonoidCyclic_k}$, and $\OperadP^{\MonoidMax_Z}$ the following morphisms.
\begin{enumerate}
    \item For any $\alpha \in \Z$, the map $\MorphismMul_\alpha \HadamardProduct \Id$ is an
    operad endomorphism of $\OperadP^\Z$.

    \item For any $k \geq 1$, the map $\MorphismRed_k \HadamardProduct \Id$ is an operad
    morphism from $\OperadP^\Z$ to $\OperadP^{\MonoidCyclic_k}$.

    \item For any subsets $Z$ and $Z'$ of $\Z$ having lower bounds and any rooted weakly
    increasing map $\theta : Z \to Z'$, the map $\MorphismIncr_\theta \HadamardProduct \Id$
    is an operad morphism from $\OperadP^{\MonoidMax_Z}$ to~$\OperadP^{\MonoidMax_{Z'}}$.

    \item For any degree monoid $D$ and $\beta \in \N$, the map $\Id \HadamardProduct
    \MorphismDil_\beta$ is an operad endomorphism of~$\OperadP^D$.
\end{enumerate}
Let us describe some morphisms involving the operad $\OperadP^D$ and the previous operads
$\OperadDP^D$ and $\OperadRP$. Let the map $\DegPatToPat{D} : \OperadDP^D \to \OperadP^D$
be defined, for any $\DegPat \in \OperadDP^D$ by
\begin{equation}
    \DegPatToPat{D}(\DegPat) := \Par{\DegPat, \Beat^{|\DegPat|}}.
\end{equation}
This map is an injective operad morphism. Let also the map $\RhyPatToPat{D} : \OperadRP \to
\OperadP^D$ be defined, for any $\RhyPat \in \OperadRP$ by
\begin{equation}
    \RhyPatToPat{D}(\RhyPat) := \Par{\MonoidUnit^{|\RhyPat|}, \RhyPat},
\end{equation}
where $\MonoidUnit$ is the unit of $D$. This map is an injective operad morphism.

\subsubsection{Operads of multi-patterns}
For any degree monoid $D$ and any positive integer $m$, let ${\OperadMP^D_m}'$ be the operad
defined as
\begin{equation}
    {\OperadMP^D_m}' :=
    \underbrace{\OperadP^D \HadamardProduct \cdots \HadamardProduct \OperadP^D}
    _{\footnotesize m \mbox{ terms}}.
\end{equation}
By definition of the Hadamard product of graded sets, the elements of ${\OperadMP^D_m}'$ are
$m$-tuples (or, equivalently, words of length $m$) of patterns. Let also $\OperadMP^D_m$ be
the graded subset of the underlying graded set of ${\OperadMP^D_m}'$ restrained to the
$m$-tuples of patterns having all an equal length.

\begin{Statement}{Theorem}{thm:operad_multi_patterns}
    For any degree monoid $D$ and any positive integer $m$, $\OperadMP^D_m$ is an operad.
\end{Statement}
\begin{Proof}
    We have to prove that the set $\OperadMP^D_m$ forms a suboperad of $\OperadMP'_m$. Let
    us denote by $\MonoidUnit$ the unit of $D$. Since $\MonoidUnit$ is also the unit of the
    operad $\OperadP^D$, the $m$-tuple $\Par{\MonoidUnit, \dots, \MonoidUnit}$ belongs to
    $\OperadMP^D_m(1)$ and is the unit of $\OperadMP^D_m$. Moreover, by
    Lemma~\ref{lem:length_operad_rhythm_patterns}, given two patterns $\Pat$ and $\Pat'$,
    for any $i \in \Han{\Brr{\Pat}}$, the length of $\Pat \circ_i \Pat'$ is
    $\Length\Par{\Pat} + \Length\Par{\Pat'} - 1$. This shows that all patterns of the
    multi-pattern resulting of a partial composition of two multi-patterns have the same
    length. This implies that the partial composition of two elements of $\OperadMP^D_m$ is
    also in $\OperadMP^D_m$ and hence, that $\OperadMP^D_m$ is an operad.
\end{Proof}

By construction, for any $n \in \N$, each $\MPat \in \OperadMP^D_m(n)$ is an $m$-tuple of
patterns all having $n$ as arity, all having elements of $D$ as degrees, and all having the
same length. For this reason, $\OperadMP^D_m$ is an operad having as underlying graded set
the graded set of the multi-patterns of multiplicity~$m$ and having elements of $D$ as
degrees. By Theorem~\ref{thm:operad_multi_patterns}, $\OperadMP^D_m$ is an operad, and we
call it the \Def{$D$-music box operad} of multiplicity $m$. This construction of
$\OperadMP^D_m$ explains why all the patterns of a multi-pattern must have the same arity.
This is a consequence of the general definition of the Hadamard product of operads.

By using the matrix notation for multi-patterns described in
Section~\ref{subsubsec:multi_patterns}, we have for instance respectively in
$\OperadMP^\Z_2$ and in $\OperadMP^\Z_3$,
\begin{subequations}
\begin{equation}
    \begin{MultiPattern}
        \Rest & \bar{2} & \bar{1} & \Rest & 0 \\
        0 & 1 & \Rest & \Rest & 1
    \end{MultiPattern}
    \circ_2
    \begin{MultiPattern}
        1 & \Rest \\
        \bar{3} & \Rest
    \end{MultiPattern}
    =
    \begin{MultiPattern}
        \Rest & \bar{2} & 0 & \Rest & \Rest & 0 \\
        0 & \bar{2} & \Rest & \Rest & \Rest & 1
    \end{MultiPattern},
\end{equation}
\begin{equation}
    \begin{MultiPattern}
        0 & \Rest & 0 \\
        2 & \Rest & 0 \\
        4 & 4 & \Rest
    \end{MultiPattern}
    \circ_2
    \begin{MultiPattern}
        7 & 7 \\
        0 & 0 \\
        \bar{7} & \bar{7}
    \end{MultiPattern}
    =
    \begin{MultiPattern}
        0 & \Rest & 7 & 7 \\
        2 & \Rest & 0 & 0 \\
        4 & \bar{3} & \bar{3} & \Rest
    \end{MultiPattern}.
\end{equation}
\end{subequations}

Due to the construction of $\OperadMP^D_m$ as a suboperad of an operad obtained by an
iterated Hadamard product of $\OperadP^m$ and to the fact that, as explained in
Section~\ref{subsubsec:operad_patterns}, $\Mirror \HadamardProduct \Mirror$ is an
anti-automorphism of $\OperadP$, the map
\begin{equation}
    \Mirror := \underbrace{\Par{\Mirror \HadamardProduct \Mirror}
    \HadamardProduct \cdots \HadamardProduct
    \Par{\Mirror \HadamardProduct \Mirror}}
    _{\footnotesize m \mbox{ terms}}
\end{equation}
is an operad anti-automorphism of $\OperadMP^D_m$. Moreover, again due to the construction
of $\OperadMP^D_m$ and the operad endomorphisms of $\OperadP^D$ presented in
Section~\ref{subsubsec:operad_patterns}, one can consider on the operads $\OperadMP^Z_m$,
$\OperadMP^{\MonoidCyclic_k}_m$, and $\OperadMP^{\MonoidMax_Z}_m$ the following morphisms.
\begin{enumerate}
    \item For any $\alpha_1, \dots, \alpha_m \in \Z$, the map
    \begin{equation}
        \MorphismMul_{\alpha_1, \dots, \alpha_m} :=
        \Par{\MorphismMul_{\alpha_1} \HadamardProduct \Id}
        \HadamardProduct \cdots \HadamardProduct
        \Par{\MorphismMul_{\alpha_m} \HadamardProduct \Id}
    \end{equation}
    is an operad endomorphism of $\OperadMP^\Z_m$.

    \item For any $k \geq 1$, the map
    \begin{equation}
        \MorphismRed_k :=
        \underbrace{\Par{\MorphismRed_k \HadamardProduct \Id}
        \HadamardProduct \cdots \HadamardProduct
        \Par{\MorphismRed_k \HadamardProduct \Id}}
        _{\footnotesize m \mbox{ terms}}
    \end{equation}
    is an operad morphism from $\OperadMP^\Z_m$ to $\OperadMP^{\MonoidCyclic_k}_m$.

    \item For any subsets $Z$ and $Z'$ of $\Z$ having lower bounds and any rooted weakly
    increasing maps $\theta_i : Z \to Z'$, $i \in [m]$, the map
    \begin{equation}
        \MorphismIncr_{\theta_1, \dots, \theta_m} :=
        \Par{\MorphismIncr_{\theta_1} \HadamardProduct \Id}
        \HadamardProduct \cdots \HadamardProduct
        \Par{\MorphismIncr_{\theta_m} \HadamardProduct \Id}
    \end{equation}
    is an operad morphism from $\OperadMP^{\MonoidMax_Z}_m$ to
    $\OperadMP^{\MonoidMax_{Z'}}_m$.

    \item For any degree monoid $D$ and $\beta \in \N$,
    \begin{equation}
        \MorphismDil_\beta :=
        \underbrace{\Par{\Id \HadamardProduct \MorphismDil_\beta}
        \HadamardProduct \cdots \HadamardProduct
        \Par{\Id \HadamardProduct \MorphismDil_\beta}}
        _{\footnotesize m \mbox{ terms}}
    \end{equation}
    is an operad endomorphism of $\OperadMP^D_m$.
\end{enumerate}
Let also, for any $m \geq 1$, the map $\MorphismCopy_m : \OperadP^D \to \OperadMP^D_m$
be defined, for any $\Pat \in \OperadP^D$, by
\begin{equation}
    \MorphismCopy_m(\Pat) :=
    \underbrace{\Pat \Conc \dots \Conc \Pat}_{\footnotesize m \mbox{ elements}}.
\end{equation}
In other words, $\MorphismCopy_m(\Pat)$ is the multi-pattern obtained by stacking the
pattern $\Pat$ with itself $m$ times. This map is an injective operad morphism.

The full diagram involving the operads $\OperadDP^D$, $\OperadRP$, $\OperadP^D$, and
$\OperadMP^D_m$ is
\begin{equation}
    \begin{tikzpicture}[Centering,xscale=1.65,yscale=1.65]
        \node(DP)at(-1,1){$\OperadDP^D$};
        \node(RP)at(3,1){$\OperadRP$};
        \node(P)at(1,1){$\OperadP^D = \OperadMP^D_1$};
        \node(Pm)at(1,2){$\OperadMP^D_m$};
        \draw[Injection](DP)edge[above,font=\footnotesize]node{$\DegPatToPat{D}$}(P);
        \draw[Injection](RP)edge[above,font=\footnotesize]node{$\RhyPatToPat{D}$}(P);
        \draw[Injection](P)edge[right,font=\footnotesize]node{$\MorphismCopy_m$}(Pm);
    \end{tikzpicture}
\end{equation}
where $D$ is a degree monoid, $m \geq 1$, and the arrows
$\tikz[scale=.5]{\draw[Injection](0,0)--(1,0);}$ are injective operad morphisms.

\subsection{Operations on musical phrases}
Thanks to the $D$-music box operads and more precisely, to the operad structures on
multi-patterns, we can see any multi-pattern as an operator acting on multi-patterns. At the
level of interpretations, a multi-pattern is hence an operator acting on musical phrases. We
first describe an application of the $D$-music box operads to decompose multi-patterns in
smaller pieces, then introduce a nomenclature for some special multi-patterns, and finish by
explaining how to express some natural transformations on musical phrases in the language of
the $D$-music box operads.

\subsubsection{Decomposing multi-patterns}
Instead of using the operads $\OperadMP^D_m$ to build multi-patterns, we can use these
structures in the opposite way to decompose multi-patterns into smaller pieces. Indeed, an
operad structure allows us to factorize its elements by means of treelike structures (see
Section~\ref{subsubsec:abstract_operators}). More precisely, given a graded subset $S$ of
$\OperadMP^D_m$ and $\MPat \in \OperadMP^D_m$, an \Def{$S$-decomposition} of $\MPat$ is a
planar rooted tree $\TreeT$ having all internal nodes decorated by elements of $S$ and such
that the evaluation (as a syntax tree) of $\TreeT$ is $\MPat$.

For instance, the pattern $\Pat := \Rest 1 \bar{1} \bar{1} 2 \Rest 1 \Rest$ decomposes in
$\OperadMP^\Z_1 = \OperadP^\Z$ as the tree
\begin{equation}
    \begin{tikzpicture}[Centering,xscale=0.32,yscale=0.2]
        \node(12)at(6.00,-11.33){};
        \node(16)at(8.00,-14.17){};
        \node(2)at(0.00,-14.17){};
        \node(5)at(2.00,-14.17){};
        \node(8)at(4.00,-8.50){};
        \node[NodeST](0)at(1.00,-5.67){$\Rest 0$};
        \node[NodeST](1)at(0.00,-11.33){$1$};
        \node[NodeST](10)at(7.00,-2.83){$1$};
        \node[NodeST](11)at(6.00,-8.50){$1$};
        \node[NodeST](13)at(7.00,-5.67){$00$};
        \node[NodeST](14)at(8.00,-8.50){$0 \Rest$};
        \node[NodeST](15)at(8.00,-11.33){$\Rest 0$};
        \node[NodeST](3)at(1.00,-8.50){$00$};
        \node[NodeST](4)at(2.00,-11.33){$\bar{1}$};
        \node[NodeST](6)at(3.00,-2.83){$00$};
        \node[NodeST](7)at(4.00,-5.67){$\bar{1}$};
        \node[NodeST](9)at(5.00,0.00){$00$};
        \draw[Edge](0)--(6);
        \draw[Edge](1)--(3);
        \draw[Edge](10)--(9);
        \draw[Edge](11)--(13);
        \draw[Edge](12)--(11);
        \draw[Edge](13)--(10);
        \draw[Edge](14)--(13);
        \draw[Edge](15)--(14);
        \draw[Edge](16)--(15);
        \draw[Edge](2)--(1);
        \draw[Edge](3)--(0);
        \draw[Edge](4)--(3);
        \draw[Edge](5)--(4);
        \draw[Edge](6)--(9);
        \draw[Edge](7)--(6);
        \draw[Edge](8)--(7);
        \node(r)at(5.00,2.12){};
        \draw[Edge](r)--(9);
    \end{tikzpicture}.
\end{equation}
This is a $\GeneratingSet$-decomposition of $\Pat$ where $\GeneratingSet$ is the on the
minimal generating set of $\OperadP^\Z$ described in
Section~\ref{subsubsec:operad_patterns}. Besides, the multi-pattern
\begin{equation}
    \MPat :=
    \begin{MultiPattern}
        0 & 0 & \Rest & 1 & \Rest & \bar{2} \\
        \Rest & 0 & 0 & 0 & 3 & \Rest
    \end{MultiPattern}
\end{equation}
admits in $\OperadMP^\Z_2$ the $\OperadMP^\Z_2$-decomposition
\begin{equation}
    \begin{tikzpicture}[Centering,xscale=0.5,yscale=0.4]
        \node(0)at(0.00,-8.00){};
        \node(11)at(10.00,-8.00){};
        \node(2)at(2.00,-8.00){};
        \node(4)at(4.00,-5.00){};
        \node(7)at(6.00,-8.00){};
        \node(9)at(8.00,-8.00){};
        \node[NodeST](1)at(1.00,-6.00){
            \begin{math}
                \begin{MultiPattern}
                    0 & 0 \\
                    0 & 0
                \end{MultiPattern}
            \end{math}};
        \node[NodeST](10)at(9.00,-6.00){
            \begin{math}
                \begin{MultiPattern}
                    \Rest & 0 \\
                    0 & \Rest
                \end{MultiPattern}
            \end{math}};
        \node[NodeST](3)at(3.00,-3.00){
            \begin{math}
                \begin{MultiPattern}
                    0 & \Rest \\
                    \Rest & 0
                \end{MultiPattern}
            \end{math}};
        \node[NodeST](5)at(5.00,0.00){
            \begin{math}
                \begin{MultiPattern}
                    0 & 0 \\
                    0 & 0
                \end{MultiPattern}
            \end{math}};
        \node[NodeST](6)at(6.00,-6.00){
            \begin{math}
                \begin{MultiPattern}
                    1 \\
                    0
                \end{MultiPattern}
            \end{math}};
        \node[NodeST](8)at(7.00,-3.00){
            \begin{math}
                \begin{MultiPattern}
                    0 & \bar{2} \\
                    0 & 3
                \end{MultiPattern}
            \end{math}};
        \draw[Edge](0)--(1);
        \draw[Edge](1)--(3);
        \draw[Edge](10)--(8);
        \draw[Edge](11)--(10);
        \draw[Edge](2)--(1);
        \draw[Edge](3)--(5);
        \draw[Edge](4)--(3);
        \draw[Edge](6)--(8);
        \draw[Edge](7)--(6);
        \draw[Edge](8)--(5);
        \draw[Edge](9)--(10);
        \node(r)at(5.00,2.00){};
        \draw[Edge](r)--(5);
    \end{tikzpicture}.
\end{equation}

At the musical level, this notion of decomposition is related to the parsing of musical
phrases (see also~\cite{LJ96}). Understanding the $\GeneratingSet$-decompositions of a
multi-pattern $\MPat$, where $\GeneratingSet$ is a minimal generating set of
$\OperadMP^D_m$, brings information about the structure of $\MPat$, including its
repetitions, its self-similarities, and its symmetries. It is possible to develop, within
this framework of the bud music box model, a notion of complexity for multi-patterns by
studying the different $\GeneratingSet$-decompositions a multi-pattern has.

Nevertheless, the description of minimal generating sets for $\OperadMP^D_m$ when $m \geq 2$
seems to be much more intricate than for $\OperadMP^D_1$ and the previous operads (than for
$\OperadDP^D$ and for $\OperadRP$). For the time being, we do not have good and usable
descriptions. Therefore, we do not have an effective way to propose
$\GeneratingSet$-decompositions of multi-patterns of multiplicity greater than $1$ for the
time being where $\GeneratingSet$ is a minimal generating set of $\OperadMP^D_m$ when $m
\geq 2$.

\subsubsection{Particular multi-patterns} \label{subsubsec:particular_multi_patterns}
We define here some multi-patterns that play special roles.
\begin{enumerate}
    \item {\bf Chords.} A multi-pattern $\MPat$ is a \Def{chord} if $\Multiplicity(\MPat)
    \geq 2$, $|\MPat|= 1$, and $\Length(\MPat) = 1$. For instance,
    \begin{equation}
        \begin{MultiPattern}
            \bar{7} \\
            0 \\
            2 \\
            4
        \end{MultiPattern}
    \end{equation}
    is a chord.

    \item {\bf Flat multi-patterns.} A multi-pattern $\MPat$ is \Def{flat} if all degrees of
    its patterns $\MPat(i)$, $i \in [\Multiplicity(\MPat)]$, are $0$. For instance,
    \begin{equation}
        \begin{MultiPattern}
            0 & \Rest & \Rest & 0 \\
            \Rest & 0 & 0 & \Rest \\
            0 & \Rest & 0 & \Rest
        \end{MultiPattern}
    \end{equation}
    is flat.

    \item {\bf Arpeggio shapes. }A multi-pattern $\MPat$ is an \Def{arpeggio shape} if
    $\MPat$ is flat, $\Multiplicity(\MPat) \geq 2$, $|\MPat| = 1$, and, for any $i, i' \in
    [\Multiplicity(\MPat)]$ and $j \in [\Length(\MPat)]$, $\MPat(i)(j) \ne \Rest \ne
    \MPat\Par{i'}(j)$ implies $i = i'$. For instance,
    \begin{equation}
        \begin{MultiPattern}
            \Rest & 0 & \Rest & \Rest \\
            0 & \Rest & \Rest & \Rest \\
            \Rest & \Rest & 0 & \Rest \\
            \Rest & \Rest & \Rest & 0
        \end{MultiPattern}
    \end{equation}
    is an arpeggio shape.

    \item {\bf Arpeggios.} A multi-pattern $\MPat$ is an \Def{arpeggio} if $\MPat$ writes as
    $\MPat' \HomogeneousComposition \MPat''$ where $\MPat'$ is a chord of multiplicity $m$,
    $\MPat''$ is an arpeggio shape of multiplicity $m$, and $\HomogeneousComposition$ is the
    homogeneous composition of the operad $\OperadMP_m^\Z$. For instance, since
    \begin{equation}
        \begin{MultiPattern}
            \Rest & \bar{7} & \Rest & \Rest \\
            0 & \Rest & \Rest & \Rest \\
            \Rest & \Rest & 2 & \Rest \\
            \Rest & \Rest & \Rest & 4
        \end{MultiPattern}
        =
        \begin{MultiPattern}
            \bar{7} \\
            0 \\
            2 \\
            4
        \end{MultiPattern}
        \HomogeneousComposition
        \begin{MultiPattern}
            \Rest & 0 & \Rest & \Rest \\
            0 & \Rest & \Rest & \Rest \\
            \Rest & \Rest & 0 & \Rest \\
            \Rest & \Rest & \Rest & 0
        \end{MultiPattern},
    \end{equation}
    the multi-pattern on left-hand side is an arpeggio.
\end{enumerate}

\subsubsection{Operations} \label{subsubsec:operations}
Let us describe here some transformations on musical phrases by using the operad
$\OperadMP^\Z_m$ and the related operad morphisms.
\begin{enumerate}
    \item {\bf Mimesis.} For any multi-patterns $\MPat$ and $\MPat'$ of the same
    multiplicity, $\MPat \HomogeneousComposition \MPat'$ is the \Def{mimesis} of $\MPat$
    according to $\MPat'$. For instance
    \begin{equation}
        \begin{MultiPattern}
            \bar{1} & \Rest & \Rest & 1
        \end{MultiPattern}
        \HomogeneousComposition
        \begin{MultiPattern}
            0 & 2 & 4 & \Rest
        \end{MultiPattern}
        =
        \begin{MultiPattern}
            \bar{1} & 1 & 3 & \Rest
            & \Rest & \Rest
            & 1 & 3 & 5 & \Rest
        \end{MultiPattern}.
    \end{equation}
    Observe that since
    \begin{equation}
        \begin{MultiPattern}
            0 & 2 & 4 & \Rest
        \end{MultiPattern}
        \HomogeneousComposition
        \begin{MultiPattern}
            \bar{1} & \Rest & \Rest & 1
        \end{MultiPattern}
        =
        \begin{MultiPattern}
            \bar{1} & \Rest & \Rest & 1
            & 1 & \Rest & \Rest & 3
            & 3 & \Rest & \Rest & 5
            & \Rest
        \end{MultiPattern},
    \end{equation}
    this operation is not commutative. Moreover, when $\MPat$ is a chord and $\MPat'$ is an
    arpeggio shape, the mimesis of $\MPat$ according to $\MPat'$ is by definition an
    arpeggio. For instance,
    \begin{equation}
        \begin{MultiPattern}
            \bar{7} \\
            0 \\
            2 \\
            4
        \end{MultiPattern}
        \HomogeneousComposition
        \begin{MultiPattern}
            \Rest & 0 & \Rest & \Rest \\
            0 & \Rest & \Rest & \Rest \\
            \Rest & \Rest & 0 & \Rest \\
            \Rest & \Rest & \Rest & 0
        \end{MultiPattern}
        =
        \begin{MultiPattern}
            \Rest & \bar{7} & \Rest & \Rest \\
            0 & \Rest & \Rest & \Rest \\
            \Rest & \Rest & 2 & \Rest \\
            \Rest & \Rest & \Rest & 4
        \end{MultiPattern}.
    \end{equation}

    \item {\bf Concatenation.} For any two multi-patterns $\MPat$ and $\MPat'$ of the same
    multiplicity $m$,
    \begin{equation}
        \Concatenation\Par{\MPat, \MPat'} := \MorphismCopy_m(00) \circ \Han{\MPat, \MPat'}
    \end{equation}
    is the concatenation of $\MPat$ and $\MPat'$. For instance,
    \begin{equation}
        \Concatenation\Par{
        \begin{MultiPattern}
            \bar{2} & \Rest & \Rest & 1 & \Rest \\
            \Rest & 2 & \Rest & 3 & \Rest
        \end{MultiPattern},
        \begin{MultiPattern}
            0 & \Rest & 0 & 1 \\
            \bar{1} & \bar{1} & \Rest & 3
        \end{MultiPattern}}
        =
        \begin{MultiPattern}
            \bar{2} & \Rest & \Rest & 1 & \Rest & 0 & \Rest & 0 & 1 \\
            \Rest & 2 & \Rest & 3 & \Rest & \bar{1} & \bar{1} & \Rest & 3
        \end{MultiPattern}.
    \end{equation}

    \item {\bf Repetition.} For any multi-pattern $\MPat$ of multiplicity $m$ and any
    $k \geq 1$,
    \begin{equation}
        \Repetition_k(\MPat) := \MorphismCopy_m\Par{0^k} \HomogeneousComposition \MPat
    \end{equation}
    is the $k$-fold
    repetition of $\MPat$. For instance,
    \begin{equation}
        \Repetition_3\Par{
        \begin{MultiPattern}
            0 & \Rest & 0 & 1 \\
            \bar{1} & \bar{1} & \Rest & 3
        \end{MultiPattern}}
        =
        \begin{MultiPattern}
            0 & \Rest & 0 & 1 & 0 & \Rest & 0 & 1 & 0 & \Rest & 0 & 1 \\
            \bar{1} & \bar{1} & \Rest & 3 & \bar{1} & \bar{1} & \Rest & 3 & \bar{1}
                & \bar{1} & \Rest & 3
        \end{MultiPattern}.
    \end{equation}

    \item {\bf Transposition.} For any multi-pattern $\MPat$ of multiplicity $m$ and any
    $d \in \Z$,
    \begin{equation}
        \Transposition_d(\MPat) := \MorphismCopy_m(d) \HomogeneousComposition \MPat
    \end{equation}
    is the transposition of $\MPat$ by $d$ degrees. For instance,
    \begin{equation}
        \Transposition_{-2}\Par{
        \begin{MultiPattern}
            \bar{2} & \Rest & \Rest & 1 & \Rest \\
            \Rest & 2 & \Rest & 3 & \Rest
        \end{MultiPattern}}
        =
        \begin{MultiPattern}
            \bar{4} & \Rest & \Rest & \bar{1} & \Rest \\
            \Rest & 0 & \Rest & 1 & \Rest
        \end{MultiPattern}.
    \end{equation}

    \item {\bf Temporization.} For any multi-pattern $\MPat$ of multiplicity $m$ and any
    $k \geq 0$,
    \begin{equation}
        \Temporization_k(\MPat) :=
        \MPat \HomogeneousComposition \MorphismCopy_m\Par{0 \Rest^k}
    \end{equation}
    is the temporization of $\MPat$ by $k$ units of time. For instance,
    \begin{equation}
        \Temporization_2\Par{
        \begin{MultiPattern}
            \bar{2} & \Rest & \Rest & 1 & \Rest \\
            \Rest & 2 & \Rest & 3 & \Rest
        \end{MultiPattern}}
        =
        \begin{MultiPattern}
            \bar{2} & \Rest & \Rest & \Rest & \Rest & 1 & \Rest & \Rest & \Rest \\
            \Rest & 2 & \Rest & \Rest & \Rest & 3 & \Rest & \Rest & \Rest
        \end{MultiPattern}.
    \end{equation}

    \item {\bf Inverse.} For any multi-pattern $\MPat$,
    \begin{equation}
        \Inverse(\MPat) := \MorphismMul_{-1, \dots, -1}(\MPat)
    \end{equation}
    is the inverse of $\MPat$. For instance,
    \begin{equation}
        \Inverse\Par{
        \begin{MultiPattern}
            \bar{2} & \Rest & \Rest & 1 & \Rest \\
            \Rest & 2 & \Rest & 3 & \Rest
        \end{MultiPattern}}
        =
        \begin{MultiPattern}
            2 & \Rest & \Rest & \bar{1} & \Rest \\
            \Rest & \bar{2} & \Rest & \bar{3} & \Rest
        \end{MultiPattern}.
    \end{equation}

    \item {\bf Retrograde.} For any multi-pattern $\MPat$, $\Mirror(\MPat)$ is the
    retrograde of $\MPat$. For instance,
    \begin{equation}
        \Mirror\Par{
        \begin{MultiPattern}
            \bar{2} & \Rest & \Rest & 1 & \Rest \\
            \Rest & 2 & \Rest & 3 & \Rest
        \end{MultiPattern}}
        =
        \begin{MultiPattern}
            \Rest & 1 & \Rest & \Rest & \bar{2} \\
            \Rest & 3 & \Rest & 2 & \Rest
        \end{MultiPattern}.
    \end{equation}

    \item {\bf Retrograde inverse.} For any multi-pattern $\MPat$,
    \begin{equation}
        \RetrogradeInverse(\MPat) := \Mirror(\Inverse(\MPat)).
    \end{equation}
    is the retrograde inverse of $\MPat$. Notice that $\RetrogradeInverse(\MPat)$ is also
    equal to $\Inverse(\Mirror(\MPat))$ since the two maps $\Mirror$ and $\Inverse$ commute.
    For instance,
    \begin{equation}
        \RetrogradeInverse\Par{
        \begin{MultiPattern}
            \bar{2} & \Rest & \Rest & 1 & \Rest \\
            \Rest & 2 & \Rest & 3 & \Rest
        \end{MultiPattern}}
        =
        \begin{MultiPattern}
            \Rest & \bar{1} & \Rest & \Rest & 2 \\
            \Rest & \bar{3} & \Rest & \bar{2} & \Rest
        \end{MultiPattern}.
    \end{equation}

    \item {\bf Harmonization.} For any pattern $\Pat$ and any chord multi-pattern $\MPat$ of
    multiplicity $m$,
    \begin{equation}
        \Harmonization(\Pat, \MPat) := \MorphismCopy_m(\Pat) \HomogeneousComposition \MPat
    \end{equation}
    is the harmonization of $\Pat$ according to $\MPat$. For instance,
    \begin{equation}
        \Harmonization\Par{
        \begin{MultiPattern}
            \bar{1} & \Rest & 1 & \Rest & 0 & 2
        \end{MultiPattern},
        \begin{MultiPattern}
            0 \\
            2 \\
            4
        \end{MultiPattern}}
        =
        \begin{MultiPattern}
            \bar{1} & \Rest & 1 & \Rest & 0 & 2 \\
            1 & \Rest & 3 & \Rest & 2 & 4 \\
            3 & \Rest & 5 & \Rest & 4 & 6
        \end{MultiPattern}.
    \end{equation}

    \item {\bf Arpeggiation.} For any pattern $\Pat$ and any arpeggio multi-pattern $\MPat$
    of multiplicity $m$,
    \begin{equation}
        \Arpeggiation(\Pat, \MPat) := \MorphismCopy_m(\Pat) \HomogeneousComposition \MPat
    \end{equation}
    is the arpeggiation of $\Pat$ according to $\MPat$. For instance,
    \begin{equation}
        \Arpeggiation\Par{
        \begin{MultiPattern}
            \bar{1} & \Rest & 1 & \Rest & 0 & 2
        \end{MultiPattern},
        \begin{MultiPattern}
            1 & \Rest & \Rest \\
            \Rest & \Rest & 4 \\
            \Rest & \bar{2} & \Rest
        \end{MultiPattern}}
        =
        \begin{MultiPattern}
            0 & \Rest & \Rest
            & \Rest
            & 2 & \Rest & \Rest
            & \Rest
            & 1 & \Rest & \Rest
            & 3 & \Rest & \Rest \\
            \Rest & \Rest & 3
            & \Rest
            & \Rest & \Rest & 5
            & \Rest
            & \Rest & \Rest & 4
            & \Rest & \Rest & 6 \\
            \Rest & \bar{3} & \Rest
            & \Rest
            & \Rest & \bar{1} & \Rest
            & \Rest
            & \Rest & \bar{2} & \Rest
            & \Rest & 0 & \Rest
        \end{MultiPattern}.
    \end{equation}
\end{enumerate}
Any operation preserving the lengths and the arities is \Def{homogeneous}. The operations
$\Transposition_d$, $d \in \Z$, $\Inverse$, $\Mirror$, and $\RetrogradeInverse$ are
homogeneous.

We are now in position to be more specific in what we explained at the end of
Section~\ref{subsubsec:strengths_music_box_model} about the fact that multi-patterns are
operations on musical phrases. On the one hand, this feature is illustrated by the existence
of the previous operations. On the other hand, observe that if $\MPat$ is a multi-pattern,
it can be seen as an operation taking as input $|\MPat|$ multi-patterns $\MPat_1, \dots,
\MPat_{|\MPat|}$ and outputting the multi-pattern $\MPat \circ \Han{\MPat_1, \dots,
\MPat_{|\MPat|}}$. For instance, the multi-pattern
\begin{equation}
    \MPat :=
    \begin{MultiPattern}
        \bar{1} & 0 & 2
    \end{MultiPattern}
\end{equation}
can be seen as an operation of arity $3$ such that, for any multi-patterns $\MPat_1$,
$\MPat_2$, and $\MPat_3$ of multiplicity $1$ as inputs, in $\OperadMP_1^\Z$, $\MPat \circ
\Han{\MPat_1, \MPat_2, \MPat_3}$ is the multi-pattern of multiplicity $1$ obtained by
concatenating a version of $\MPat_1$ transposed one degree down, with a copy of $\MPat_2$,
and with a version of $\MPat_3$ transposed two degrees up. Similarly, the multi-pattern
\begin{equation}
    \MPat :=
    \begin{MultiPattern}
        0 & \Rest \\
        \Rest & 0
    \end{MultiPattern}
\end{equation}
can be seen an operation of arity $1$ such that, for any multi-pattern $\MPat_1$ of
multiplicity $2$ as input, in $\OperadMP_2^\Z$, $\MPat \circ \Han{\MPat_1}$ is the
multi-pattern obtained by adding a final rest to the first pattern of $\MPat_1$ and by
adding an initial rest to the second pattern of $\MPat_1$. The interpretation of this
result is a musical phrase wherein the second voice has been shifted one unit of time. A
lot of other operations modifying both the degrees and the rhythm can be constructed in this
way.

\subsubsection{Some benefits}
Let us conclude this section by highlighting two additional significant benefits of the
music box model, in addition to the ones presented in
Section~\ref{subsubsec:strengths_music_box_model} for multi-patterns:
\begin{enumerate}
    \item the model is homogeneous;
    \item the composition of multi-patterns is flexible.
\end{enumerate}

The first benefit concerns the homogeneity of the music box model. Operations on musical
phrases are represented by using the same language as the musical phrases themselves, due to
the fact that multi-patterns are operations (as mentioned in
Section~\ref{subsubsec:particular_multi_patterns}). This is in contrast to other systems
where musical phrases and operations are defined using different languages, as seen in some
other models (see for instance~\cite{FV13}).

The second benefit, about the flexibility of the composition, is a consequence of the fact
that the $D$-bud music box operad is parameterizable with different degree monoids $D$. By
considering different degree monoids, it is possible to drastically change the behavior of
the operations on multi-patterns and, as a result, the musical phrases generated by the
model. While this section has mainly focused on $D$-music box operads with $D := (\Z, +, 0)$
as the degree monoid, other monoids may provide interesting operations on multi-patterns and
musical phrases.

\section{Generation and random generation} \label{sec:random_generation}
Now we exploit the music box operads introduced in the previous section to design three
random generation algorithms devoted to generate new musical phrases from a finite set of
multi-patterns. This relies on colored operads and bud generating systems, a sort of formal
grammars introduced in~\cite{Gir19}.

\subsection{Colored operads and bud operads}
We provide here the elementary notions about colored operads~\cite{Yau16} used in this work.
We also explain how to build colored operads from an operad.

\subsubsection{Colored operads}
A \Def{set of colors} is any nonempty finite set
\begin{math}
    \SetColors := \Bra{\Color_1, \dots, \Color_r}
\end{math}
wherein elements are called \Def{colors}. A \Def{$\SetColors$-colored set} is a set
$\ColoredOperad$ decomposing as a disjoint union
\begin{equation}
    \ColoredOperad
    :=
    \bigsqcup_{\substack{
        a \in \SetColors \\
        u \in \SetColors^*}}
    \ColoredOperad(a, u),
\end{equation}
where the $\ColoredOperad(a, u)$, $a \in \SetColors$, $u \in \SetColors^*$, are sets. For
any $x \in \ColoredOperad$, there is by definition a unique pair $(a, u) \in \SetColors
\times \SetColors^*$ such that $x \in \ColoredOperad(a, u)$. The \Def{arity} $|x|$ of $x$ is
the length $\Length(u)$ of $u$ as a word, the \Def{output color} $\Out(x)$ of $x$ is $a$,
and for any $i \in [|x|]$, the \Def{$i$-th input color} $\In_i(x)$ of $x$ is the $i$-th
letter $u(i)$ of $u$. We also denote, for any $n \in \N$, by $\ColoredOperad(n)$ the set of
the elements of $\ColoredOperad$ of arity $n$. Therefore, a colored set is in particular a
graded set.

A \Def{$\SetColors$-colored operad} is a triple $\Par{\ColoredOperad, \circ_i, \Unit}$ such
that $\ColoredOperad$ is a $\SetColors$-colored set, $\circ_i$ is a map
\begin{equation}
    \circ_i : \ColoredOperad(a, u) \times \ColoredOperad\Par{u(i), v}
    \to \ColoredOperad\Par{a, u \circ_i v},
    \qquad 1 \leq i \leq \Length(u),
\end{equation}
called a \Def{partial composition} map, where $u \circ_i v$ is the word on $\SetColors$
obtained by replacing the $i$-th letter $u(i)$ of $u$ by $v$, and $\Unit$ is a map
\begin{equation}
    \Unit : \SetColors \to \ColoredOperad(a, a),
\end{equation}
such that for any $a \in \SetColors$, $\Unit(a) \in \ColoredOperad(a, a)$, called
a \Def{colored unit} map. This data has to satisfy Relations~\eqref{equ:operad_axiom_1}
and~\eqref{equ:operad_axiom_2} when their left and right members are both well-defined, and,
for any $x \in \ColoredOperad$, the relation
\begin{equation}
    \Unit\Par{\Out(x)} \circ_1 x = x = x \circ_i \Unit\Par{\In_i(x)},
    \qquad 1 \leq i \leq |x|.
\end{equation}

Intuitively, an element $x$ of a colored operad having $a$ as output color and $u(i)$ as
$i$-th input color for any $i \in [|x|]$ can be seen as an abstract operator wherein colors
are assigned to its output and to each of its inputs. Such an operator is depicted as
\begin{equation}
    \begin{tikzpicture}[Centering,xscale=.35,yscale=.4,font=\scriptsize]
        \node[NodeST](x)at(0,0){$x$};
        \node(r)at(0,2.75){};
        \node(x1)at(-3,-2){};
        \node(xn)at(3,-2){};
        \node[below of=x1,node distance=1mm](ex1){$1$};
        \node[below of=xn,node distance=1mm](exn){$|x|$};
        \draw[Edge](r)edge[]node[EdgeLabel,inner sep=1pt]{$a$}(x);
        \draw[Edge](x)edge[]node[EdgeLabel,inner sep=1pt]{$u(1)$}(x1);
        \draw[Edge](x)edge[]node[EdgeLabel,inner sep=1pt]{$u(|x|)$}(xn);
        \node[below of=x,node distance=9mm]{$\dots$};
    \end{tikzpicture},
\end{equation}
where the colors of the output and inputs are the squares put on the corresponding edges.
The partial composition of two elements $x$ and $y$ of a colored operad expresses
pictorially as
\begin{equation}
    \begin{tikzpicture}[Centering,xscale=.5,yscale=.5,font=\scriptsize]
        \node[NodeST](x)at(0,0){$x$};
        \node(r)at(0,2.25){};
        \node(x1)at(-3,-2){};
        \node(xn)at(3,-2){};
        \node(xi)at(0,-2){};
        \node[below of=x1,node distance=1mm](ex1){$1$};
        \node[below of=xn,node distance=1mm](exn){$|x|$};
        \node[below of=xi,node distance=1mm](exi){$i$};
        \draw[Edge](r)edge[]node[EdgeLabel,inner sep=1pt]{$a$}(x);
        \draw[Edge](x)edge[]node[EdgeLabel,inner sep=1pt]{$u(1)$}(x1);
        \draw[Edge](x)edge[]node[EdgeLabel,inner sep=1pt]{$u(|x|)$}(xn);
        \draw[Edge](x)edge[]node[EdgeLabel,inner sep=1pt]{$u(i)$}(xi);
        \node[right of=ex1,node distance=7mm]{$\dots$};
        \node[left of=exn,node distance=7mm]{$\dots$};
    \end{tikzpicture}
    \enspace \circ_i \enspace
    \begin{tikzpicture}[Centering,xscale=.35,yscale=.4,font=\scriptsize]
        \node[NodeST](x)at(0,0){$y$};
        \node(r)at(0,2.75){};
        \node(x1)at(-3,-2){};
        \node(xn)at(3,-2){};
        \node[below of=x1,node distance=1mm](ex1){$1$};
        \node[below of=xn,node distance=1mm](exn){$|y|$};
        \draw[Edge](r)edge[]node[EdgeLabel,inner sep=1pt]{$u(i)$}(x);
        \draw[Edge](x)edge[]node[EdgeLabel,inner sep=1pt]{$v(1)$}(x1);
        \draw[Edge](x)edge[]node[EdgeLabel,inner sep=1pt]{$v(|y|)$}(xn);
        \node[below of=x,node distance=9mm]{$\dots$};
    \end{tikzpicture}
    \enspace = \enspace
    \begin{tikzpicture}[Centering,xscale=.6,yscale=.5,font=\scriptsize]
        \node[NodeST](x)at(0,0){$x$};
        \node(r)at(0,2){};
        \node(x1)at(-3,-2){};
        \node(xn)at(3,-2){};
        \node[below of=x1,node distance=1mm](ex1){$1$};
        \node[below of=xn,node distance=1mm](exn){$|x| + |y| - 1$};
        \node[right of=ex1,node distance=8mm]{$\dots$};
        \node[left of=exn,node distance=9mm]{$\dots$};
        \draw[Edge](r)edge[]node[EdgeLabel,inner sep=1pt]{$a$}(x);
        \draw[Edge](x)edge[]node[EdgeLabel,inner sep=1pt]{$u(1)$}(x1);
        \draw[Edge](x)edge[]node[EdgeLabel,inner sep=1pt]{$u(|x|)$}(xn);
        \node[NodeST](y)at(0,-2.5){$y$};
        \node(y1)at(-1.6,-4.5){};
        \node(yn)at(1.6,-4.5){};
        \node[below of=y1,node distance=1mm](ey1){$i$};
        \node[below of=yn,node distance=1mm](eyn){$i + |y| - 1$};
        \draw[Edge](y)edge[]node[EdgeLabel,inner sep=1pt]{$v(1)$}(y1);
        \draw[Edge](y)edge[]node[EdgeLabel,inner sep=1pt]{$v(|y|)$}(yn);
        \node[below of=y,node distance=11mm]{$\dots$};
        \draw[Edge](x)edge[]node[EdgeLabel,inner sep=1pt]{$u(i)$}(y);
    \end{tikzpicture}.
\end{equation}
Most of the definitions about operads recalled in
Section~\ref{subsubsec:elementary_definitions_operads} generalize straightforwardly to
colored operads. In particular, one can consider the full composition map of a colored
operad defined by~\eqref{equ:full_composition_maps} when its right member is well-defined.

The situation is specific for the homogeneous composition for colored operads. Let
$\Par{\ColoredOperad, \circ_i, \Unit}$ be a colored operad. The \Def{homogeneous
composition} map of $\ColoredOperad$ is the map
\begin{equation}
    \HomogeneousComposition :
    \ColoredOperad(a, u) \times \ColoredOperad(b, v) \to \ColoredOperad,
    \qquad
    a, b \in \SetColors,
    \enspace
    u, v \in \SetColors^*,
\end{equation}
defined, for any $x \in \ColoredOperad(a, u)$ and $y \in \ColoredOperad(b, v)$, by using the
full composition map, by
\begin{equation}
    x \HomogeneousComposition y := x \circ \Han{y_1, \dots, y_{|x|}},
\end{equation}
where for any $i \in [|x|]$,
\begin{equation}
    y_i :=
    \begin{cases}
        y & \mbox{if } \In_i(x) = \Out(y), \\
        \Unit\Par{\In_i(x)} & \mbox{otherwise}.
    \end{cases}
\end{equation}
Intuitively, $x \HomogeneousComposition y$ is obtained by grafting simultaneously the
outputs of copies of $y$ onto the inputs of $x$ having the same color as the output color
of~$y$. If the set of colors of $\ColoredOperad$ is a singleton, the homogeneous composition
map of $\ColoredOperad$ is the homogeneous composition map of operads described in
Section~\ref{subsubsec:elementary_definitions_operads}.

\subsubsection{Bud operads}
Let us describe a general construction building a colored operad from a noncolored one
introduced in~\cite{Gir19}. Given a noncolored operad $\Par{\Operad, \circ_i, \Unit}$ and a
set $\SetColors$ of colors, the \Def{$\SetColors$-bud operad} of $\Operad$ is the
$\SetColors$-colored operad $\BudOperad_\SetColors(\Operad)$ defined in the following way.
First, $\BudOperad_\SetColors(\Operad$) is the $\SetColors$-colored set defined, for any $a
\in \SetColors$ and $u \in \SetColors^*$, by
\begin{equation}
    \BudOperad_\SetColors(\Operad)(a, u) := \Bra{(a, x, u) : x \in \Operad(\Length(u))}.
\end{equation}
Second, the partial composition maps $\circ_i$ of $\BudOperad_\SetColors(\Operad)$ are
defined, for any $(a, x, u), \Par{u(i), y, v} \in \BudOperad_\SetColors(\Operad)$, and $i
\in [|u|]$, by
\begin{equation} \label{equ:partial_composition_map_bud_operad}
    (a, x, u) \circ_i \Par{u_i, y, v} := \Par{a, x \circ_i y, u \circ_i v}
\end{equation}
where the first occurrence of $\circ_i$ in the right member
of~\eqref{equ:partial_composition_map_bud_operad} is the partial composition map of
$\Operad$ and the second one is a substitution of words: $u \circ_i v$ is the word obtained
by replacing in $u$ the $i$-th letter $u(i)$ of $u$ by $v$.  Finally, the colored unit map
$\Unit$ of $\BudOperad_\SetColors(\Operad)$ is defined by $\Unit(a) := \Par{a, \Unit, a}$
for any $a \in \SetColors$, where $\Unit$ is the unit of $\Operad$. The \Def{pruning}
$\Prune((a, x, u))$ of an element $(a, x, u)$ of $\BudOperad_\SetColors(\Operad)$ is the
element $x$ of~$\Operad$.

Intuitively, this construction consists in forming a colored operad
$\BudOperad_\SetColors(\Operad)$ out of $\Operad$ by surrounding its elements with an output
color and input colors coming from $\SetColors$ in all possible ways.

For a fixed degree monoid $D$, we apply this construction to the $D$-music box operad by
setting, for any set $\SetColors$ of colors,
\begin{equation}
    \BudOperadMP^D_{m, \SetColors} := \BudOperad_\SetColors\Par{\OperadMP^D_m}.
\end{equation}
We call $\BudOperadMP^D_{m, \SetColors}$ the \Def{$\SetColors$-bud $D$-music box operad}.
The elements of $\BudOperadMP^D_{m, \SetColors}$ are called \Def{$\SetColors$-colored
multi-patterns}. For instance, for $\SetColors := \Bra{\Color_1, \Color_2, \Color_3}$,
\begin{equation}
    \Par{\Color_1,
    \begin{MultiPattern}
        1 & \Rest & 0 & \Rest & 1 \\
        \bar{7} & \Rest & 0 & 0 & \Rest
    \end{MultiPattern},
    \Color_2 \Color_2 \Color_1}
\end{equation}
is a $\SetColors$-colored multi-pattern. Moreover, in the colored operad
$\BudOperadMP^\Z_{2, \SetColors}$, one has
\begin{equation}
    \Par{\Color_3,
    \begin{MultiPattern}
        0 & 1 & \Rest \\
        \bar{1} & \Rest & 0
    \end{MultiPattern},
    \Color_2 \Color_1}
    \circ_2
    \Par{\Color_1,
    \begin{MultiPattern}
        1 & 1 & 2 \\
        2 & \bar{1} & \bar{2}
    \end{MultiPattern},
    \Color_3 \Color_3 \Color_2}
    =
    \Par{\Color_3,
    \begin{MultiPattern}
        0 & 2 & 2 & 3 & \Rest \\
        \bar{1} & \Rest & 2 & \bar{1} & \bar{2}
    \end{MultiPattern},
    \Color_2\Color_3 \Color_3 \Color_2}.
\end{equation}

The intuition that justifies the introduction of these colored versions of multi-patterns
and of the $D$-music box operad is that colors restrict the right to perform the composition
of two given multi-patterns. In this way, on a small scale, one can for instance forbid some
intervals in the musical phrases specified by the multi-patterns of a suboperad of
$\BudOperadMP^D_{m, \SetColors}$ generated by a given set of $\SetColors$-colored
multi-patterns. On a larger scale, the colors allow us to build complex phrases by imposing
some specific parts (for instance, a color might be used for the beginning of a piece,
another for the middle, and a third one for the end).

Besides, given a set $\GeneratingSet$ of $\SetColors$-colored multi-patterns, the elements
of the suboperad of $\BudOperadMP^D_{m, \SetColors}$ generated by $\GeneratingSet$ are
obtained by composing elements of $\GeneratingSet$. Therefore, in some sense, these elements
inherit from properties of the multi-patterns of~$\GeneratingSet$. The next section uses
these ideas to propose random generation algorithms outputting new multi-patterns from
existing ones in a controlled way.

\subsection{Bud generating systems} \label{subsec:bud_generating_systems}
We describe here a sort of generating systems using operads and colored operads introduced
in~\cite{Gir19}. Slight variations are considered in this present work.

\subsubsection{Bud generating systems} \label{subsubsec:bud_generating_systems}
A \Def{bud generating system}~\cite{Gir19} is a tuple $\Par{\Operad, \SetColors, \SetRules,
\Color}$ where
\begin{enumerate}[label={\em (\roman*)}]
    \item $\Par{\Operad, \circ_i, \Unit}$ is an operad, called \Def{ground operad};
    \item $\SetColors$ is a set of colors;
    \item $\SetRules$ is a finite subset of $\BudOperad_\SetColors(\Operad)$, called
    \Def{set of rules};
    \item $\Color$ is a color of $\SetColors$, called \Def{initial color}.
\end{enumerate}
For any color $a \in \SetColors$, we shall denote by $\SetRules(a)$ the set of the rules of
$\SetRules$ having $a$ as output color.

For instance,
\begin{math}
    \BudSystem_{\mathrm{ex}} :=
    \Par{\OperadMP^\Z_2, \Bra{\Color_1, \Color_2, \Color_3},
    \Bra{\CPat_1, \CPat_2, \CPat_3, \CPat_4, \CPat_5}, \Color_1}
\end{math}
where
\begin{multline}
    \CPat_1 :=
    \Par{\Color_1,
    \begin{MultiPattern}
        0 & 2 & \Rest & 1 & \Rest & 0 & 4 \\
        \bar{5} & \Rest & \Rest & 0 & 0 & 0 & 0
    \end{MultiPattern},
    \Color_3 \Color_2 \Color_1 \Color_1 \Color_3},
    \qquad
    \CPat_2 := \Par{\Color_1,
    \begin{MultiPattern}
        1 & \Rest & 0 \\
        0 & \Rest & 1
    \end{MultiPattern},
    \Color_1 \Color_1},
    \\
    \CPat_3 := \Par{\Color_2,
    \begin{MultiPattern}
        \bar{1} \\
        \bar{1}
    \end{MultiPattern},
    \Color_1},
    \qquad
    \CPat_4 := \Par{\Color_2,
    \begin{MultiPattern}
        0 & 0 \\
        0 & 0
    \end{MultiPattern},
    \Color_1 \Color_1},
    \qquad
    \CPat_5 := \Par{\Color_3,
    \begin{MultiPattern}
        0 \\
        0
    \end{MultiPattern},
    \Color_3}
\end{multline}
is a bud generating system.

Bud generating systems are devices similar to context-free formal grammars~\cite{HMU06}
wherein colors play the role of nonterminal symbols. Each element $(a, x, u) \in \SetRules$
can be seen as a production rule of the form $a \to (x, u)$. The color $a$ plays the role of
a nonterminal symbol, $u$ is a word of symbols wherein each letter can be seen as a terminal
or a nonterminal symbol, and $x$ denotes additional information. As we shall see, this
information is very important because bud generating systems are intended to generate
elements of $\Operad$.

If context-free grammars are devices designed to generate sets of words, bud generating
systems are designed to generate more general combinatorial objects. More precisely, a bud
generating system $\Par{\Operad, \SetColors, \SetRules, \Color}$ allows us to build elements
of $\Operad$ by following three different operating modes. We describe in the next sections
the three corresponding random generation algorithms. These algorithms are in particular
intended to work with $\OperadMP^D_m$ as ground operad in order to generate multi-patterns.

\subsubsection{Random generation}
In the next three sections, we shall design three random generation algorithms with the goal
to randomly generate elements of $\Operad$ given a bud generating system having $\Operad$ as
ground operad. In these sections, $\BudSystem := (\Operad, \SetColors, \SetRules, \Color)$
is a bud generating system. We shall also consider the bud generating system
$\BudSystem_{\mathrm{ex}}$ introduced in Section~\ref{subsubsec:bud_generating_systems} to
provide some concrete examples. Besides, for any finite and nonempty set $S$, $\Random(S)$
is an element of $S$ picked uniformly at random among the elements of~$S$.

\subsubsection{Partial generation}
Let $\PartialDerivation$ be the binary relation on $\BudOperad_\SetColors(\Operad)$ such
that $(a, x, u) \PartialDerivation (a, y, v)$ if there is a rule $r\in \SetRules$ and $i \in
[|u|]$ such that
\begin{equation}
    (a, y, v) = (a, x, u) \circ_i r.
\end{equation}
An element $x$ of $\Operad$ is \Def{partially generated} by $\BudSystem$ if there is an
element $(\Color, x, u)$ such that $\Par{\Color, \Unit, \Color}$ is in relation with
$(\Color, x, u)$ w.r.t.\ the reflexive and transitive closure of~$\PartialDerivation$.

For instance, by considering the bud generating system $\BudSystem_{\mathrm{ex}}$, since
\begin{equation} \begin{split}
    \Par{\Color_1,
    \begin{MultiPattern}
        0 \\
        0
    \end{MultiPattern},
    \Color_1}
    & \PartialDerivation
    \Par{\Color_1,
    \begin{MultiPattern}
        1 & \Rest & 0 \\
        0 & \Rest & 1
    \end{MultiPattern},
    \Color_1 \Color_1} \\
    & \PartialDerivation
    \Par{\Color_1,
    \begin{MultiPattern}
        1 & \Rest & 0 & 2 & \Rest & 1 & \Rest & 0 & 4 \\
        0 & \Rest & \bar{4} & \Rest & \Rest & 1 & 1 & 1 & 1
    \end{MultiPattern},
    \Color_1 \Color_3 \Color_2 \Color_1 \Color_1 \Color_3}
    \\
    & \PartialDerivation
    \Par{\Color_1,
    \begin{MultiPattern}
        1 & \Rest & 0 & 2 & 2 & \Rest & 1 & \Rest & 0 & 4 \\
        0 & \Rest & \bar{4} & \Rest & \Rest & 1 & 1 & 1 & 1 & 1
    \end{MultiPattern},
    \Color_1 \Color_3 \Color_1 \Color_1 \Color_1 \Color_1 \Color_3},
\end{split} \end{equation}
the multi-pattern
\begin{equation}
    \begin{MultiPattern}
        1 & \Rest & 0 & 2 & 2 & \Rest & 1 & \Rest & 0 & 4 \\
        0 & \Rest & \bar{4} & \Rest & \Rest & 1 & 1 & 1 & 1 & 1
    \end{MultiPattern}
\end{equation}
is partially generated by~$\BudSystem_{\mathrm{ex}}$.

Algorithm~\ref{algo:partial_random_generation} returns an element partially generated by
$\BudSystem$ obtained by applying at most $k$ rules to the initial element $(\Color, \Unit,
\Color)$. The execution of the algorithm builds a syntax tree of elements of $\SetRules$
with at most $k$ internal nodes.
\begin{center}
    \begin{Page}{.83}
        \begin{algorithm}[H]
            \KwData{A bud generating system
                $\BudSystem := (\Operad, \SetColors, \SetRules, \Color)$ and an integer
                $k \in \N$.}
            \KwResult{A randomly generated element of $\Operad$.}
            \Begin{
                $x \gets (\Color, \Unit, \Color)$ \\
                \For{$j \in [k]$}{
                    $i \gets \Random([|x|])$ \\
                    \If{$\SetRules\Par{\In_i(x)} \ne \emptyset$}{
                        $r \gets \Random\Par{\SetRules\Par{\In_i(x)}}$ \\
                        $x \gets x \circ_i r$ \\
                    }
                }
                \Return{$\Prune(x)$}
            }
            \caption{Partial random generation.}
            \label{algo:partial_random_generation}
        \end{algorithm}
    \end{Page}
\end{center}

For instance, by considering the previous bud generating system $\BudSystem_{\mathrm{ex}}$,
this algorithm run with $k := 5$ builds the syntax tree of colored multi-patterns
\begin{equation}
    \begin{tikzpicture}[Centering,xscale=0.24,yscale=0.18]
        \node(0)at(-1.00,-3.5){};
        \node(10)at(7.00,-11.0){};
        \node(11)at(8.00,-11.0){};
        \node(12)at(8.00,-3.5){};
        \node(14)at(11.00,-7.25){};
        \node(2)at(2.00,-7.25){};
        \node(4)at(4.00,-7.25){};
        \node(6)at(4.00,-11.0){};
        \node(7)at(5.00,-11.0){};
        \node(9)at(6.00,-11.0){};
        \node[NodeST](1)at(2.00,-3.75){$\CPat_3$};
        \node[NodeST](13)at(11.00,-3.75){$\CPat_5$};
        \node[NodeST](3)at(5.00,0.00){$\CPat_1$};
        \node[NodeST](5)at(5.00,-3.75){$\CPat_2$};
        \node[NodeST](8)at(6.00,-7.50){$\CPat_1$};
        \draw[Edge](0)--(3);
        \draw[Edge](1)--(3);
        \draw[Edge](10)--(8);
        \draw[Edge](11)--(8);
        \draw[Edge](12)--(3);
        \draw[Edge](13)--(3);
        \draw[Edge](14)--(13);
        \draw[Edge](2)--(1);
        \draw[Edge](4)--(5);
        \draw[Edge](5)--(3);
        \draw[Edge](6)--(8);
        \draw[Edge](7)--(8);
        \draw[Edge](8)--(5);
        \draw[Edge](9)--(8);
        \node(r)at(5.00,2.81){};
        \draw[Edge](r)--(3);
    \end{tikzpicture}
\end{equation}
which produces, when evaluated, the multi-pattern
\begin{equation}
    \begin{MultiPattern}
        0 & 1 & \Rest & 2 & \Rest & 1 & 3 & \Rest & 2 & \Rest & 1 & 5 & \Rest & 0 & 4 \\
        \bar{5} & \Rest & \Rest & \bar{1} & 0 & \Rest & \bar{4} & \Rest & \Rest & 1 & 1 & 1
            & 1 & 0 & 0
    \end{MultiPattern}.
\end{equation}

\subsubsection{Full generation}
Let $\FullDerivation$ be the binary relation on $\BudOperad_\SetColors(\Operad)$ such
that $(a, x, u) \FullDerivation (a, y, v)$ if there are rules $r_1, \dots, r_{|x|}
\in \SetRules$ such that
\begin{equation}
    (a, y, v) = (a, x, u) \circ \Han{r_1, \dots, r_{|x|}}.
\end{equation}
An element $x$ of $\Operad$ is \Def{fully generated} by $\BudSystem$ if there is an element
$(\Color, x, u)$ such that $\Par{\Color, \Unit, \Color}$ is in relation with $(\Color, x,
u)$ w.r.t.\ the reflexive and transitive closure of~$\FullDerivation$.

For instance, by considering the bud generating system $\BudSystem_{\mathrm{ex}}$, since
\begin{equation} \begin{split}
    \Par{\Color_1,
    \begin{MultiPattern}
        0 \\
        0
    \end{MultiPattern},
    \Color_1}
    & \FullDerivation
    \Par{\Color_1,
    \begin{MultiPattern}
        0 & 2 & \Rest & 1 & \Rest & 0 & 4 \\
        \bar{5} & \Rest & \Rest & 0 & 0 & 0 & 0
    \end{MultiPattern},
    \Color_3 \Color_2 \Color_1 \Color_1 \Color_3}
    \\
    & \FullDerivation
    \Par{\Color_1,
    \begin{MultiPattern}
        0 & 1 & \Rest & 2 & \Rest & 1 & \Rest & 0 & 2 & \Rest & 1 & \Rest & 0 & 4 & 4 \\
        \bar{5} & \Rest & \Rest & \bar{1} & 0 & \Rest & 1 & \bar{5} & \Rest & \Rest & 0 & 0
            & 0 & 0 & 0
    \end{MultiPattern},
    \Color_3 \Color_1 \Color_1 \Color_1 \Color_3 \Color_2 \Color_1 \Color_1 \Color_3
        \Color_3},
\end{split} \end{equation}
the multi-pattern
\begin{equation}
    \begin{MultiPattern}
        0 & 1 & \Rest & 2 & \Rest & 1 & \Rest & 0 & 2 & \Rest & 1 & \Rest & 0 & 4 & 4 \\
        \bar{5} & \Rest & \Rest & \bar{1} & 0 & \Rest & 1 & \bar{5} & \Rest & \Rest & 0 & 0
            & 0 & 0 & 0
    \end{MultiPattern}
\end{equation}
is fully generated by $\BudSystem_{\mathrm{ex}}$.

Bud generating systems together with this scheme for generation are very similar to
Lindenmayer systems, which are sorts of formal grammars~\cite{Lin68}. Such systems lead to
frameworks to generate musical phrases~\cite{MS94,McC96,HQ18}.

Algorithm~\ref{algo:full_random_generation} returns an element synchronously generated by
$\BudSystem$ obtained by applying at most $k$ rules to the initial element $(\Color, \Unit,
\Color)$. The execution of the algorithm builds a syntax tree of elements of $\SetRules$ of
height at most~$k + 1$ wherein the leaves are all at the same distance from the root.
\begin{center}
    \begin{Page}{.83}
        \begin{algorithm}[H]
            \KwData{A bud generating system
                $\BudSystem := (\Operad, \SetColors, \SetRules, \Color)$ and an integer
                $k \in \N$.}
            \KwResult{A randomly generated element of $\Operad$.}
            \Begin{
                $x \gets (\Color, \Unit, \Color)$ \\
                \For{$j \in [k]$}{
                    $R \gets
                    \SetRules\Par{\In_1(x)} \times \dots \times \SetRules\Par{\In_{|x|}(x)}$
                        \\
                    \If{$R \ne \emptyset$}{
                        $\Par{r_1, \dots, r_{|x|}} \gets \Random(R)$ \\
                        $x \gets x \circ \Han{r_1, \dots, r_{|x|}}$ \\
                    }
                }
                \Return{$\Prune(x)$}
            }
            \caption{Full random generation.}
            \label{algo:full_random_generation}
        \end{algorithm}
    \end{Page}
\end{center}
Observe that when for all colors $\Color \in \SetColors$, the sets $\SetRules(\Color)$ have
no more than one element, this algorithm is deterministic.

For instance, by considering the previous bud generating system $\BudSystem_{\mathrm{ex}}$,
this algorithm run with $k := 2$ builds the syntax tree of colored multi-patterns
\begin{equation}
    \begin{tikzpicture}[Centering,xscale=0.24,yscale=0.12]
        \node(0)at(0.00,-17.00){};
        \node(1)at(1.00,-17.00){};
        \node(12)at(9.00,-17.00){};
        \node(14)at(11.00,-17.00){};
        \node(16)at(12.50,-17.00){};
        \node(18)at(13.50,-17.00){};
        \node(19)at(14.50,-17.00){};
        \node(21)at(15.50,-17.00){};
        \node(23)at(17.00,-17.00){};
        \node(3)at(2.00,-17.00){};
        \node(4)at(3.00,-17.00){};
        \node(5)at(4.00,-17.00){};
        \node(7)at(5.50,-17.00){};
        \node(9)at(6.50,-17.00){};
        \node[NodeST](10)at(9.00,0.00){$\CPat_2$};
        \node[NodeST](11)at(9.00,-12.00){$\CPat_5$};
        \node[NodeST](13)at(11.00,-12.00){$\CPat_3$};
        \node[NodeST](15)at(13.00,-6.00){$\CPat_1$};
        \node[NodeST](17)at(13.00,-12.00){$\CPat_2$};
        \node[NodeST](2)at(2.00,-12.00){$\CPat_1$};
        \node[NodeST](20)at(15.00,-12.00){$\CPat_2$};
        \node[NodeST](22)at(17.00,-12.00){$\CPat_5$};
        \node[NodeST](6)at(4.00,-6.00){$\CPat_2$};
        \node[NodeST](8)at(6.00,-12.00){$\CPat_2$};
        \draw[Edge](0)--(2);
        \draw[Edge](1)--(2);
        \draw[Edge](11)--(15);
        \draw[Edge](12)--(11);
        \draw[Edge](13)--(15);
        \draw[Edge](14)--(13);
        \draw[Edge](15)--(10);
        \draw[Edge](16)--(17);
        \draw[Edge](17)--(15);
        \draw[Edge](18)--(17);
        \draw[Edge](19)--(20);
        \draw[Edge](2)--(6);
        \draw[Edge](20)--(15);
        \draw[Edge](21)--(20);
        \draw[Edge](22)--(15);
        \draw[Edge](23)--(22);
        \draw[Edge](3)--(2);
        \draw[Edge](4)--(2);
        \draw[Edge](5)--(2);
        \draw[Edge](6)--(10);
        \draw[Edge](7)--(8);
        \draw[Edge](8)--(6);
        \draw[Edge](9)--(8);
        \node(r)at(9.00,4.00){};
        \draw[Edge](r)--(10);
    \end{tikzpicture}
\end{equation}
which produces, when evaluated, the multi-pattern
\begin{equation}
    \begin{MultiPattern}
        2 & 4 & \Rest & 3 & \Rest & 2 & 6 & \Rest & 2 & \Rest & 1 & \Rest & 0 & 1 & \Rest
            & 2 & \Rest & 1 & \Rest & 1 & \Rest & 0 & 4 \\
        \bar{5} & \Rest & \Rest & 0 & 0 & 0 & 0 & \Rest & 1 & \Rest & 2 & \Rest & \bar{4}
            & \Rest & \Rest & 0 & 1 & \Rest & 2 & 1 & \Rest & 2 & 1
    \end{MultiPattern}.
\end{equation}

\subsubsection{Homogeneous generation}
Let $\HomogeneousDerivation$ be the binary relation on $\BudOperad_\SetColors(\Operad)$ such
that $(a, x, u) \HomogeneousDerivation (a, y, v)$ if there is a rule $r \in \SetRules$ such
that
\begin{equation}
    (a, y, v) = (a, x, u) \HomogeneousComposition r.
\end{equation}
An element $x$ of $\Operad$ is \Def{homogeneously generated} by $\BudSystem$ if there is an
element $(\Color, x, u)$ such that $\Par{\Color, \Unit, \Color}$ is in relation with
$(\Color, x, u)$ w.r.t.\ the reflexive and transitive closure of~$\HomogeneousDerivation$.

For instance, by considering the bud generating system $\BudSystem_{\mathrm{ex}}$, since
\begin{equation} \begin{split}
    \Par{\Color_1,
    \begin{MultiPattern}
        0 \\
        0
    \end{MultiPattern},
    \Color_1}
    & \HomogeneousDerivation
    \Par{\Color_1,
    \begin{MultiPattern}
        0 & 2 & \Rest & 1 & \Rest & 0 & 4 \\
        \bar{5} & \Rest & \Rest & 0 & 0 & 0 & 0
    \end{MultiPattern},
    \Color_3 \Color_2 \Color_1 \Color_1 \Color_3}
    \\
    & \HomogeneousDerivation
    \Par{\Color_1,
    \begin{MultiPattern}
        0 & 2 & \Rest & 2 & \Rest & 1 & \Rest & 1 & \Rest & 0 & 4 \\
        \bar{5} & \Rest & \Rest & 0 & 0 & \Rest & 1 & 0 & \Rest & 1 & 0
    \end{MultiPattern},
    \Color_3 \Color_2 \Color_1 \Color_1 \Color_1 \Color_1 \Color_3}
    \\
    & \HomogeneousDerivation
    \Par{\Color_1,
    \begin{MultiPattern}
        0 & 1 & \Rest & 2 & \Rest & 1 & \Rest & 1 & \Rest & 0 & 4 \\
        \bar{5} & \Rest & \Rest & \bar{1} & 0 & \Rest & 1 & 0 & \Rest & 1 & 0
    \end{MultiPattern},
    \Color_3 \Color_1 \Color_1 \Color_1 \Color_1 \Color_1 \Color_3},
\end{split} \end{equation}
the multi-pattern
\begin{equation}
    \begin{MultiPattern}
        0 & 1 & \Rest & 2 & \Rest & 1 & \Rest & 1 & \Rest & 0 & 4 \\
        \bar{5} & \Rest & \Rest & \bar{1} & 0 & \Rest & 1 & 0 & \Rest & 1 & 0
    \end{MultiPattern}
\end{equation}
is homogeneously generated by $\BudSystem_{\mathrm{ex}}$.

Algorithm~\ref{algo:homogeneous_random_generation} returns an element homogeneously
generated by $\BudSystem$ obtained by applying at most $k$ rules to the initial element
$(\Color, \Unit, \Color)$. The execution of the algorithm builds a syntax tree of elements
of height at most~$k + 1$.
\begin{center}
    \begin{Page}{.83}
        \begin{algorithm}[H]
            \KwData{A bud generating system
                $\BudSystem := (\Operad, \SetColors, \SetRules, \Color)$ and an integer
                $k \in \N$.}
            \KwResult{A randomly generated element of $\Operad$.}
            \Begin{
                $x \gets (\Color, \Unit, \Color)$ \\
                \For{$j \in [k]$}{
                    \If{$\SetRules \ne \emptyset$}{
                        $r \gets \Random(\SetRules)$ \\
                        $r \gets x \HomogeneousComposition r$
                    }
                }
                \Return{$\Prune(x)$}
            }
            \caption{Homogeneous random generation.}
            \label{algo:homogeneous_random_generation}
        \end{algorithm}
    \end{Page}
\end{center}
Observe that when the set of rules $\SetRules$ has no more than one element, this algorithm
is deterministic.

For instance, by considering the previous bud generating system $\BudSystem_{\mathrm{ex}}$,
this algorithm run with $k := 3$ builds the syntax tree of colored multi-patterns
\begin{equation}
    \begin{tikzpicture}[Centering,xscale=0.4,yscale=0.12]
        \node(1)at(0.00,-12.0){};
        \node(10)at(4.00,-12.0){};
        \node(11)at(4.50,-12.0){};
        \node(13)at(5.50,-18.25){};
        \node(15)at(6.50,-18.25){};
        \node(17)at(7.50,-18.25){};
        \node(19)at(8.00,-12.25){};
        \node(20)at(8.50,-12.25){};
        \node(22)at(9.50,-18.5){};
        \node(24)at(10.50,-12.25){};
        \node(3)at(1.50,-12.25){};
        \node(6)at(2.50,-18.5){};
        \node(8)at(3.50,-18.5){};
        \node[NodeST](0)at(0.00,-6.25){$\CPat_5$};
        \node[NodeST](12)at(5.50,-12.50){$\CPat_5$};
        \node[NodeST](14)at(6.50,-12.50){$\CPat_5$};
        \node[NodeST](16)at(7.50,-12.50){$\CPat_3$};
        \node[NodeST](18)at(8.00,-6.25){$\CPat_1$};
        \node[NodeST](2)at(1.50,-6.25){$\CPat_3$};
        \node[NodeST](21)at(9.50,-12.50){$\CPat_5$};
        \node[NodeST](23)at(10.50,-6.25){$\CPat_5$};
        \node[NodeST](4)at(4.00,0.00){$\CPat_1$};
        \node[NodeST](5)at(2.50,-12.50){$\CPat_5$};
        \node[NodeST](7)at(3.50,-12.50){$\CPat_3$};
        \node[NodeST](9)at(4.00,-6.25){$\CPat_1$};
        \draw[Edge](0)--(4);
        \draw[Edge](1)--(0);
        \draw[Edge](10)--(9);
        \draw[Edge](11)--(9);
        \draw[Edge](12)--(9);
        \draw[Edge](13)--(12);
        \draw[Edge](14)--(18);
        \draw[Edge](15)--(14);
        \draw[Edge](16)--(18);
        \draw[Edge](17)--(16);
        \draw[Edge](18)--(4);
        \draw[Edge](19)--(18);
        \draw[Edge](2)--(4);
        \draw[Edge](20)--(18);
        \draw[Edge](21)--(18);
        \draw[Edge](22)--(21);
        \draw[Edge](23)--(4);
        \draw[Edge](24)--(23);
        \draw[Edge](3)--(2);
        \draw[Edge](5)--(9);
        \draw[Edge](6)--(5);
        \draw[Edge](7)--(9);
        \draw[Edge](8)--(7);
        \draw[Edge](9)--(4);
        \node(r)at(4.00,4.69){};
        \draw[Edge](r)--(4);
    \end{tikzpicture}
\end{equation}
which produces, when evaluated, the multi-pattern
\begin{equation}
    \begin{MultiPattern}
        0 & 1 & \Rest & 1 & 2 & \Rest & 2 & \Rest & 1 & 5 & \Rest & 0 & 1 & \Rest & 1
            & \Rest & 0 & 4 & 4  \\
        \bar{5} & \Rest & \Rest & \bar{1} & \bar{5} & \Rest & \Rest & \bar{1} & 0 & 0 & 0
            & \bar{5} & \Rest & \Rest & \bar{1} & 0 & 0 & 0 & 0
    \end{MultiPattern}.
\end{equation}

\section{Applications} \label{sec:applications}
In this part, we introduce some basic notions about an implementation of the bud music box
model. Next, we construct three particular bud generating systems devoted to work with the
algorithms introduced in Section~\ref{subsec:bud_generating_systems}. They generate
variations of multi-patterns placed at input, with possibly some auxiliary data. All these
constructions are illustrated with some small examples. Finally, we use these constructions
to propose a bud generating system for the random generation of a complete musical piece.

\subsection{The Bud Music Box system}
The music box model, the related operads, and the related random generation algorithms have
been implemented by the author. The program, named {\Code{Bud Music Box}}, as well as its
complete documentation are available at~\cite{Gir23}. We review some of the main
instructions of the language thus designed, useful to understand the examples of the next
sections.

\subsubsection{Context instructions}
Here are the main instructions of the language used to specify the context in which the
multi-patterns will be interpreted:
\begin{itemize}
    \item \Code{{\bf scale} {\it i\_1} \dots\ {\it i\_k}} sets the scale specified by the
    sequence \Code{{\it i\_1} \dots\ {\it i\_k}} designating the distances in semitones
    between two consecutive notes. The multi-patterns of the program are interpreted through
    this scale.
    \item \Code{{\bf root} {\it note}} sets the root note specified by the MIDI code
    \Code{{\it note}}. The multi-patterns of the program are interpreted so that the degree
    $0$ corresponds with the specified MIDI note.
    \item \Code{{\bf tempo} {\it v}} sets the tempo specified by the value \Code{{\it v}}
    designating a tempo in beats per minute. The multi-patterns of the program are
    interpreted through this specified tempo.
    \item \Code{{\bf sounds} {\it s\_1} \dots\ {\it s\_k}} sets the sounds specified by the
    sequence \Code{{\it s\_1} \dots\ {\it s\_k}} of General MIDI programs (which are
    nonnegative integers). The code \Code{{\it s\_i}} designates the MIDI instrument for the
    $i$-th voice of each multi-pattern of the program.
    \item \Code{{\bf monoid} {\it dm}} set \Code{{\it dm}} as the degree monoid on which the
    multi-patterns of the program are defined. The possible values for \Code{{\it dm}} are
    \Code{{\bf add}} for the degree monoid $\Z$, \Code{{\bf cyclic} {\it k}} for the degree
    monoid $\MonoidCyclic_k$, and \Code{{\bf max} {\it z}} for the degree monoid
    $\MonoidMax_Z$ where $Z$ is the set of the integers greater than or equal to~$z$.
\end{itemize}

\subsubsection{Multi-pattern operation instructions}
Here are the main instructions of the language used to define, modify, and generate
multi-patterns:
\begin{itemize}
    \item \Code{{\bf multi-pattern} {\it mpat} {\it description}} creates a multi-pattern
    identified by \Code{\it mpat}. It is specified by \Code{\it description} which encodes a
    multi-pattern by the sequence of its patterns separated by \Code{+} symbols, where
    degrees are signed integers and rests are encoded by \Code{.} symbols.
    \item \Code{{\bf stack} {\it mpat} {\it mpat\_1} \dots\ {\it mpat\_r}} creates a
    multi-pattern identified by \Code{\it mpat} obtained by stacking the existing
    multi-patterns identified by \Code{\it mpat\_1}, \dots, \Code{\it mpat\_r}.
    \item \Code{{\bf mirror} {\it mpat} {\it mpat'}} creates a multi-pattern identified by
    \Code{\it mpat} and whose value is the retrograde of an existing multi-pattern
    identified by \Code{\it mpat'}.
    \item \Code{{\bf inverse} {\it mpat} {\it mpat'}} creates a multi-pattern identified by
    \Code{\it mpat} and whose value is the inverse image of an existing multi-pattern
    identified by \Code{\it mpat'}.
    \item \Code{{\bf colorize {\it cmpat}} {\it \%col\_out} {\it mpat}
        {\it \%col\_in\_1} \dots\ {\it \%col\_in\_n}}
    creates a colored multi-pattern identified by \Code{\it cmpat} and whose value is the
    triple having \Code{\it \%col\_out} as output color, the existing multi-pattern
    identified by \Code{\it mpat} as multi-pattern, and the sequence \Code{{\it
    \%col\_in\_1} \dots\ {\it \%col\_in\_n}} as input colors.
    \item \Code{{\bf mono-colorize {\it cmpat}} {\it \%col\_out} {\it mpat}
        {\it \%col\_in}}
    creates a colored multi-pattern identified by \Code{\it cmpat} and whose value is the
    triple having \Code{\it \%col\_out} as output color, the existing multi-pattern
    identified by \Code{\it mpat} as multi-pattern, and the sequence made of the single
    color \Code{\it \%col\_in} as input colors. This sequence has automatically the right
    size according to the arity of the multi-pattern identified by \Code{\it mpat}.
    \item \Code{{\bf generate} {\it mpat} {\it mode} {\it k} {\it \%col\_init}
        {\it cmpat\_1} \dots\ {\it cmpat\_r}}
    creates a multi-pattern identified by \Code{\it mpat} and whose value is a multi-pattern
    randomly generated by the partial (resp.\ full, homogeneous) generation algorithm if
    \Code{\it mode} is equal to \Code{{\bf partial}} (resp.\ \Code{{\bf full}}, \Code{{\bf
    homogeneous}}), run with its parameter $k$ equal to \Code{\it k} on the bud generating
    system $\Par{\OperadMP^D_m, \SetColors, \SetRules, \Color}$ defined as follows. The
    degree monoid $D$ is the one which has been specified in the context. The set of rules
    $\SetRules$ consists of the existing colored multi-patterns identified by \Code{\it
    cmpat\_1}, \dots, \Code{\it cmpat\_r}, $m$ is the common multiplicity of the
    multi-patterns appearing in $\SetRules$, $\Color$ is the color \Code{\it \%col\_init},
    and $\SetColors$ is the smallest set of colors containing the involved colors.
\end{itemize}

\subsection{Particular bud generating systems} \label{subsec:particular_systems}
We introduce here monochrome bud generating systems, which are basically bud generating
systems where colors do not play any role, and use these to construct a bud generating
system devoted to randomly generate mixes of multi-patterns. We describe also three bud
generating systems devoted to randomly generate multi-patterns from a single pattern and
some auxiliary data.

\subsubsection{Monochrome bud generating systems}
A bud generating system $\BudSystem$ is \Def{monochrome} if its set of colors is a singleton
$\{\Color\}$. In this case, the fact that there is only one color implies that there is no
constraint for the application of the rules of $\BudSystem$. Any pair $(\Operad, R)$ where
$\Operad$ is an operad and $R$ is a finite subset of $\Operad$ specifies the monochrome bud
generating system $\Par{\Operad, \{\Color\}, \SetRules, \Color}$ where $\SetRules$ is the
set of the rules $\Par{\Color, x, \Color^{|x|}}$ for all $x \in R$.

\subsubsection{Style emulation}
Given a set $M$ of multi-patterns all of the same multiplicity $m$, the \Def{mix bud
generating system} $\BudSystemMix_M$ of $M$ is the monochrome bud generating system
specified by the pair $\Par{\OperadMP^\Z_m, M}$. The partial, full, and homogeneous random
generation algorithms run with $\BudSystemMix_M$ randomly generate multi-patterns obtained
by composing elements of $M$ together. Such a multi-pattern $\MPat$ potentially inherits
characteristics from the multi-patterns of $M$. Hence, $\BudSystemMix_M$ can be seen as a
style emulation device.

For instance, consider the multi-patterns
\begin{equation}
    \MPat_1 :=
    \begin{MultiPattern}
        0 & \Rest & 0 & 1 & 2 & \bar{1} & \Rest & \bar{1} & 0 & 1 & \bar{2} & \Rest & 0 \\
        \bar{3} & \Rest & 0 & 1 & 2 & \bar{4} & \Rest & \bar{1} & 0 & 1 & 2 & \bar{3} &
            \Rest
    \end{MultiPattern}
    \quad
    \mbox{and}
    \quad
    \MPat_2 :=
    \begin{MultiPattern}
        0 & 4 & 0 & 5 & \Rest & 1 & 2 & 0 \\
        0 & 4 & 0 & \Rest & \bar{2} & 1 & 2 & 0
    \end{MultiPattern}.
\end{equation}
The partial random generation algorithm run with $\BudSystemMix_M$ where $M := \Bra{\MPat_1,
\MPat_2}$ and $k := 3$ as inputs produces the multi-pattern
\begin{equation}
    \begin{MultiPattern}
        0 & 4 & 0 & 5 & \Rest & 1 & 2 & 0 & \Rest & 0 & 1 & 2 & \bar{1} & \Rest & \bar{1} &
        0 & \Rest & 0 & 1 & 2 & \bar{1} & \Rest & \bar{1} & 0 & 1 & \bar{2} & \Rest & 0 & 1
        & \bar{2} & \Rest & 0 \\
        0 & 4 & 0 & \Rest & \bar{2} & 1 & 2 & \bar{3} & \Rest & 0 & 1 & 2 & \bar{4} & \Rest
        & \bar{1} & \bar{3} & \Rest & 0 & 1 & 2 & \bar{4} & \Rest & \bar{1} & 0 & 1 & 2 &
        \bar{3} & \Rest & 1 & 2 & \bar{3} & \Rest
    \end{MultiPattern}.
\end{equation}
This multi-pattern interprets as the musical phrase
\begin{abc}[name=ExampleMix,width=0.85\abcwidth]
X:
T:
C:
M: 4/4
K: Am
Q: 1/4=120
V: 1 clef=treble
L: 1/4
A, E A, F z B, C A, z A, B, C G, z G, A, z A, B, C G, z G, A, B, F, z A, B, F, z A,
V: 2 clef=bass
L: 1/4
A, E A, z F, B, C E, z A, B, C D, z G, E, z A, B, C D, z G, A, B, C E, z B, C E, z
\end{abc}
Here is the bud music box program associated with this example:
\begin{center}
    \fcolorbox{ColA}{ColA!10}{%
    \Code{
        \ {\bf scale} 2 1 2 2 1 2 2 \\
        \ {\bf root} 57 \\
        \ {\bf tempo} 120 \\
        \ {\bf sounds} 0 0 \\
        \ {\bf monoid} {\bf add} \\
        \ {\bf multi-pattern} m1
            0 .\ 0 1 2 -1 .\ -1 0 1 -2 .\ 0 + -3 .\ 0 1 2 -4 .\ -1 0 1 2 -3 .\ \\
        \ {\bf multi-pattern} m2 0 4 0 5 .\ 1 2 0 + 0 4 0 .\ -2 1 2 0 \\
        \ {\bf mono-colorize} c1 \%b m1 \%b \\
        \ {\bf mono-colorize} c2 \%b m2 \%b \\
        \ {\bf generate} res {\bf partial} 3 \%b c1 c2
    }}
\end{center}

\subsubsection{Horizontal transformations}
Given a pattern $\Pat$ and a sequence $\Psi$ of length $q \geq 1$ of maps on the set of the
patterns, the \Def{$\Psi$-horizontal bud generating system} of $\Pat$ and $\Psi$ is the
monochrome bud generating system
\begin{math}
    \BudSystemHorizontal_{\Pat, \Psi} := \BudSystemMix_M
\end{math}
where $M$ is the set of multi-patterns
\begin{equation}
    \Bra{\Psi(1)(\Pat), \dots, \Psi(q)(\Pat)}.
\end{equation}

For instance, consider the pattern
\begin{math}
    \Pat :=
    \begin{MultiPattern}
        0 & 1 & \Rest & \bar{1} & 0 & \Rest & 2 & 0 & \Rest
    \end{MultiPattern}
\end{math}
and the sequence $\Psi$ of length $3$ of maps such that $\Psi(1)$ is the identity map,
$\Psi(2)$ is the retrograde map, and $\Psi(3)$ is the inverse map (see
Section~\ref{subsubsec:operations}). The partial random generation algorithm run with
$\BudSystemHorizontal_{\Pat, \Psi}$ and $k := 4$ as inputs produces the multi-pattern
\begin{equation}
    \begin{MultiPattern}
        0 & \bar{1} & \Rest & 1 & 2 & \Rest & 0 & 1 & \Rest & 3 & 1 & \Rest & 0 & \Rest &
        \bar{2} & 0 & 1 & \Rest & \bar{1} & \Rest & 0 & 2 & \Rest & 0 & \bar{1} & \Rest & 1
        & 0 & \Rest & 2 & 0 & \Rest & \Rest
    \end{MultiPattern}.
\end{equation}
This multi-pattern interprets as the musical phrase
\begin{abc}[name=ExampleHorizontal,width=0.85\abcwidth]
X:
T:
C:
M: 4/4
K: Am
Q: 1/4=120
V: 1 clef=treble
L: 1/4
A, G, z B, C z A, B, z D B, z A, z F, A, B, z G, z A, C z A, G, z B, A, z C A, z z
\end{abc}
Here is the bud music box program associated with this example:
\begin{center}
    \fcolorbox{ColA}{ColA!10}{%
    \Code{
        \ {\bf scale} 2 1 2 2 1 2 2 \\
        \ {\bf root} 57 \\
        \ {\bf tempo} 120 \\
        \ {\bf sounds} 0 0 \\
        \ {\bf monoid} {\bf add} \\
        \ {\bf multi-pattern} p 0 1 .\ -1 0 .\ 2 0 .\ \\
        \ {\bf mirror} p1 p \\
        \ {\bf inverse} p2 p \\
        \ {\bf mono-colorize} c \%b p \%b \\
        \ {\bf mono-colorize} c1 \%b p1 \%b \\
        \ {\bf mono-colorize} c2 \%b p2 \%b \\
        \ {\bf generate} res {\bf partial} 4 \%b c c1 c2
    }}
\end{center}

\subsubsection{Vertical transformations}
Given a pattern $\Pat$ and a sequence $\Lambda$ of length $q \geq 1$ of homogeneous
operations (see Section~\ref{subsubsec:operations}) on the set of the patterns, the
\Def{$\Lambda$-vertical bud generating system} of $\Pat$ and $\Lambda$ is the monochrome bud
generating system
\begin{math}
    \BudSystemVertical_{\Pat, \Lambda} := \BudSystemMix_M
\end{math}
where $M$ is the singleton containing the multi-pattern
\begin{equation} \label{equ:homogeneous_map_application}
    \Lambda(\Pat) := \Lambda(1)(\Pat) \Conc \dots \Conc \Lambda(q)(\Pat).
\end{equation}

For instance, consider the pattern
\begin{math}
    \Pat :=
    \begin{MultiPattern}
        0 & 1 & \Rest & \bar{1} & 0 & \Rest & 2 & 0 & \Rest
    \end{MultiPattern}
\end{math}
and the sequence $\Lambda$ of length $2$ of homogeneous operations such that $\Lambda(1)$ is
the identity map and $\Lambda(2)$ is the retrograde inverse map. The partial random
generation algorithm run with $\BudSystemVertical_{\Pat, \Lambda}$ and $k := 4$ as inputs
produces the multi-pattern
\begin{equation}
    \begin{MultiPattern}
        0 & 1 & 2 & \Rest & 0 & 1 & 2 & \Rest & 0 & 1 & \Rest & 3 & 4 & \Rest & 2 & 3 &
        \Rest & 5 & 3 & \Rest & 1 & \Rest & \Rest & 3 & 1 & \Rest & \Rest & \bar{1} & 0 &
        \Rest & 2 & 0 & \Rest \\
        \Rest & 0 & \Rest & \bar{2} & \bar{4} & \Rest & \bar{2} & \Rest & \bar{1} & \bar{3}
        & \Rest & \bar{1} & 0 & \Rest & \Rest & \bar{2} & \bar{4} & \Rest & \bar{2} &
        \bar{1} & \Rest & \bar{3} & \bar{2} & \bar{1} & \Rest & \bar{3} & \bar{2} & \Rest &
        0 & 1 & \Rest & \bar{1} & 0
    \end{MultiPattern}.
\end{equation}
This multi-pattern interprets as the musical phrase
\begin{abc}[name=ExampleVertical,width=.85\abcwidth]
X:
T:
C:
M: 4/4
K: Am
Q: 1/4=120
V: 1 clef=treble
L: 1/4
A, B, C z A, B, C z A, B, z D E z C D z F D z B, z z D B, z z G, A, z C A, z
V: 2 clef=bass
L: 1/4
z A, z F, D, z F, z G, E, z G, A, z z F, D, z F, G, z E, F, G, z E, F, z A, B, z G, A,
\end{abc}
Here is the bud music box program associated with this example:
\begin{center}
    \fcolorbox{ColA}{ColA!10}{%
    \Code{
        \ {\bf scale} 2 1 2 2 1 2 2 \\
        \ {\bf root} 57 \\
        \ {\bf tempo} 120 \\
        \ {\bf sounds} 0 0 \\
        \ {\bf monoid} {\bf add} \\
        \ {\bf multi-pattern} p 0 1 .\ -1 0 .\ 2 0 .\ \\
        \ {\bf mirror} p1 p \\
        \ {\bf inverse} p2 p \\
        \ {\bf mirror} p3 p2 \\
        \ {\bf stack} m p p3 \\
        \ {\bf mono-colorize} c \%b m \%b \\
        \ {\bf generate} res {\bf partial} 4 \%b c
    }}
\end{center}

\subsubsection{Local variations}
Given a pattern $\Pat$ and a set $M := \Bra{\MPat_1, \dots, \MPat_q}$, $q \geq 0$, of
multi-patterns of multiplicity $m$, the \Def{variation bud generating system} of $\Pat$ and
$M$ is the bud generating system
\begin{math}
    \BudSystemVariation_{\Pat, M}
    := \Par{\OperadMP_m^\Z, \SetColors, \SetRules, \Color_1}
\end{math}
where $\SetColors$ is the set of colors $\Bra{\Color_1, \Color_2, \Color_3}$ and $\SetRules$
is the set of rules containing the colored multi-patterns
\begin{equation}
    \Par{\Color_1, \MorphismCopy_m(\Pat), \Color_2^{|\Pat|}},
    \Par{\Color_2, \MorphismCopy_m(\Pat), \Color_2^{|\Pat|}},
    \Par{\Color_2, \MPat_1, \Color_3^{\Brr{\MPat_1}}},
    \dots,
    \Par{\Color_2, \MPat_q, \Color_3^{\Brr{\MPat_q}}},
    \Par{\Color_3, \MorphismCopy_m(0), \Color_3}.
\end{equation}
The partial, full, and homogeneous random generation algorithms run with
$\BudSystemVariation_{\Pat, M}$ randomly generate multi-patterns of multiplicity $m$
obtained by self-composing $\MorphismCopy_m(\Pat)$ and then, by transforming some beats of
the obtained multi-pattern by composing them with elements of $M$. The color $\Color_1$ is
the initial color in order to start the generation with $\MorphismCopy_m(\Pat)$, $\Color_2$
is an intermediate color, and the color $\Color_3$ prevents compositions of patterns of~$M$
with $\MorphismCopy_m(\Pat)$ or with other patterns of $M$ in order to not create
degenerated results. The rule $\Par{\Color_3, \MorphismCopy_m(0), \Color_3}$ is important
since this rule admits $\Color_3$ as output color, what is required in the case where this
bud generating system is used with the full random generation algorithm.

Observe that a chord of $M$ acts by changing one beat of the current generated pattern by an
harmonized version of it. Similarly, a flat multi-pattern of $M$ acts by changing one beat
of the current generated pattern by a scheme of beats having the same degree. Finally, an
arpeggio of $M$ acts by changing one beat of the current generated pattern by an arpeggiated
version of it.

For instance, consider the multi-patterns
\begin{equation}
    \Pat :=
    \begin{MultiPattern}
        0 & 1 & \Rest & \bar{1} & 0 & \Rest & 2 & 0 & \Rest
    \end{MultiPattern},
    \qquad
    \MPat_1 :=
    \begin{MultiPattern}
        0 & 0 & \Rest \\
        \Rest & 0 & 0
    \end{MultiPattern},
    \quad \mbox{and} \quad
    \MPat_2 :=
    \begin{MultiPattern}
        0 \\
        4
    \end{MultiPattern}.
\end{equation}
The full random generation algorithm run with $\BudSystemVariation_{\Pat, M}$ where $M :=
\Bra{\MPat_1, \MPat_2}$ and $k := 2$ as inputs produces the multi-pattern
\begin{equation}
    \begin{MultiPattern}
        0 & 1 & 2 & \Rest & 0 & 1 & \Rest & 3 & 1 & \Rest & \Rest & \bar{1} & \bar{1} &
        \Rest & 0 & \Rest & 2 & 3 & \Rest & 1 & 2 & \Rest & 4 & 2 & \Rest & 0 & 0 & \Rest &
        \Rest \\
        4 & 1 & 2 & \Rest & 0 & 1 & \Rest & 3 & 1 & \Rest & \Rest & \Rest & \bar{1} &
        \bar{1} & 4 & \Rest & 2 & 3 & \Rest & 1 & 2 & \Rest & 4 & 2 & \Rest & \Rest & 0 & 0
        & \Rest
    \end{MultiPattern}.
\end{equation}
This multi-pattern interprets as the musical phrase
\begin{abc}[name=ExampleVariation,width=.85\abcwidth]
X:
T:
C:
M: 4/4
K: Am
Q: 1/4=120
V: 1 clef=treble
L: 1/4
A, B, C z A, B, z D B, z z G, G, z A, z C D z B, C z E C z A, A, z z
V: 2 clef=treble
L: 1/4
E B, C z A, B, z D B, z z z G, G, E z C D z B, C z E C z z A, A, z
\end{abc}
Here is the bud music box program associated with this example:
\begin{center}
    \fcolorbox{ColA}{ColA!10}{%
    \Code{
        \ {\bf scale} 2 1 2 2 1 2 2 \\
        \ {\bf root} 57 \\
        \ {\bf tempo} 120 \\
        \ {\bf sounds} 0 0 \\
        \ {\bf monoid} {\bf add} \\
        \ {\bf multi-pattern} p 0 1 .\ -1 0 .\ 2 0 .\ \\
        \ {\bf stack} m p p \\
        \ {\bf multi-pattern} m1 0 0 .\ + .\ 0 0 \\
        \ {\bf multi-pattern} m2 0 + 4 \\
        \ {\bf multi-pattern} u 0 + 0 \\
        \ {\bf mono-colorize} c1 \%b1 m \%b2 \\
        \ {\bf mono-colorize} c2 \%b2 m \%b2 \\
        \ {\bf mono-colorize} cm1 \%b2 m1 \%b3 \\
        \ {\bf colorize} cm2 \%b2 m2 \%b3 \\
        \ {\bf colorize} cu \%b3 u \%b3 \\
        \ {\bf generate} res {\bf full} 2 \%b1 c1 c2 cm1 cm2 cu
    }}
\end{center}

\subsection{Generating structured pieces}
With the aim to use bud generating systems to produce complete pieces, we first define a
notion of composition of bud generating systems. Then, we use the constructions presented in
Section~\ref{subsec:particular_systems} to propose an example of a piece generated by the
tools presented by this paper.

\subsubsection{Composition of bud generating systems}
Let $\Operad$ be an operad, $x$ be an element of arity $n \in \N$ of $\Operad$, and
$\BudSystem_i := \Par{\Operad, \SetColors_i, \SetRules_i, \Color_i}$, $i \in [n]$, be bud
generating systems such that for all $i \ne i' \in [n]$, $\SetColors_i$ and
$\SetColors_{i'}$ are disjoint. The \Def{composition} of $x$ with $\BudSystem_1$, \dots,
$\BudSystem_n$ is the bud generating system
\begin{math}
    x \circ \Han{\BudSystem_1, \dots, \BudSystem_n}
    := \Par{\Operad, \SetColors, \SetRules, \Color}
\end{math}
such that
\begin{equation}
    \SetColors
    := \Par{\bigsqcup_{i \in [n]} \SetColors_i} \sqcup \Bra{\Color}
    \quad \mbox{and} \quad
    \SetRules
    := \Par{\bigsqcup_{i \in [n]} \SetRules_i} \sqcup
    \Bra{\Par{\Color, x, \Color_1 \dots \Color_n}}.
\end{equation}
It is easy to see that any element partially (resp.\ fully, homogeneously) generated by $x
\circ \Han{\BudSystem_1, \dots, \BudSystem_n}$ is of the form $x \circ \Han{x_1, \dots,
x_n}$ where for any $i \in [n]$, $x_i$ is an element partially (resp.\ fully, homogeneously)
generated by $\BudSystem_i$.

In this context of bud generating systems having $\OperadMP^Z_m$, $m \geq 1$, as ground
operad, this construction is useful to generate complete musical pieces. Indeed, given a
multi-pattern $\MPat \in \OperadMP^\Z_m$ of arity $n \in \N$ and bud generating systems
$\BudSystem_i$, $i \in [n]$, having $\OperadMP^\Z_m$ as ground operads, the partial, full,
and homogeneous random generation algorithms run with $\MPat \circ \Han{\BudSystem_1, \dots,
\BudSystem_n}$ generate multi-patterns formed of $n$ parts such that each $\BudSystem_i$
directs the $i$-th part, and such that $\MPat$ connects all the parts together.

\subsubsection{A complete example} \label{subsubsec:complete_example}
Let us use the composition of bud generating systems and some of the previous constructions
to provide a complete example of the generation of a musical piece. Let $\Lambda$ be the
sequence of length $2$ of homogeneous operations such that $\Lambda(1)$ is the identity map
and $\Lambda(2)$ is the retrograde inverse map. Let the multi-patterns
\begin{equation}
    \MPat :=
    \begin{MultiPattern}
        7 & \bar{1} & 7 & 14 & 2 & 14 \\
        0 & \bar{8} & 0 & 0 & \bar{5} & 0
    \end{MultiPattern},
    \qquad
    \Pat_1 :=
    \begin{MultiPattern}
        0 & \Rest 0 & \bar{2} & 2 & \Rest & \bar{1} & 0
    \end{MultiPattern},
    \qquad
    \Pat_2 :=
    \begin{MultiPattern}
        0 & 2 & \Rest & \bar{2} & 0 & 2 & 0 & \Rest
    \end{MultiPattern},
\end{equation}
and, by using the notation introduced in~\eqref{equ:homogeneous_map_application},
$\Pat_1' := \Lambda\Par{\Pat_1}$ and $\Pat_2' := \Lambda\Par{\Pat_2}$. Let also the bud
generating systems $\BudSystem_1 := \BudSystem_2 := \BudSystemVertical_{\Pat_1, \Lambda}$,
\begin{math}
    \BudSystem_3 := \BudSystem_4
    := \BudSystemMix_{\Bra{\Pat_1', \Pat_2'}},
\end{math}
$\BudSystem_5 := \BudSystem_6 := \BudSystemVertical_{\Pat_2, \Lambda}$,
and
\begin{math}
    \BudSystem :=
    \MPat \circ
    \Han{\BudSystem_1, \BudSystem_2, \BudSystem_3,
    \BudSystem_4, \BudSystem_5, \BudSystem_6}.
\end{math}

Let us discuss about the structure of a multi-pattern generated by any of the three random
generation algorithm run with $\BudSystem$. Since $\MPat$ is of arity $6$, such a
multi-pattern consists of six parts, having the following properties:
\begin{enumerate}
    \item The first one consists in self-composing $\Pat_1'$, where the first voice is
    transposed one octave up (since $\MPat(1)(1) = 7$). 
    \item The second one consists in self-composing $\Pat_1'$ where the first voice is
    transposed one degree down (since $\MPat(2)(1) = \bar{1}$) and the second voice is
    transposed one degree and one octave down (since $\MPat(2)(2) = \bar{8}$).
    \item The third one consists in composing freely $\Pat_1'$ and $\Pat_2'$ together where
    the first voice is transposed one octave up (since
    $\MPat(3)(1) = 7$).
    \item The fourth one consists in composing freely $\Pat_1'$ and $\Pat_2'$ together where
    the first voice is transposed two octaves up (since $\MPat(4)(1) = 14$).
    \item The fifth one consists in self-composing $\Pat_2'$ where the first voice is
    transposed two degrees up (since $\MPat(5)(1) = 2$) and the second voice is transposed
    five degrees down (since $\MPat(5)(2) = \bar{5}$).
    \item The sixth one consists in self-composing $\Pat_2'$ where the first voice is
    transposed two octaves up (since $\MPat(6)(1) = 14$).
\end{enumerate}
Here is the bud music box program associated with this example:
\begin{center}
    \fcolorbox{ColA}{ColA!10}{%
    \Code{
        \ {\bf scale} 2 2 1 2 2 2 1 \\
        \ {\bf root} 60 \\
        \ {\bf tempo} 180 \\
        \ {\bf sounds} 0 0 \\
        \ {\bf monoid} {\bf add} \\
        \ {\bf multi-pattern} p1 0 .\ 0 -2 2 .\ -1 0 + 0 1 .\ -2 2 0 .\ 0 \\
        \ {\bf multi-pattern} p2 0 2 .\ -2 0 2 0 .\ + .\ 0 -2 0 2 .\ -2 0 \\
        \ {\bf multi-pattern} p3 7 -1 7 14 2 14 + 0 -8 0 0 -5 0 \\
        \ {\bf mono-colorize} c1 \%ph\_1 p1 \%ph\_1 \\
        \ {\bf mono-colorize} c2 \%ph\_2 p1 \%ph\_2 \\
        \ {\bf mono-colorize} c3 \%ph\_2 p2 \%ph\_2 \\
        \ {\bf mono-colorize} c4 \%ph\_3 p2 \%ph\_3 \\
        \ {\bf colorize} c5 \%start p3
            \%ph\_1 \%ph\_1 \%ph\_2 \%ph\_2 \%ph\_3 \%ph\_3 \\
        \ {\bf generate} res {\bf full} 3 \%start c1 c2 c3 c4 c5
    }}
\end{center}

\section{Evaluation of the generation algorithms} \label{sec:evaluation}
We have developed a listening study including a questionnaire, intended to gain a clearer
understanding of the musical scope of our generation algorithms. We begin by describing the
questionnaire, then present its results, and finally provide an interpretation.

\subsection{Description of the study}
Let us now offer a detailed description of the questionnaire, outline its modalities, and
present some pertinent data about its participants.

\subsubsection{Practical details, objectives, and participants}
The questionnaire was made available for a month starting from December 2023, and was hosted
on Google Forms. It was broadcast through several channels, including the \textit{Creative
Code Paris} community, the \textit{International Society for Music Information Retrieval}
community, the \textit{Society for Mathematics and Computation in Music} community, the
\textit{Institut de recherche et coordination acoustique/musique}, and the
\textit{musiSorbonne} mailing list. The duration of the questionnaire is approximately ten
minutes. After giving their consent for this study, the participants were asked to specify
their level of expertise and then their areas of expertise. Subsequently, ten tracks were
presented to them for listening.

We have conceived five bud music box programs generating some tracks, and selected five
works composed by humans. Following the listening of each of these pieces, we ask the
following four identical questions to evaluate how they are received:
\begin{enumerate}[label=\ColB{(Q\arabic*)}]
    \item \label{question:appeal}
    \textit{How would you rate the aesthetic appeal of this track? Range: 0 is ``very ugly''
    and 10 is ``very beautiful''.}
    \item \label{question:complexity}
    \textit{How would you rate the general complexity of this track? Range: 0 is ``extremely
    simplistic'' and 10 is ``extremely complex''.}
    \item \label{question:willingness_listen}
    \textit{Would you have liked the track to have lasted longer? For instance, 4 or 5 means
    that listening was a good experience over the proposed duration, but that a longer
    duration might have been less interesting. Range: 0 is ``absolutely not'' and 10 is
    ``definitively''.}
    \item \label{question:human_composed}
    \textit{How likely do you think this track was composed by a human (instead of an
    algorithm)? Range: 0 is ``most likely by an algorithm'' and 10 is ``most likely by a
    human''.}
\end{enumerate}
In addition, a comment space is provided to collect the comments of each participant for
each piece. The scores out of $10$ obtained for each track provide us with an indication of
their perceived quality according to various aspects. The participants, not knowing which
work is composed according to our method or by a human, results in the scores obtained on
all five human works serving as a benchmark to evaluate the works generated by our method.

The questionnaire was filled out by $70$ participants. Among them, $29$ declared having a
level of expertise equal to or greater than $8$ out of $10$. Our analysis of the results is
based solely on the opinions of these $29$ experts. Upon analyzing the declared areas of
expertise, the most frequently occurring fields of expertise are piano practice (nine
times), electronic music (six times), and popular music (four times).

\subsubsection{Musical tracks}
Each offered track lasts $30$ seconds and in any case, the sound begins with a gradual
increase in the first few seconds, then stabilizes, and diminishes in the last few seconds.
In order to standardize the interpretation, each track is played from a MIDI file. Here is
the list of the considered tracks:
\begin{enumerate}[label=\textbullet \ColA{T\arabic*}, ref=\ColA{T\arabic*}]
    \item \label{track:1}
    Computer-generated, \texttt{P\_2\_4.bmb -\,-seed 8}.
    \item \label{track:2}
    Human-composed, Erik Satie, \textit{Vexations}, 1893.
    \item \label{track:3}
    Computer-generated, \texttt{H\_2\_5-2-1-1.bmb -\,-seed 4}.
    \item \label{track:4}
    Computer-generated, \texttt{F\_2\_6-6-6.bmb -\,-seed 0}.
    \item \label{track:5}
    Human-composed, Johann Sebastian Bach, \textit{Duetto in E minor, BWV 802}, 1739.
    \item \label{track:6}
    Computer-generated, \texttt{P\_2\_15-3-1.bmb -\,-seed 5}.
    \item \label{track:7}
    Human-composed, Radan Papezik, \textit{12-tone blues}, 2006.
    \item \label{track:8}
    Human-composed, Erik Satie, \textit{Españaña, Croquis et Agaceries d'un gros bonhomme en
    bois}, 1913.
    \item \label{track:9}
    Computer-generated, \texttt{H\_2\_3-2-1-1.bmb -\,-seed 3}.
    \item \label{track:10}
    Human-composed, Johann Sebastian Bach, \textit{Fugue in C minor, BWV 906b}, 1738.
\end{enumerate}
They are presented in this exact order to all participants. This order was selected at
random when the questionnaire was designed.

As explained in Section~\ref{sec:random_generation}, the bud music box algorithms work
partly in generating musical phrases through performing some mechanical operations on them,
reminiscent of some works by Bach. For this reason, we chose two excerpts~\ref{track:5} and
\ref{track:10} from this composer. Additionally, since some generated phrases sometimes
recall passages of minimal music, we have chosen~\ref{track:2} and~\ref{track:8}, extracted
from works of Satie who is often considered one of the precursors of this trend, although
this is certainly debatable. We have also included~\ref{track:7}, a more recent and
royalty-free composition of Papezik which falls within the serialism movement. Under
particular settings, our generation method can produce passages that are close to this
style.

Our five computer-generated tracks are based upon small sets of small multi-patterns. For
instance, \ref{track:4} results from the program presented in
Section~\ref{subsubsec:complete_example}. We made this paradigmatic choice because, in the
extreme, with very long multi-patterns taken from existing compositions, our algorithms
would generate pieces too close to the original ones. This approach, which prioritizes
distinctiveness over replication, is intentional for the scope of our current evaluation,
although this stance is open to reconsideration, as will be discussed in
Section~\ref{subsubsec:limitations}. Consequently, we aimed to find a middle ground between
the simplicity of the multi-patterns and our subjective perception of the outcomes.
Additionally, it is important to note that our expertise does not lie in the field of
musical composition.

Here is a very important remark: our objective is not to claim that our method can, even
remotely, produce pieces comparable to those ones of human composers. Rather, our aim is to
have standard works that can be used to detect the quality of responses, and whose metrics
can be used as benchmarks to estimate the quality of the computer-generated creations.

\subsection{Results and interpretation}
Here we present some significant data collected during the study. We then offer an
interpretation of this data before presenting the limitations of the study and a conclusion.

\subsubsection{some statistics}
The following histograms present some distributions of responses, ranging from $0$ to $10$,
for each question among~\ref{question:appeal}, \ref{question:complexity},
\ref{question:willingness_listen}, and~\ref{question:human_composed} merged for the
human-composed pieces on the left and for the computer-generated tracks on the right. Orange
thick lines are positioned on the means of the distributions. Approximations of these values
are denoted by ``$m$'' under the histograms. Transparent orange rectangles around the means,
having widths of twice the standard deviations, are also depicted. Approximations of these
values are denoted by ``$\sigma$'' under the histograms.
\begin{itemize}[fullwidth]
    \item Distributions of the aesthetic appeal evaluation:
    \smallbreak

    \Graph{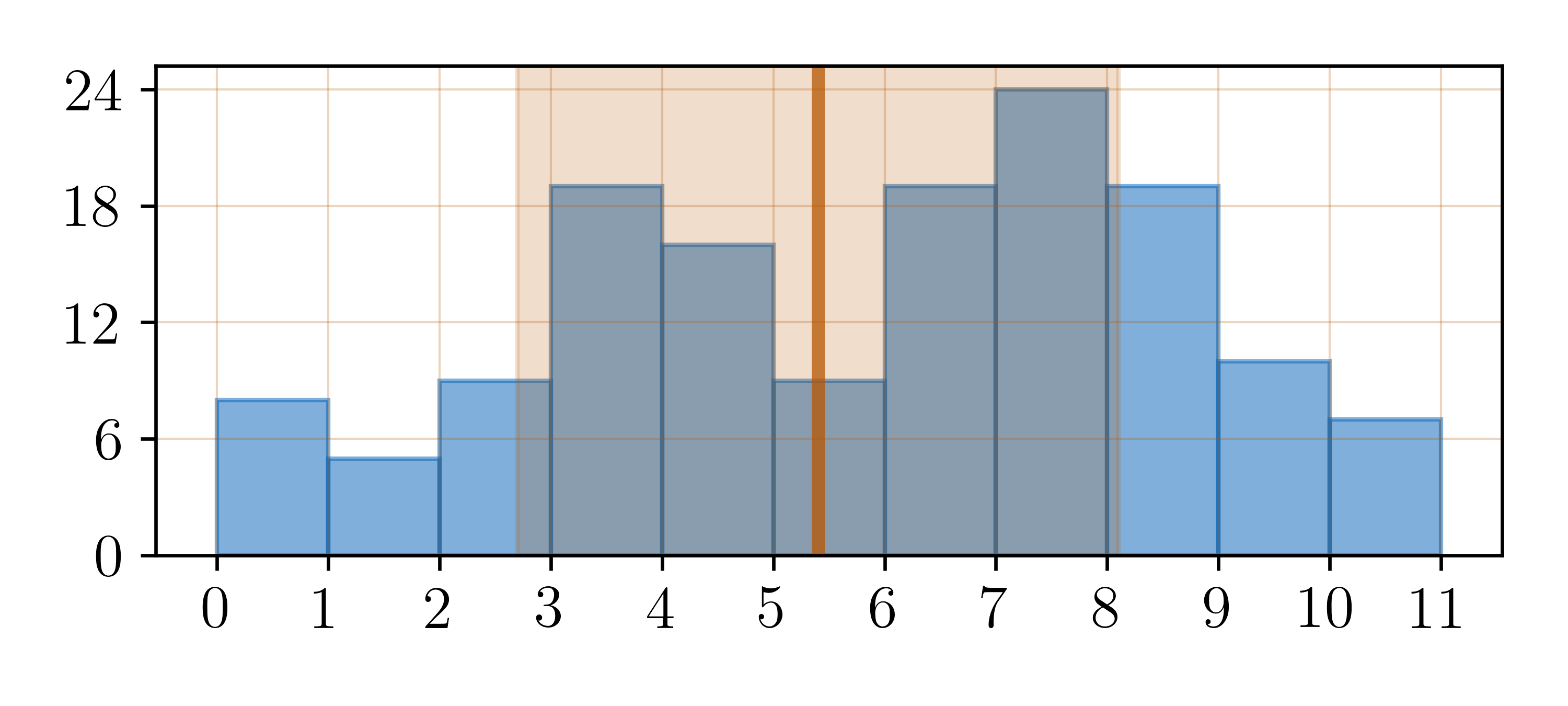}{$m \simeq 5.40$, $\sigma \simeq 2.70$}
    \hfill
    \Graph{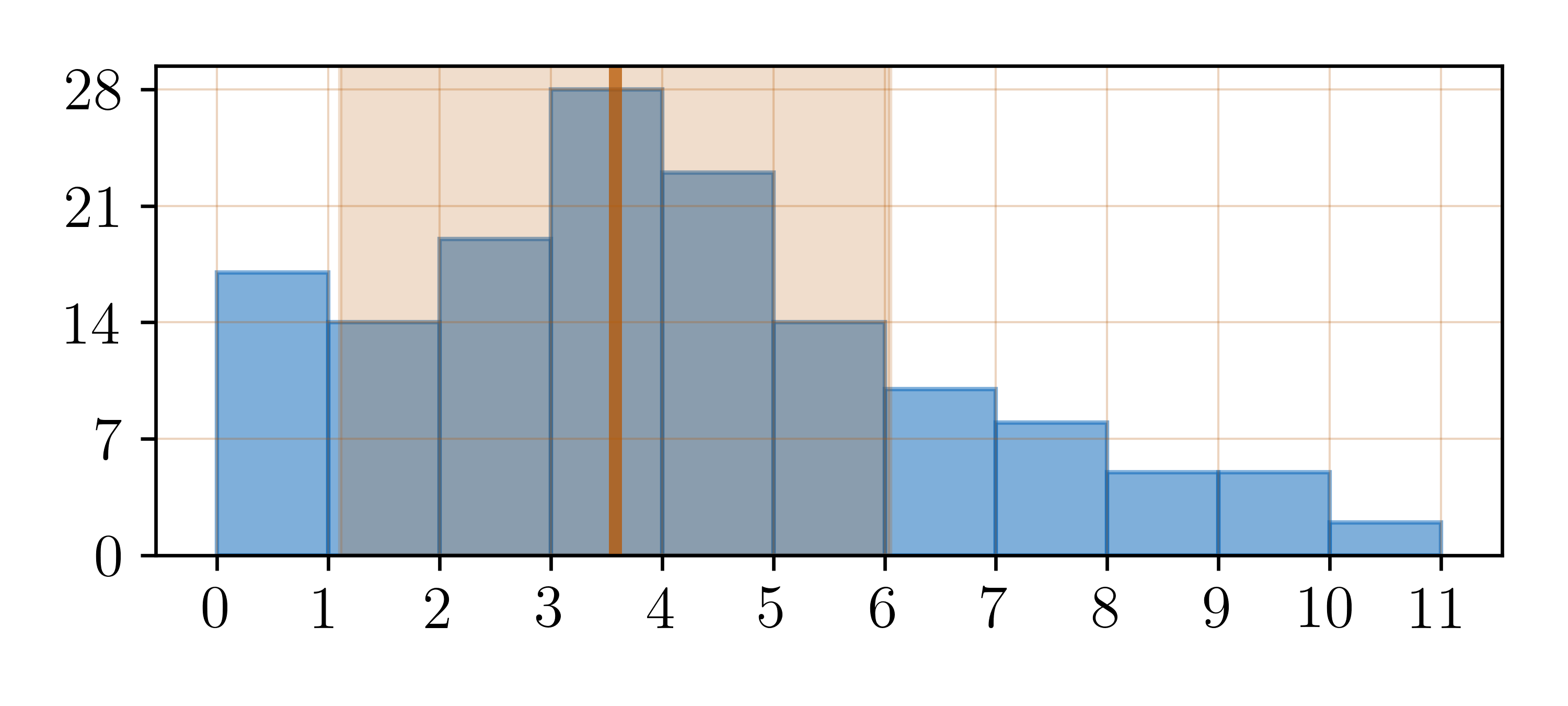}{$m \simeq 3.58$, $\sigma \simeq 2.47$}
    \medbreak
    \item Distributions of the complexity evaluations:
    \smallbreak

    \Graph{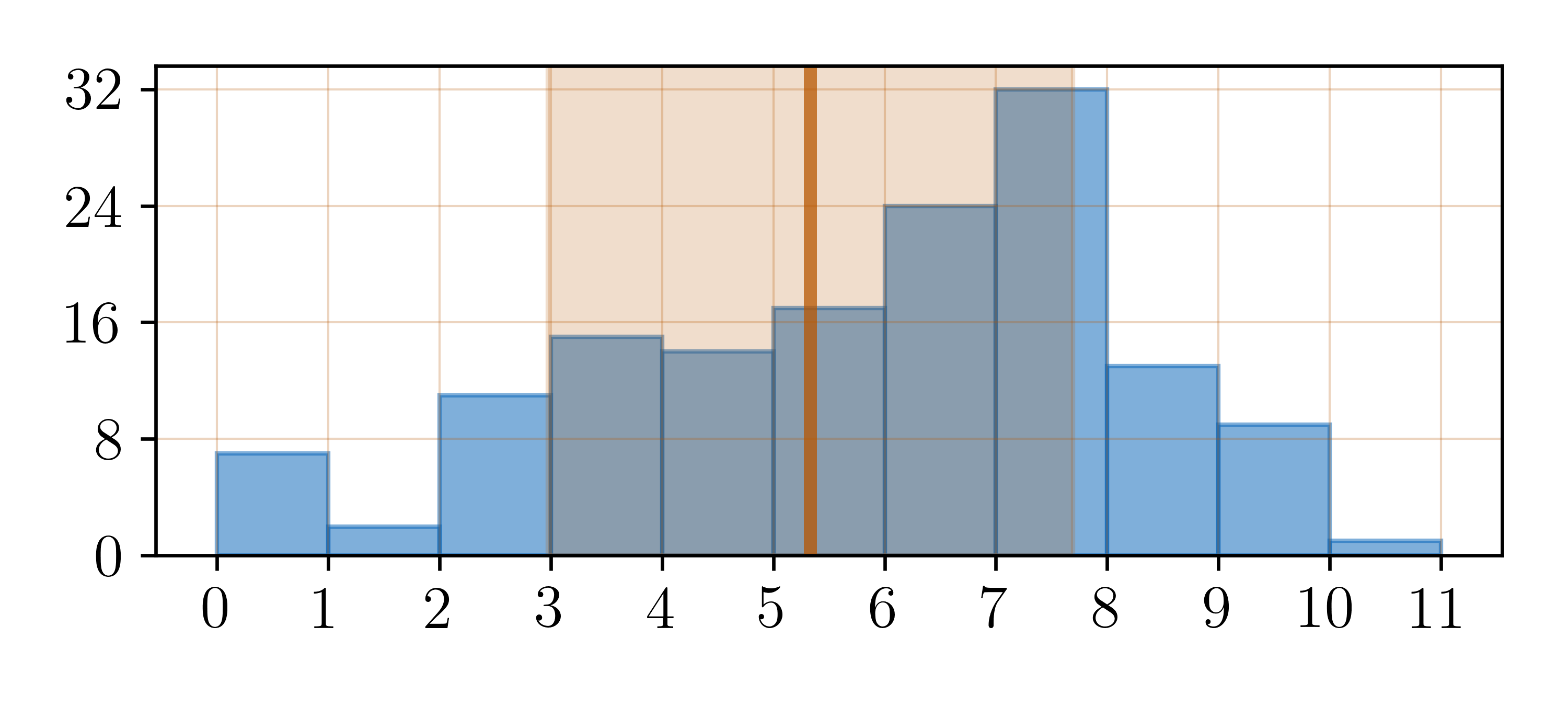}{$m \simeq 5.33$, $\sigma \simeq 2.36$}
    \hfill
    \Graph{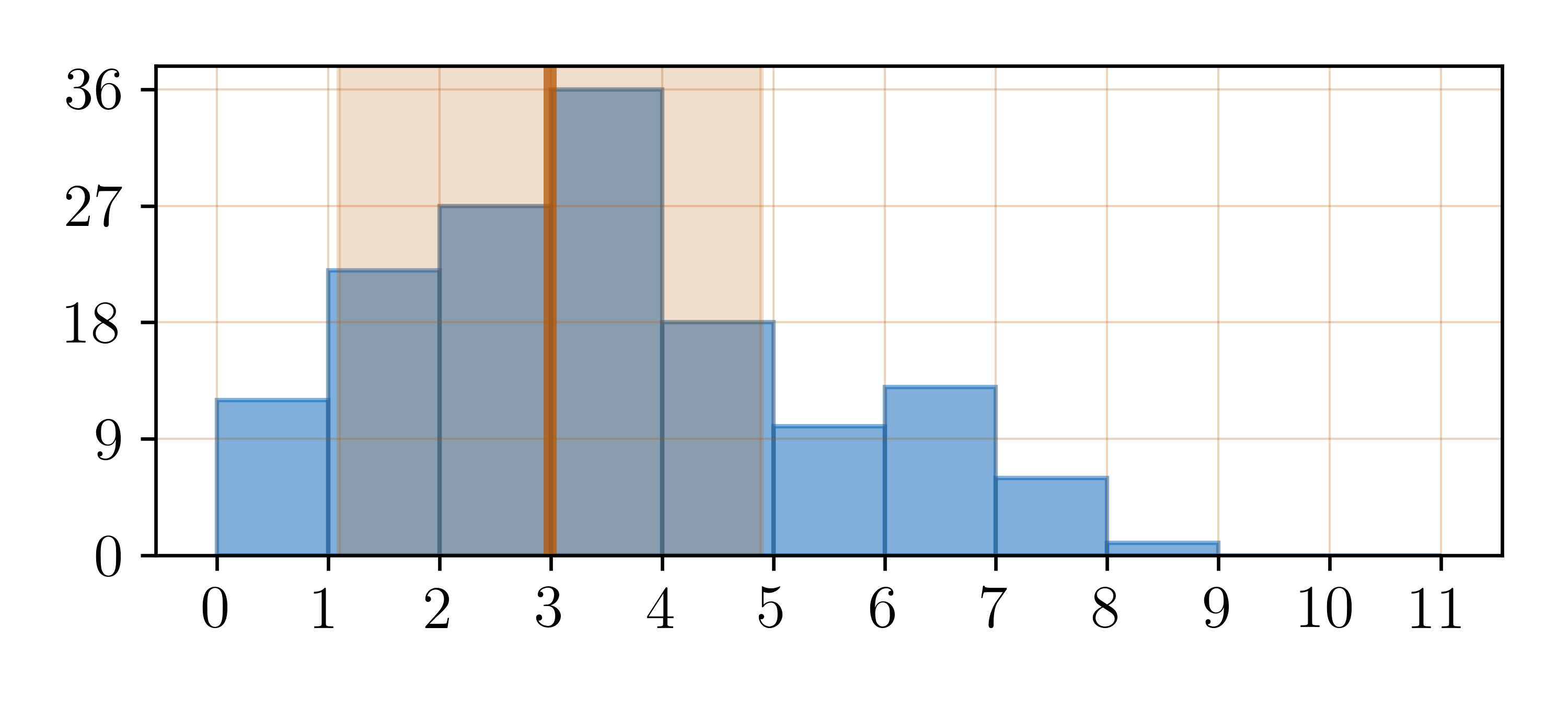}{$m \simeq 2.99$, $\sigma \simeq 1.90$}
    \medbreak
    \item Distributions of the willingness to listen evaluation:
    \smallbreak

    \Graph{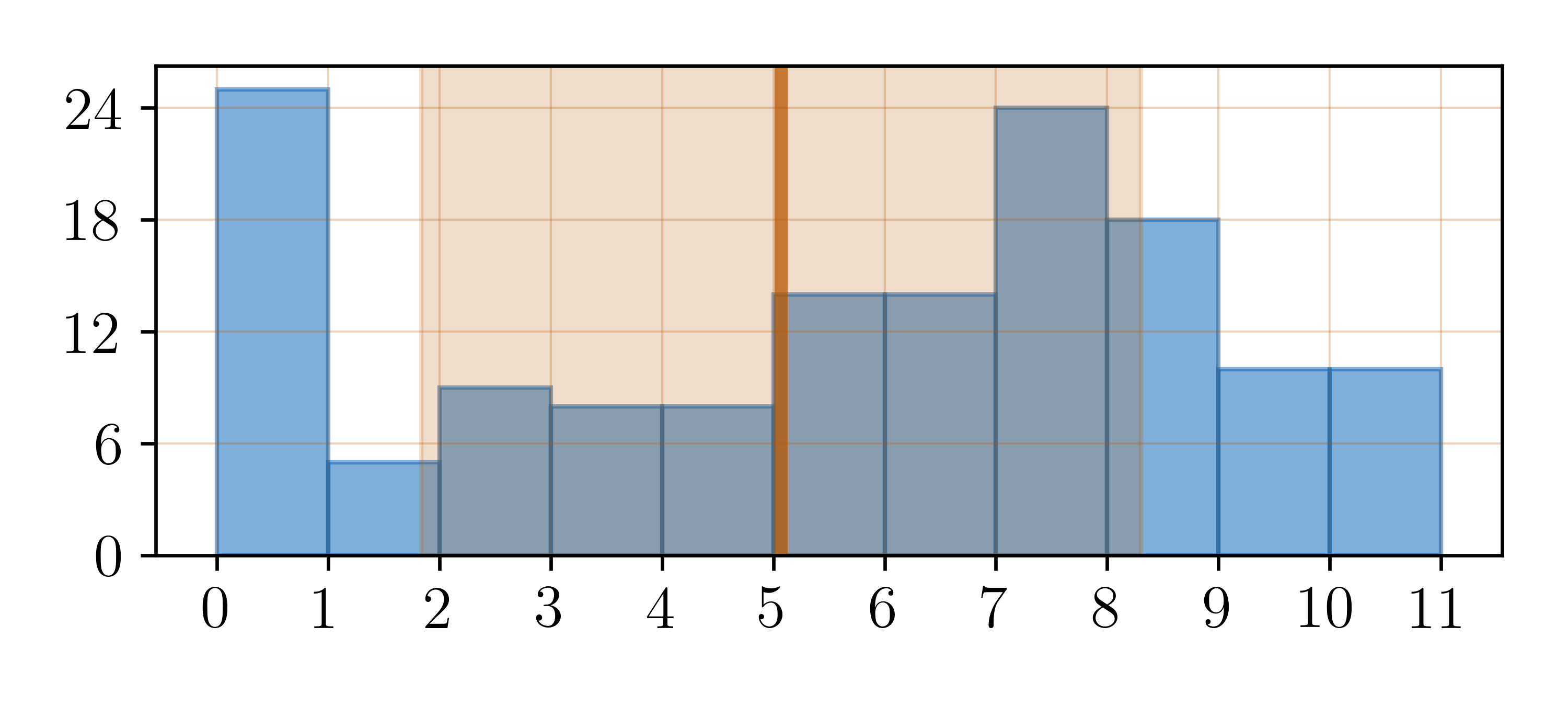}{$m \simeq 5.07$, $\sigma \simeq 3.24$}
    \hfill
    \Graph{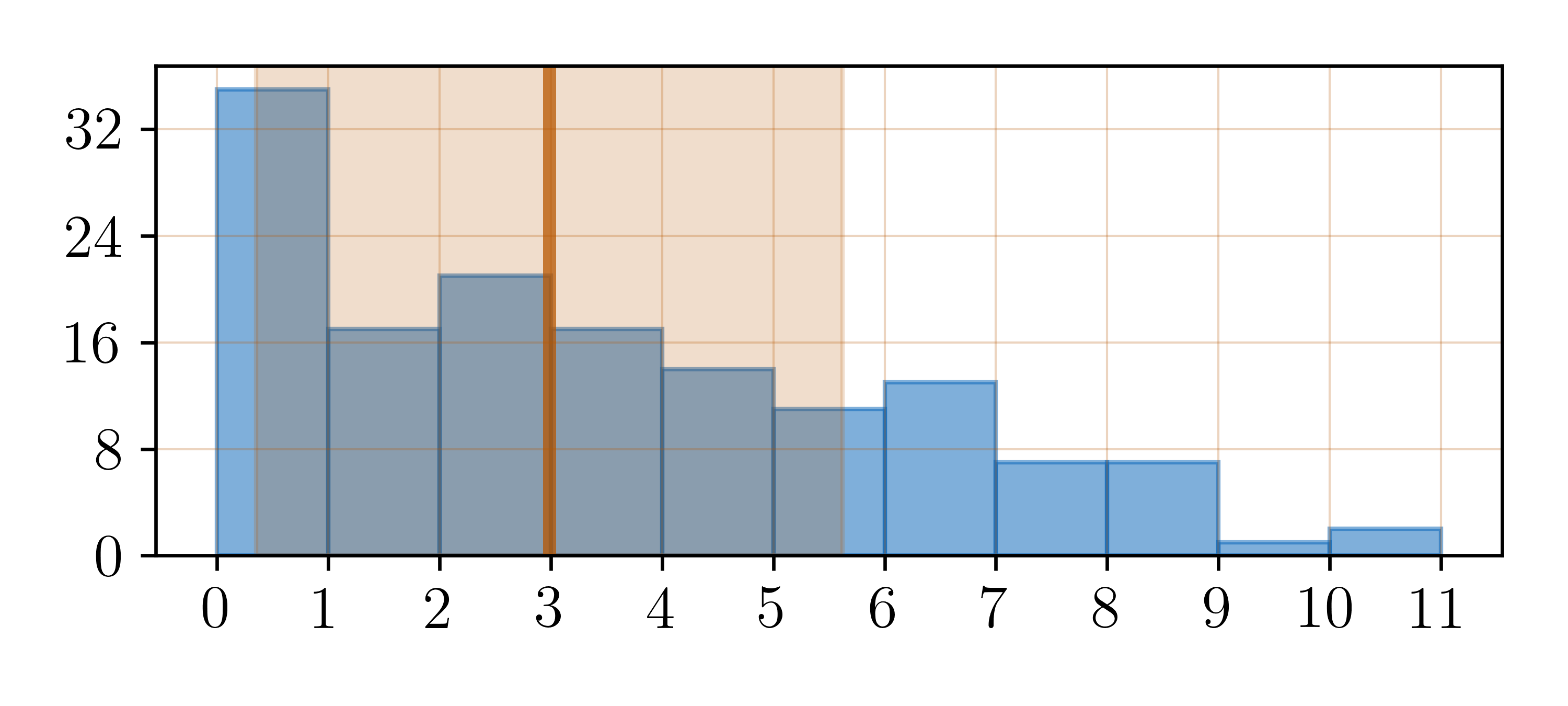}{$m \simeq 2.99$, $\sigma \simeq 2.64$}
    \medbreak
    \item Distributions of the human composition likelihood evaluation:
    \smallbreak

    \Graph{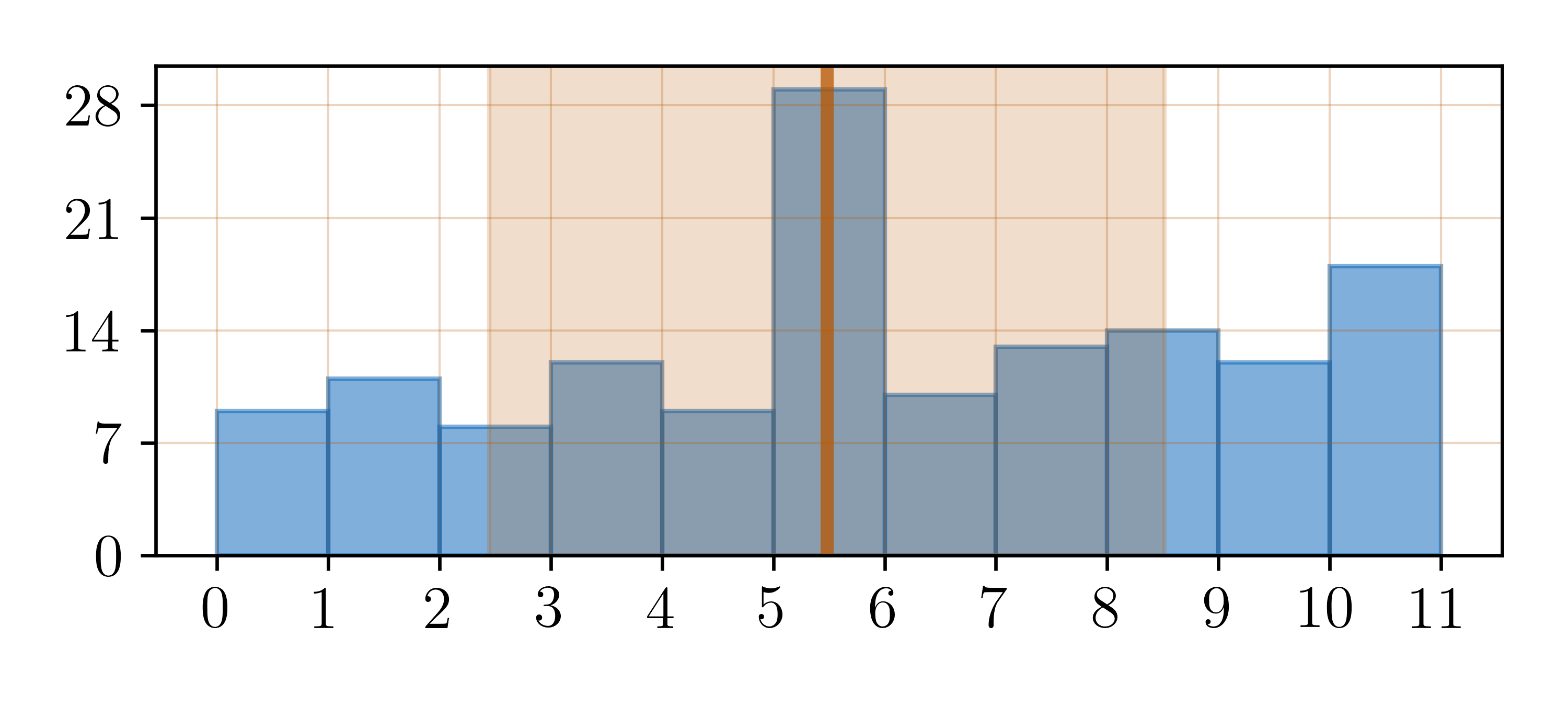}{$m \simeq 5.48$, $\sigma \simeq 3.03$}
    \hfill
    \Graph{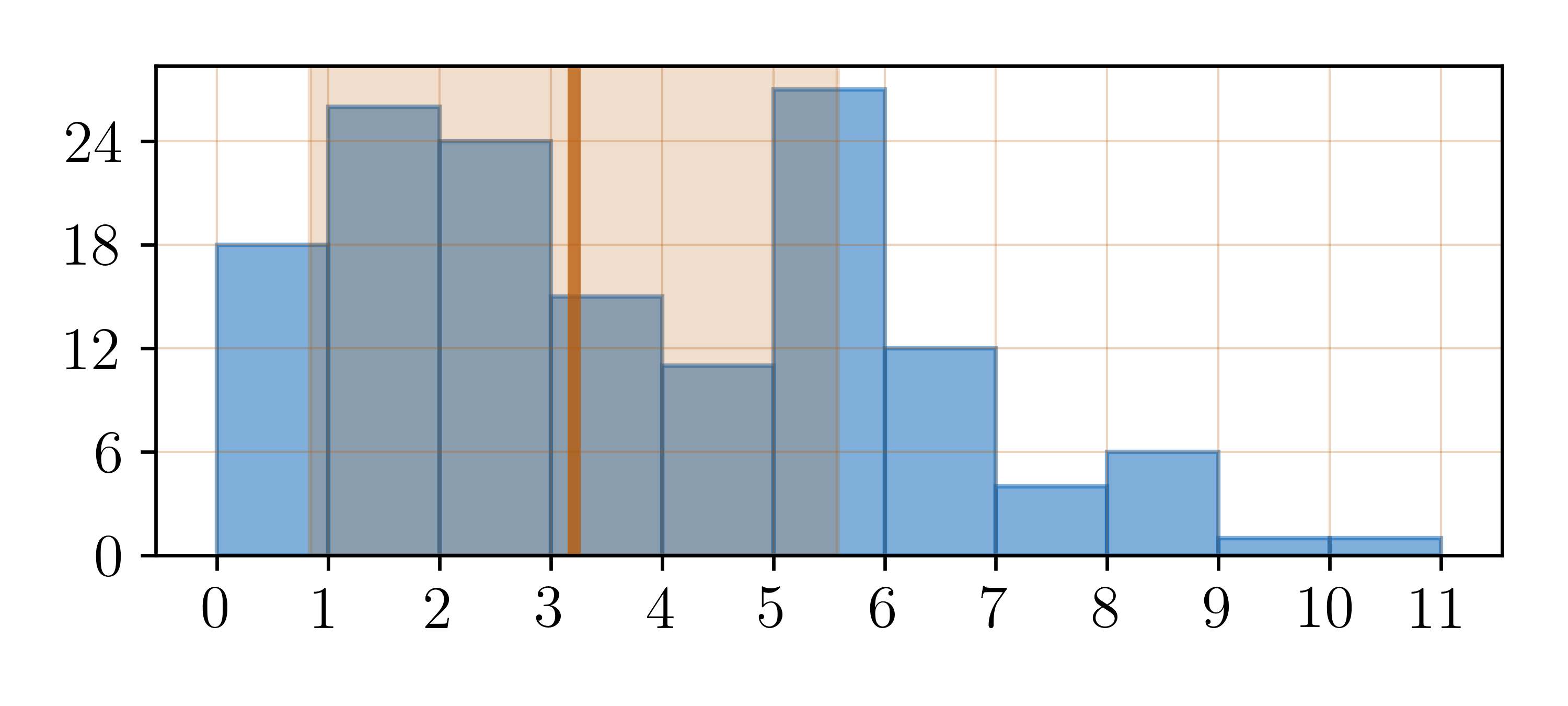}{$m \simeq 3.21$, $\sigma \simeq 2.37$}
\end{itemize}
\medbreak

Let us consider the mean absolute error regarding Question~\ref{question:human_composed},
which asks to classify the tracks according to their origin. By interpreting it as an
accuracy rate, this value is about $55 \% \simeq 5.48 / 10$ on human-composed pieces (since
the expected correct response is $10 / 10$) and about $68 \% \simeq 1 - 3.21 / 10$ on
computer-generated tracks (since the expected correct response is~$0 / 10$).

The following histograms present the ten tracks ranked w.r.t.\ the average rating assigned
by participants for each question among~\ref{question:appeal}, \ref{question:complexity},
\ref{question:willingness_listen}, and~\ref{question:human_composed}:
\begin{multicols}{2}
    \begin{itemize}[fullwidth]
        \item Aesthetic appeal ranking:
        \smallbreak

        \Graph{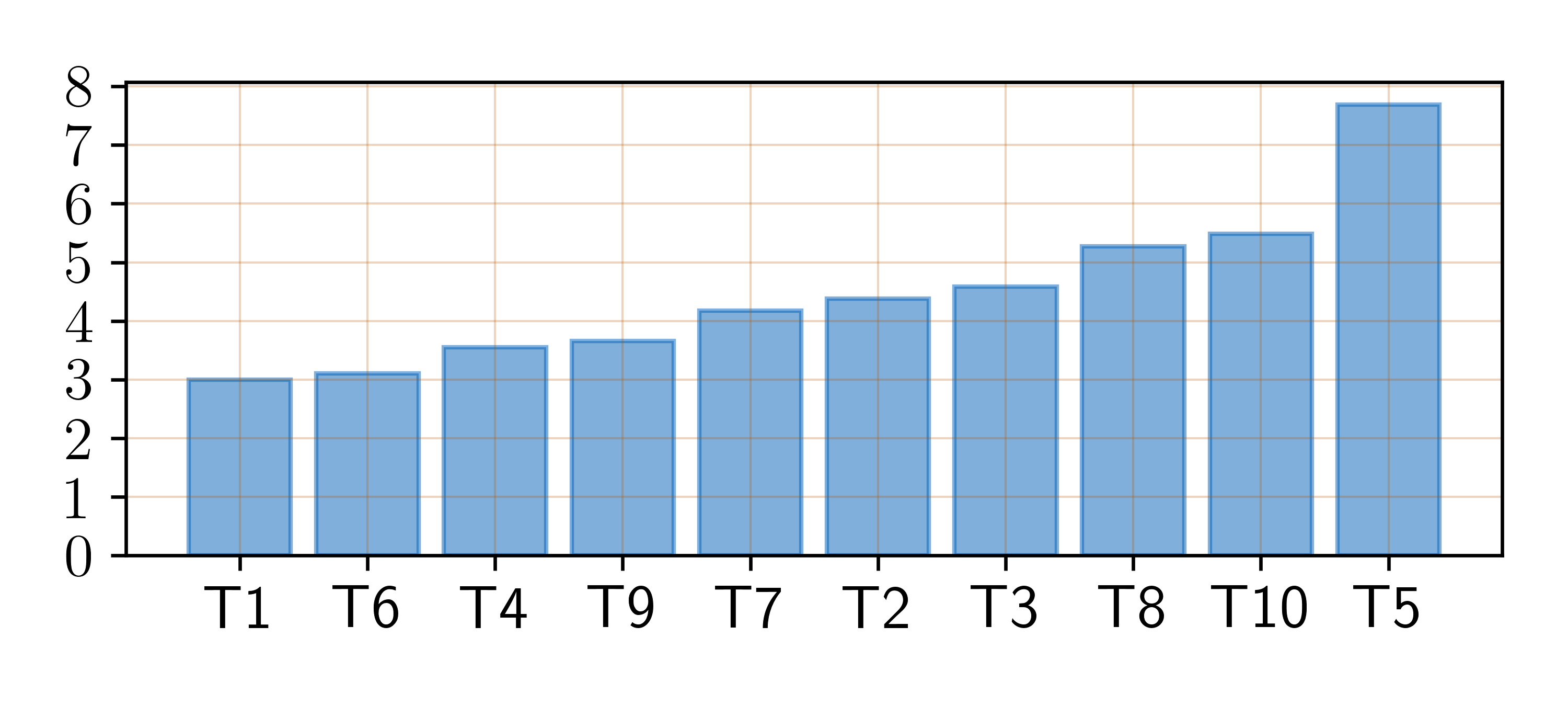}{}
        \medbreak

        \item Complexity ranking:
        \smallbreak

        \Graph{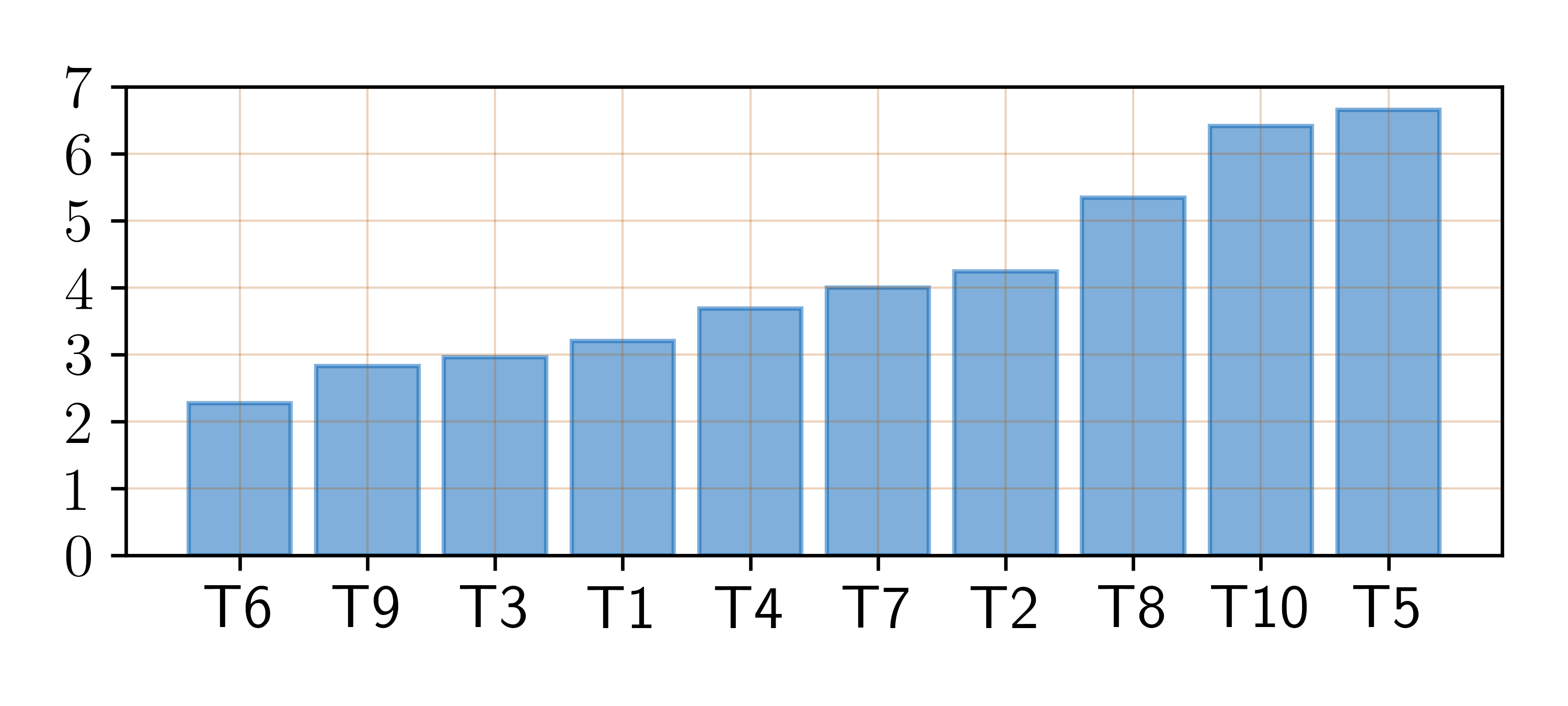}{}

        \item Willingness to listen ranking:
        \smallbreak

        \Graph{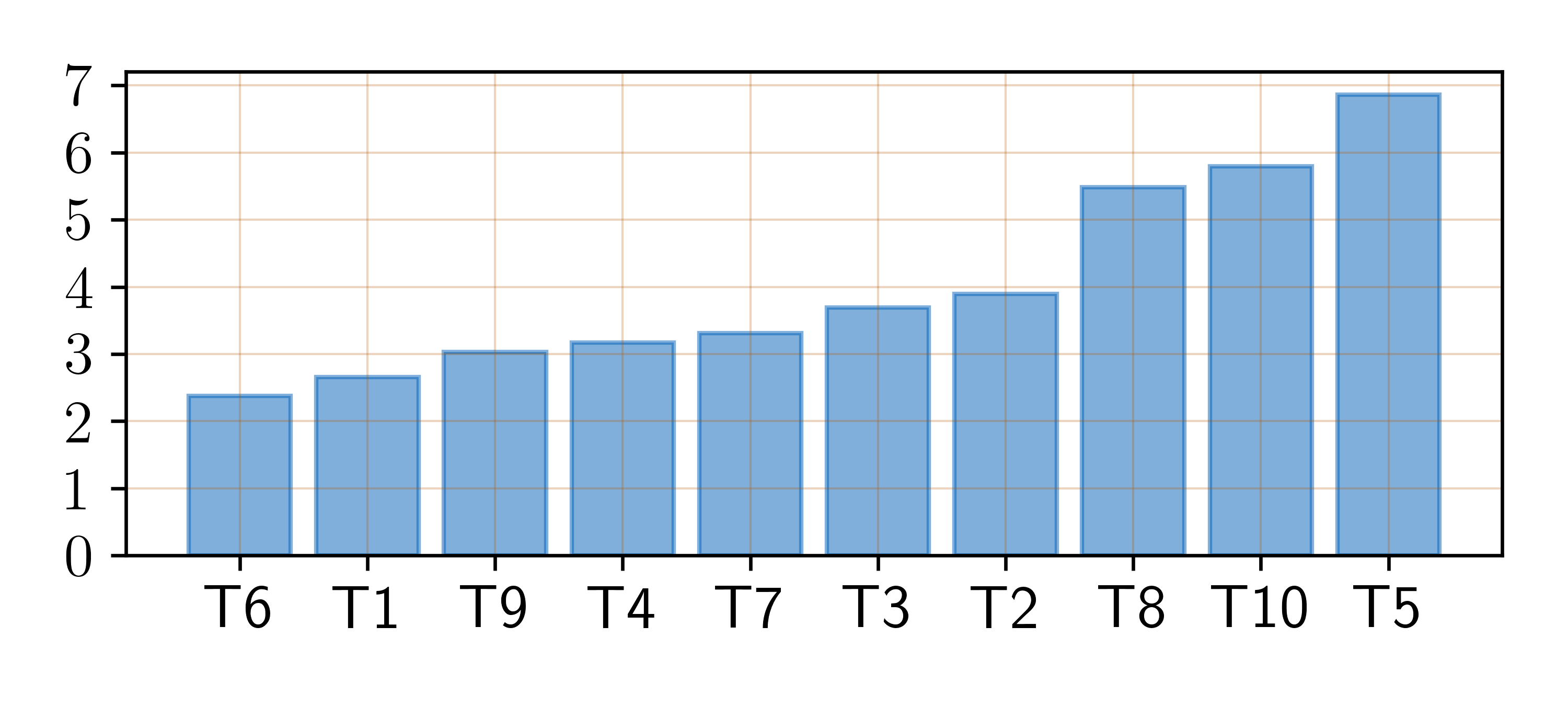}{}
        \medbreak

        \item Human composition likelihood ranking:
        \smallbreak

        \Graph{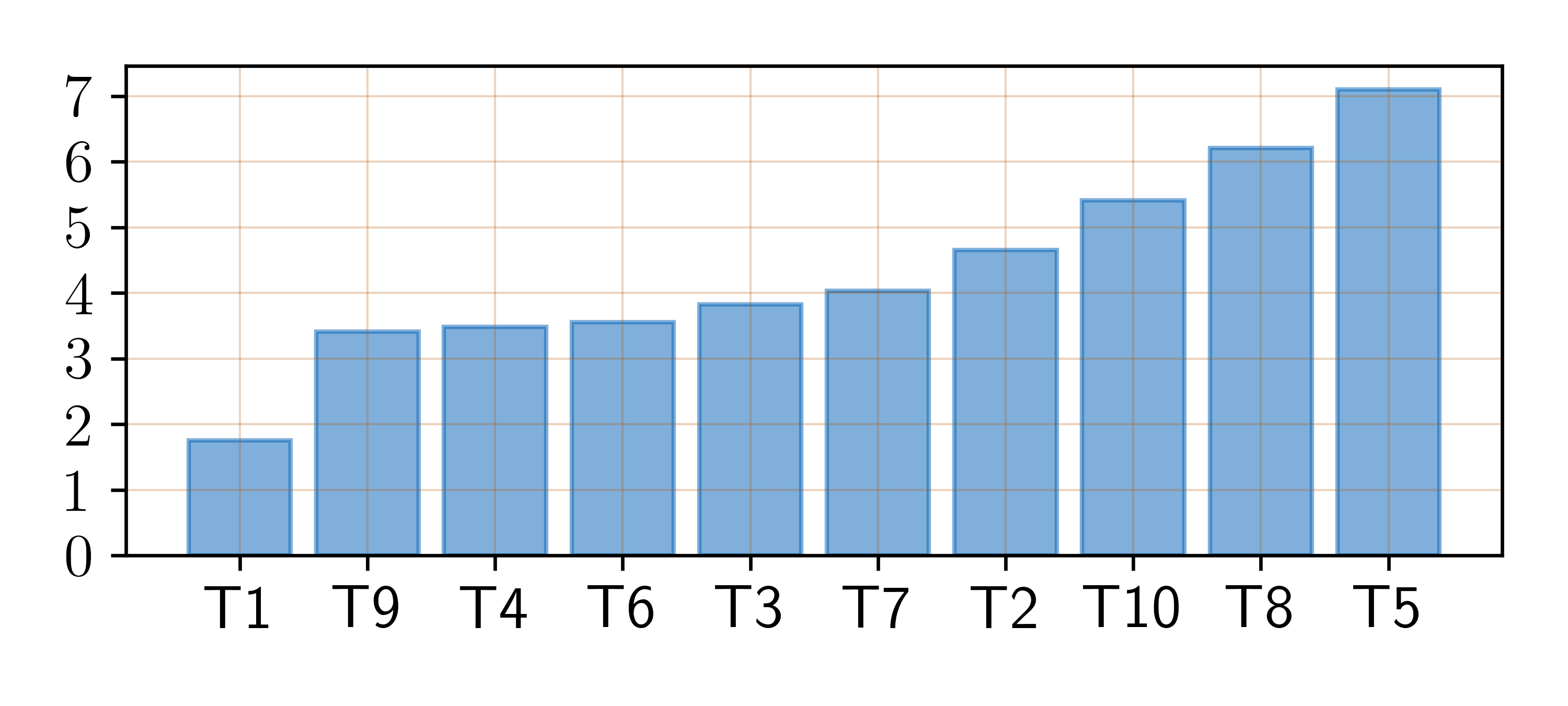}{}
    \end{itemize}
\end{multicols}

Here is the ranking w.r.t.\ the means of the sums of the evaluations of
Questions~\ref{question:appeal}, \ref{question:complexity},
\ref{question:willingness_listen}, and~\ref{question:human_composed}:
\begin{center}
    \Graph{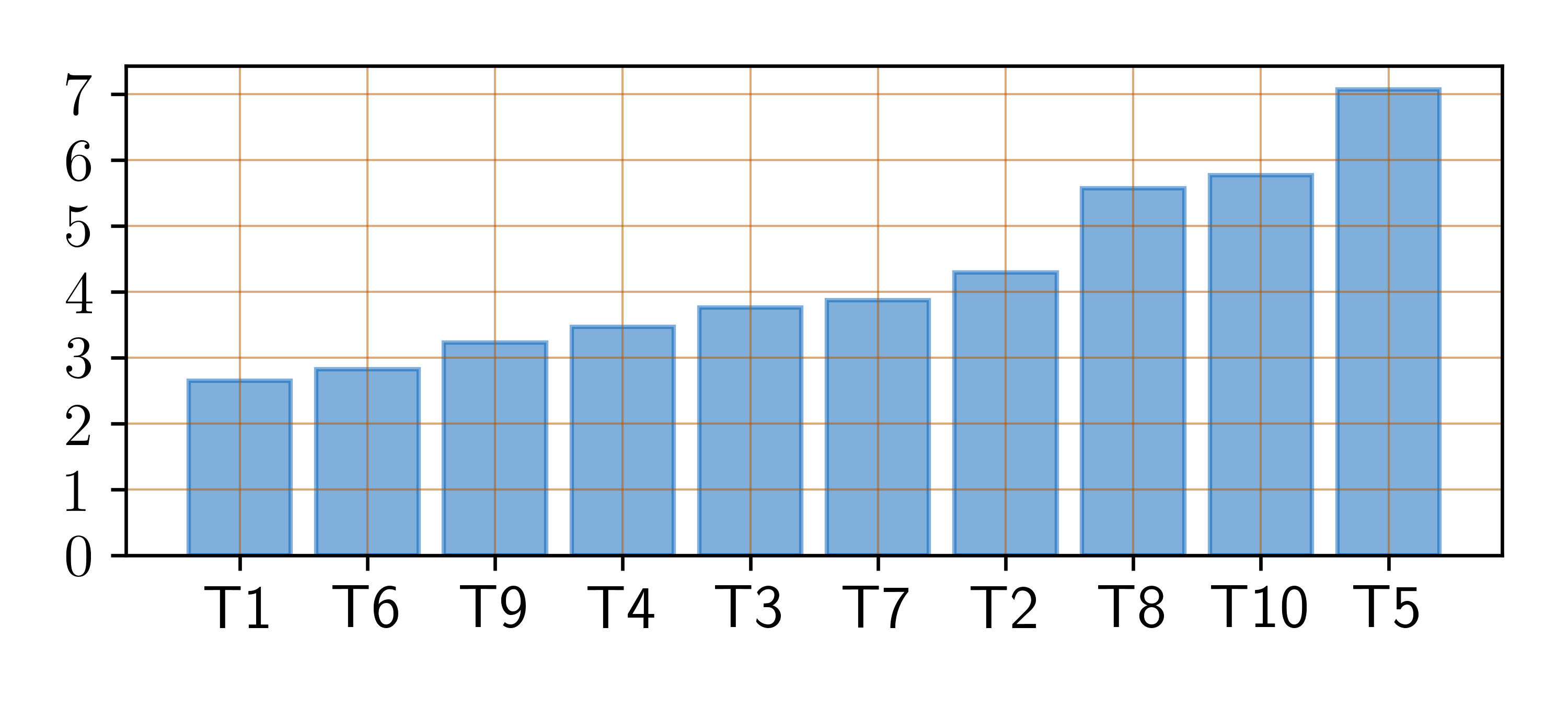}{}
\end{center}
\medbreak

Let us now present a synthesis of the observations of the participants. The total number of
comments left for human-composed pieces is $69$ (approximately $48 \%$ of participants
commented on average on each such track) and the total number of comments left for
computer-generated tracks is $61$ (approximately of $42 \%$ participants commented on
average on each such track).

In evaluating the five human-composed pieces, participants noted contrasts between simple
rhythms and complex harmonies. There was frequent speculation about whether the pieces were
human-composed or algorithmically generated, particularly due to their mechanical tones and
lack of expressive variation. Certain pieces were appreciated for their clear direction and
contrapuntal styles, drawing comparisons to baroque composers like Bach. Opinions remained
divided regarding the authenticity of the compositions, with some tracks being praised for
their sophistication and others critiqued as generic or excessively mechanical in their
emulation of classical styles.

The evaluations of the five tracks generated by our method indicate a general perception of
simplicity and predictability in melody and harmonic structure. The mechanical and synthetic
quality was often highlighted, with listeners expressing a need for more variety and depth,
often finding the compositions too repetitive. The repetitive and predictable nature of some
pieces led to speculation about the possibilities for algorithmic composition. However,
positive elements such as interesting rhythmic patterns, minimalist aspects, gamelan-like
elements, and intriguing intervals were also noted.

\subsubsection{Interpretation}
Let us provide an interpretation of the previous results.

\paragraph{About classification accuracy}
The classification accuracy rate on computer-generated tracks suggests that roughly one
out of every three tracks generated by the bud music box algorithms is perceived as having a
reasonable chance of being composed by a human. Similarly, one out of every two tracks
composed by a human is perceived as having been generated by a computer. This relatively
high rate might be explained by the choice of the sounds forming the tracks, as discussed
further in Section~\ref{subsubsec:limitations}. Therefore, although the classification
accuracy is not exceptional, it is encouraging for our method that nearly a third of the
tracks generated by the machine could be reasonably perceived as human-composed. This
outcome is particularly significant considering the limitations in sound quality.

\paragraph{About the general aesthetic appeal of the tracks}
We observed that the mean aesthetic appeal for all five computer-generated tracks is about
$3.6 / 10$. Conversely, the aesthetic appeal for all five human-composed tracks is about
$5.4 / 10$. This notable quasi two-points difference demonstrates consistency in the
responses since the human-composed tracks served as benchmarks for evaluating the perceived
quality of the tracks generated by our algorithms. This relatively modest rating of $5.4 /
10$ reflects a level of severity and stringency in the responses to
Question~\ref{question:appeal}. Based on participant comments, this severity can be partly
attributed to the quality of the MIDI sounds and automatic interpretations. Nevertheless,
the aesthetic appeal of the tracks generated by our method is reasonably good. It was
particularly noteworthy to find~\ref{track:3} in the fourth position in the overall rankings
of tracks w.r.t.\ their aesthetic appeals.

\paragraph{About the general ranking of the tracks}
To derive a general evaluation of a track, we calculated a score based on the mean of the
sums of the responses received for Questions~\ref{question:appeal},
\ref{question:complexity}, \ref{question:willingness_listen},
and~\ref{question:human_composed}. This approach is deemed relevant as all four dimensions
highlight positive attributes of the tracks. Particularly for the last question, it was
considered coherent to assign a value to a piece proportional to the impression it creates
of being human-composed. All computer-generated tracks were ranked lower than human-composed
pieces. Track~\ref{track:1} ranked the lowest, but notably, \ref{track:3} was very
close to the human-composed track at fifth position. This appears to be influenced by the
relatively low perceived complexity in response to Question~\ref{question:complexity}.
Overall, the best tracks generated by our method are~\ref{track:3} and~\ref{track:4}.

\subsubsection{Limitations} \label{subsubsec:limitations}
Let us state some of the limitations of this study.

\paragraph{On the order of the presented tracks}
A primary concern pertains to the questionnaire design, especially concerning the order in
which the tracks were presented. As mentioned above, the order was randomly selected once
the questionnaire was designed. However, to more effectively mitigate potential bias arising
from the order in which tracks are presented, it would have been preferable to provide each
participant with a uniquely randomized order of tracks.

\paragraph{On the identification of expert participants}
We paid particular attention to the responses from experts in various fields of music, which
was crucial to get significant responses. The identification of experts was based on
self-assessment of their expertise. This approach, however, introduces at least two
significant issues. Firstly, self-assessment is inherently subjective and can be influenced
by an individual level of humility or self-confidence. Consequently, it is probable that
some individuals with expert-level skills were categorized as non-experts. Additionally,
setting a minimum threshold of $8$ out of $10$ for expert classification is arbitrary. While
efforts were made to calibrate responses to this question, the process was not foolproof.

\paragraph{On the sound quality and interpretation of the tracks}
A significant challenge in this study was the quality of the interpretation of the tracks.
As outlined in the questionnaire description, we opted to use MIDI-generated tracks to
ensure that the quality of their interpretation did not overshadow their intrinsic
compositional qualities. However, as highlighted in numerous participant comments, this
decision significantly impacted the reception of the pieces. It is important to note that
due to this limitation, the scores obtained from the study should not be interpreted in
absolute terms, but rather in relative terms. Despite this constraint, the approach was
adequate for the objectives of the current study.

\paragraph{On the musical quality of the computer-generated tracks}
The tempered reception of the participants towards the tracks generated by our algorithms
was previously discussed. However, as highlighted in earlier sections, some characteristics
of the computer-generated pieces were better received than certain human-composed works
(like for instance for~\ref{track:3} which is ranked ahead of~\ref{track:2}
and~\ref{track:7} for Question~\ref{question:appeal}, and ahead of~\ref{track:7} for
Question~\ref{question:willingness_listen}). Of course, this does not necessarily imply
intrinsic value in the algorithmically generated pieces. A preferable approach would have
been to offer tracks of significantly higher quality. As mentioned, the pieces produced by
our algorithms were constructed from small sets of multi-patterns. An advantageous
modification could have involved relaxing this somewhat arbitrary constraint slightly,
allowing the inclusion of phrases from works in the styles we aimed to emulate as
multi-patterns. This approach, being easily implementable, would have yielded a distinctly
different set of pieces.

\subsubsection{Conclusion}
The objective of this evaluation study was successfully achieved. We garnered very consistent
responses from participants, which provided a clearer understanding of the quality of music
generated by the bud music box algorithms. As anticipated, the reception of the
algorithm-generated pieces was less favorable compared to those composed by humans. However,
we now possess a valuable point of comparison, highlighting that certain rhythmic and
harmonic elements were effectively captured. Interestingly, the results revealed not only
expected elements of minimal music but also surprising instances of gamelan music
influences. The primary criticism revolved around the simplicity and repetitiveness of the
pieces. To further advance this research, it would be beneficial for a composer to directly
engage with our algorithm. By merging their expertise with the capabilities of the
algorithm, we could explore high-quality compositions and fully realize the potential of the
method.

\section{Conclusion and perspectives}
In this work, we have introduced the music box model, a framework to represent musical
phrases as multi-patterns, perform computations on these, and various random generation
algorithms. This framework has some strengths and weaknesses as discussed in
Sections~\ref{subsubsec:limitations_music_box_model},
\ref{subsubsec:strengths_music_box_model}, and~\ref{subsubsec:operations}. Our generative
algorithms, derived from this framework, were also the subject of an evaluation, reported in
Section~\ref{sec:evaluation}. Here are some perspectives raised by this work.

The first one consists in the description of a minimal generating set for the operad
$\OperadMP^D_m$, $m \geq 2$, even just for the additive degree monoid $D = \Z$. The
knowledge of such a minimal generating set would provide a way to decompose a multi-pattern
as a syntax tree decorated by generators. This means we can parse a musical phrase into
primitive parts. By using some algorithms coming from operad theory, this would lead to
applications such as the automatic discovery of repeated parts of musical phrases (up to
some elementary transformations like transpositions or more complex ones involving rhythmic
motives), or the compression of musical data.

Another perspective is to consider variations of the operad $\OperadRP$ controlling the
rhythmic part of the multi-patterns. Such a variation can potentially produce very different
results from the present ones when used with bud generating systems. In the present work,
the operad $\OperadRP$ is constructed as the image by the construction $\ConstructionU$ of
the additive monoid on $\N$. It is envisaged to see if it is possible to use other monoids
and to what extent this can lead to new ways of composing rhythm patterns.

\paragraph{Acknowledgments}
We would like to express our gratitude to all the participants who contributed to the study.
Special thanks go to Moreno Andreatta, Grégory Châtel, Chiara Giraudo, and Grégory Jarrige
for their help in distributing the questionnaire.

\MakeReferences

\end{document}